\documentclass{egpubl}
\usepackage{eg2024}
 
\ConferenceSubmission   
\usepackage[T1]{fontenc}
\usepackage{dfadobe}

\usepackage[ruled]{algorithm2e} 

\SetAlFnt{\small}
\SetAlCapFnt{\small}
\SetAlCapNameFnt{\small}
\SetAlCapHSkip{0pt}

\usepackage{cite}  
\BibtexOrBiblatex
\electronicVersion
\PrintedOrElectronic
\ifpdf \usepackage[pdftex]{graphicx} \pdfcompresslevel=9
\else \usepackage[dvips]{graphicx} \fi

\usepackage{egweblnk} 

\usepackage[super]{nth}
\usepackage{amsmath}
\usepackage{amssymb}
\usepackage{bm}
\usepackage{cases}
\usepackage{threeparttable}
\usepackage{enumitem}
\usepackage{subcaption}
\usepackage{wrapfig}
\usepackage{tikz}
\usepackage{tikz-imagelabels}
\usetikzlibrary{arrows}
\usetikzlibrary{math}
\usetikzlibrary{intersections}
\usepackage{siunitx}
\usepackage[inline]{asymptote}
\usepackage{pgfplots}
\pgfplotsset{compat=1.18}
\usepgfplotslibrary{patchplots}

\title[Simulating Parametric Thin Shells by Bicubic Hermite Elements]%
      {Simulating Parametric Thin Shells by Bicubic Hermite Elements}

\author[X.\ Ni et al.]{Xingyu Ni$^*$, Xuwen Chen$^*$, Cheng Yu, Bin Wang, Baoquan Chen\\
\parbox{\textwidth}{\centering $^*$Joint first authors}}

\begin{document}

\teaser{
 \centering
  \includegraphics[trim=0cm 0cm 0cm 0cm,clip,width=1\textwidth]{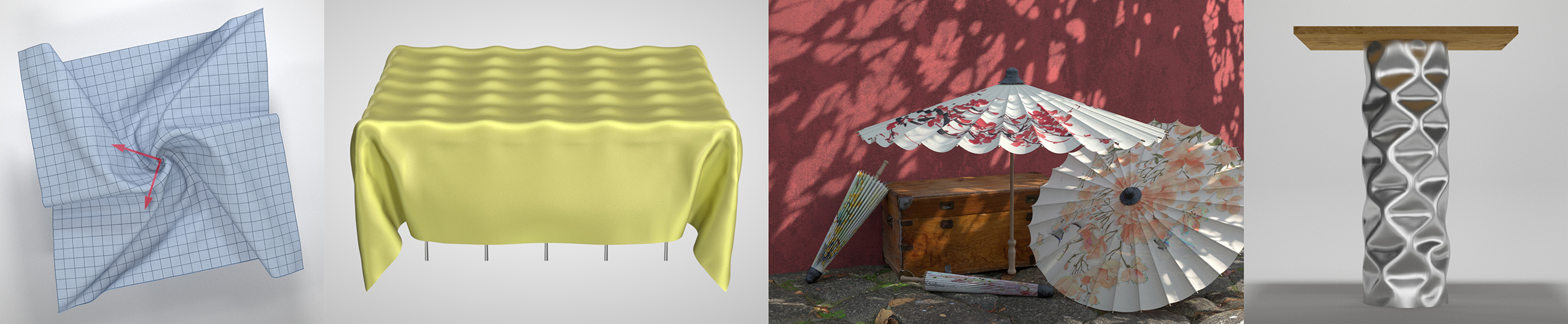}
  \caption{Our computational framework, the Bicubic-Hermite-Element Method (BHEM), can simulate the dynamics of thin shells under various external boundary conditions. From left to right: the sheet twisted through rotating its center point's first derivation (red arrows), with its $30\times30$ simulated patches illustrated; a piece of cloth falls on a needle array and exhibits small bulges pushed out by the needle tips ($30\times30$ patches); Oriental paper parasol  ($280\times5$ patches) folds driven by the prescripted rib motion; Diamond-like buckles manifests over the entire shell surface under the axial compression ($30\times30$ patches).}
\label{fig:teaser}
}

\maketitle

\begin{abstract}
In this study, we present the bicubic Hermite element method (BHEM), a new computational framework devised for the elastodynamic simulation of parametric thin-shell structures. 
The BHEM is constructed based on parametric quadrilateral Hermite patches, which serve as a unified representation for shell geometry, simulation, collision avoidance, as well as rendering.
Compared with the commonly utilized linear FEM, the BHEM offers higher-order solution spaces, enabling the capture of more intricate and smoother geometries while employing significantly fewer finite elements. 
In comparison to other high-order methods, the BHEM achieves conforming $\mathcal{C}^1$ continuity for Kirchhoff–Love (KL) shells with minimal complexity.
Furthermore, by leveraging the subdivision and convex hull properties of Hermite patches, we develop an efficient algorithm for ray-patch intersections, facilitating collision handling in simulations and ray tracing in rendering. 
This eliminates the need for laborious remodeling of the pre-existing parametric surface as the conventional approaches do. 
We substantiate our claims with comprehensive experiments, which demonstrate the high accuracy and versatility of the proposed method.


\end{abstract}  

\section{Introduction}
The mechanical characteristics of thin-shell structures are commonly characterized by the Kirchhoff--Love (KL) theory \cite{Ciarlet2000}, which assumes that transverse shear is negligible and describes the kinetics of a shell through in-plane stretching and lateral bending of its midsurface. 
Under this KL assumption, the energy density function derived from the elastic strain incorporates second-order derivatives of displacements (as elaborated in \S\ref{sec:theory}). This requirement necessitates the $\mathcal{H}^2$ regularity of the geometric representation of midsurface, which implies the $\mathcal{C}^1$ continuity, to ensure a well-defined analysis of the elastic energy.

To effectively address the second-order thin-shell energy,
many endeavors have been
developed, tested, and honed. In the realm of linear finite-element analysis, which is highly favored within the field of computer graphics, a customary solution is to reformulate the bending energy as a specialized function of the dihedral angle based on the discrete differential geometry principle \cite{Bridson2002,Grinspun2003,Bergou2006}. Nevertheless, this discretized edge-based energy allows bending motion along the common edges of elements only and eventually fails to converge towards the shape operator of the smooth surface at the element interfaces, irrespective of the chosen discretization scheme and mesh resolution \cite{Gingold2004}.
Isogeometric Analysis (IGA) \cite{Hughes2005, Cottrell2009} is an appealing alternative to attaining high-order continuity solution space. It utilizes various basis functions emanating from computer-aided geometric design (CAGD) in finite element analysis for their global smoothness.
One of the earliest depictions of this approach was presented by Cirak et al.\ \cite{Cirak2000, Cirak2001}, who devised a finite element formulation grounded in Loop subdivision surfaces for Kirchhoff--Love thin shells simulation. 
Loop subdivision scheme can easily represent smooth surfaces of arbitrary topology with polygonal mesh data structure. By employing the same convergent shape function for displacement field interpolation, the subdivision finite element scheme \cite{Cirak2000, Cirak2001} requires only nodal displacement degrees of freedom while retaining $\mathcal{C}^1$ continuity across elements, which is necessary for thin shell simulation. 
The use of more general non-uniform rational B-splines (NURBS) basis functions in the finite element context was also proposed by Hughes et al.\ \cite{Hughes2005} in 2005. NURBS patch is more memory-friendly than subdivision surface, but it needs additional constraints to maintain conforming $\mathcal{C}^1$ continuity within multi-patch models \cite{Lu2014}. 
Since the smooth surface do not pass through the coarse control mesh nodes for both NURBS and subdivision FEM scheme, applying boundary condition and resolving contacts on control nodes need complicated treatment. 
Hermite elements stand out for their inherent guarantee of $\mathcal{C}^1$ continuity among patches by incorporating shared derivatives as degrees of freedom. This particular type of element has been
demonstrated to be advantageous when solving Kirchhoff plate problems \cite{Phusakulkajorn2013, GRECO2021113476, GRECO2019, Zienkiewicz2005}. 
Even though, the curved Hermite surface exactly goes through the nodal position, existing studies still heavily resort to surface triangulation for collision handling and rendering, at the expense of diminishing the value of high-order methods.

In this paper, in order to fully leverage the precision advantages provided by the high-order method, while concurrently minimizing the modifications to the traditional FEM simulation pipeline, we develop a novel framework for thin-shell simulation. The most crucial aspect of our framework lies in the utilization of a unified bicubic-Hermite-patches-based representation, which serves as the foundation of midsurface geometric modeling, dynamics simulation, and ray-tracing rendering purposes.
Specifically, we present three main contributions to achieving this goal.

First, we develop a $\mathcal{C}^1$-continuous finite element solution for dynamic simulation of Kirchhoff--Love thin shells, dubbed BHEM (Bicubic Hermite Element Method). 
We listed the derivation process of the governing equations and its weak form  (\S\ref{sec:bhem}) of KL thin shells from the first principles of continuum mechanics. In the Hermit polynomial space, the discretized form of the governing equations, along with the gradient and the Hessian matrix (\S\ref{apx:derivative}) of the hyperelastic energy, are provided to facilitate a seamless implicit solve (\S\ref{sec:solver}). Furthermore, the BHEM also incorporates a tailored Hessian matrix which entails much less computational overhead, to cater to applications with restricted time constraints. 
Further consolidating the integration between geometry representation and simulation, and bypassing mesh-based collision detection, our second main contribution is a BH-surface intersection detection algorithm that serves for both CCD and rendering tasks (\S\ref{sec:detection}).
In our method, the bicubic Hermite surfaces are transformed into their equivalent B\'{e}zier form.
Based on the convex hull property of the B\'{e}zier surface, a bounding-volume-hierarchy (BVH) tree is constructed through dynamic subdivision of the surface and then pruned using Newton's method to enhance the computation efficiency.
Our method 
ensures the discovery of the first intersection point in both static and dynamic settings.
It can also be extended to other spline-based surface intersection detection, such as rational B\'{e}zier patches and NURBS,
which we believe is a critical missing piece of current IGA research.

Last but not least, we conduct a broad array of experiments meticulously designed to showcase the fidelity and efficiency of the BHEM framework.
We first test our method under typical quasi-static settings. The simulation results are highly consistent with the theoretical solutions. 
Compared with traditional linear FEM, BHEM can capture rich geometric features with much fewer DoFs, better convergence speed, and less time cost.
Furthermore, we demonstrate the applicability of the framework by applying it to a variety of graphics scenarios with complex collisions and diverse boundary conditions. We emphasize the smoothness and highlight the characteristic wrinkles and folds of cloth geometries during dynamic simulations in all these experiments.


\begin{figure}[t]
    \newcommand{\formattedgraphics}[1]{\includegraphics[trim=5cm 0cm 5cm 0cm,clip,width=0.326\linewidth]{#1}}
    \centering
    \formattedgraphics{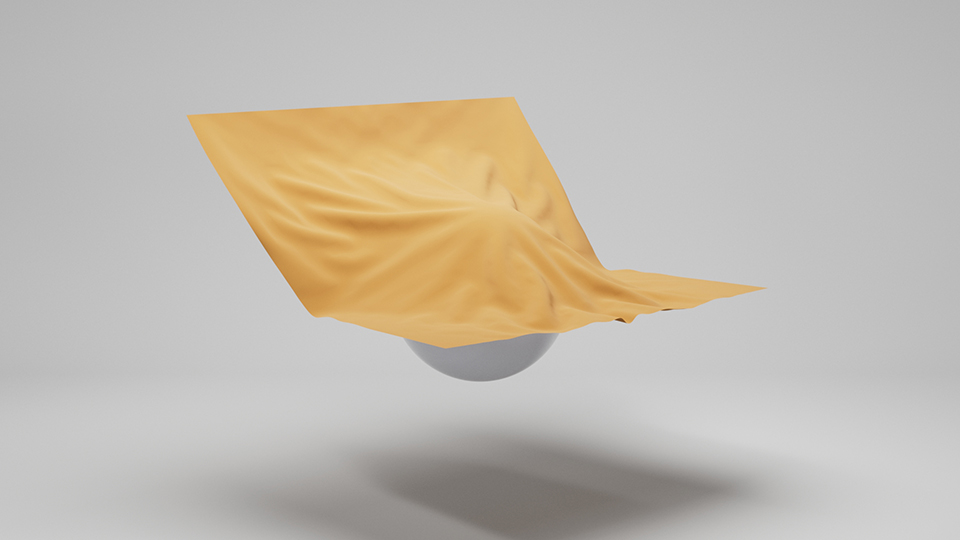}
    \formattedgraphics{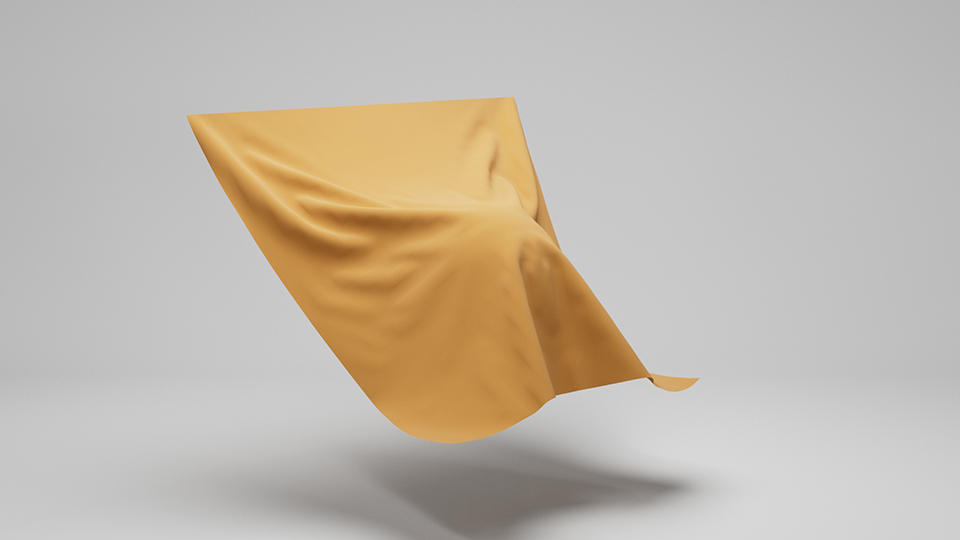}
    \formattedgraphics{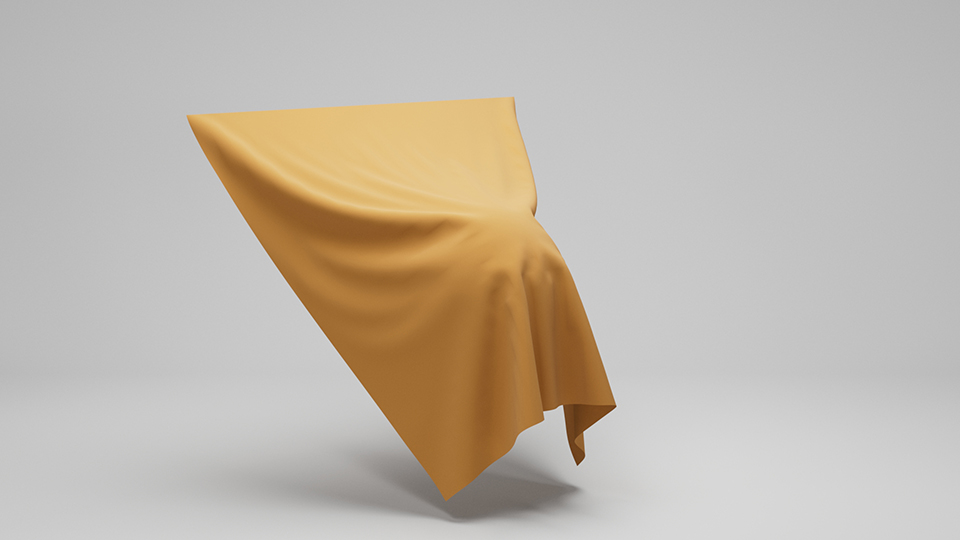}
    \caption{A sheet of square cloth ($30\times30$ patches) drapes on a ball, exhibiting rich wrinkle patterns in the process of reaching a steady state.}
    \label{fig:ball}
\end{figure}

\begin{figure}
    \newcommand{\formattedgraphics}[1]{\includegraphics[trim=8cm 0.5cm 8cm 3.5cm,clip,width=0.326\linewidth]{#1}}
    \centering
    \formattedgraphics{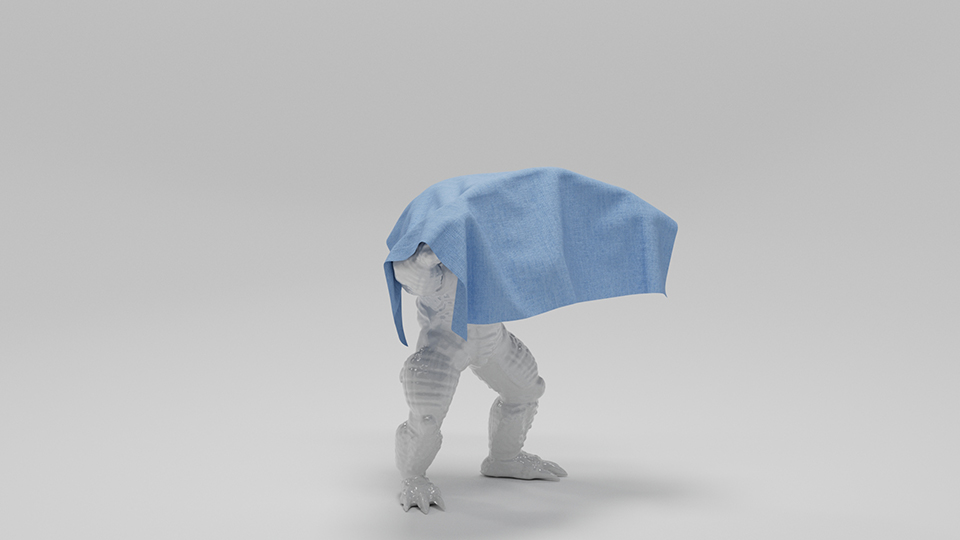}
    \formattedgraphics{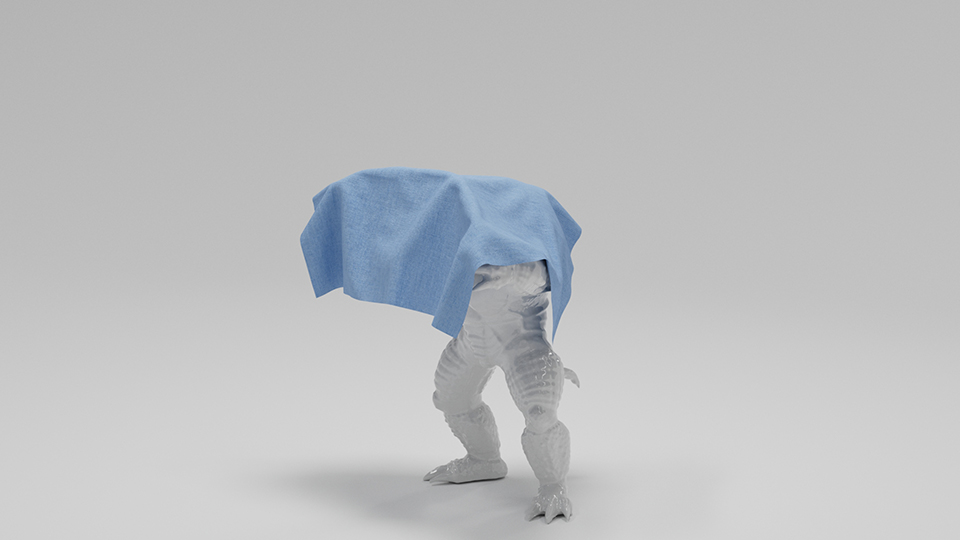}
    \formattedgraphics{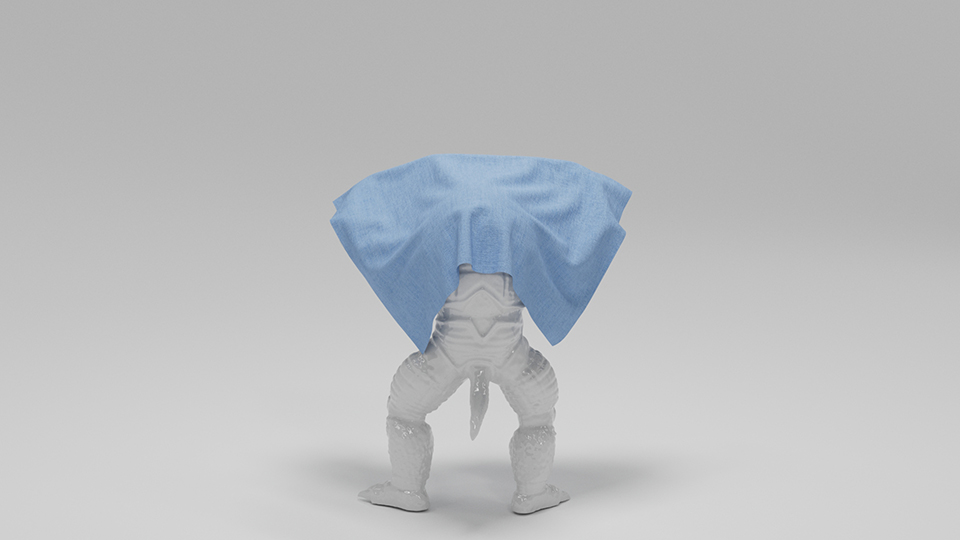}
    \caption{A sheet of square cloth ($30\times30$ patches) drapes on an armadillo model, where the sharp edges of the model are reflected by the cloth deformation.}
    \label{fig:armadillo}
\end{figure}

\section{Related Work}

\subsection{Thin-shell Simulation}
Physics-based thin shell simulation, such as cloth \cite{Baraff1998, Choi2002, Tamstorf2015, Chen2021, Wang2021, Zhang2022}, paper
 \cite{Narain2013, Chen2018, Burgoon2006, Pfaff2014}, and skin \cite{remillard2013, Li2018}, is a long-standing topic in computer graphics owing to their visually appealing geometry and desirable mechanical properties. 
Following the seminal work by Terzopoulos et al.\ \cite{terzopoulos1987},
a series of discrete constitutive models \cite{Grinspun2003, Bridson2003, Bergou2006, Gingold2004} have been developed for the elastic strain energy by applying geometric operators over the piecewise-linear surface, but with the bending energy being a non-integrable function of the dihedral angle.
Researchers have taken many measures to circumvent the issue, such as subdivision FEM \cite{Cirak2000, Thomaszewski2006, Clyde2017, Guo2018} and discontinuous Galerkin (DG) FEM \cite{Kaufmann2009}, but both with numerical issues due to their unavoidable special treatments to approximate or meet with the continuity requirement of a Kirchhoff--Love shell.

Another branch of the solution follows the route of isogeometric analysis, in which the basis functions that express geometric shapes are utilized in FE analysis for the physical field interpolation \cite{Wawrzinek2011,Lu2014}.
After the first attempt \cite{Qin1997} for using NURBS surface to represent a deformable thin object, NURBS has been widely exploited in cloth motion to describe the characteristic wrinkles or folds \cite{Lu2014} and in volumetric object simulation \cite{Trusty2021}.
However, due to its complex math formulation, the NURBS surfaces have involved derivation and high computational cost in graphics applications.

Studies have explored Hermite-interpolated surfaces on triangular elements \cite{Tocher1965,Bell1969}, quadrangular elements \cite{Bogner1965, Petera1994, Beheshti2018}, and hybrid elements \cite{SOLIN2007} in engineering and numerical analysis. 
Triangular Hermite elements use higher-order polynomials as basis functions so as to achieve $C^1$ continuity, leading to more computational cost. 
By contrast, quadrangular Hermite elements can have simpler but accurate expressions, though fall short in geometric flexibility.
In graphics, Hermite patches were originally adopted for freeform surface modelling \cite{Farin2001,Salomon2005}, but they lack applications in deformable objects.
In this work, we try to establish a complete framework that enables dynamic simulation and rendering of Hermite-interpolated surfaces.

\subsection{Rendering of Parametric Surfaces}
The calculation of ray tracing a parametric surface is to find the \emph{first} intersection point between a ray and a surface, which is equivalent to solving a system of nonlinear equations obtained by combining the parametric equations of the two geometric primitives.
A commonly adopted approach to ray-trace a high-order parametric surface is to triangulate the surface and render the approximated mesh. But to reserve the special advantages of high-order representation, it is beneficial to explore an efficient and robust rendering method that solves the problem directly on the parametric surface.

Existing studies have developed various numerical methods for this problem. Resultant elimination \cite{Kajiya1982, Manocha1994} transforms the problem into finding the minimum of a uni-variant polynomial, which works but requires heavy numerical computation with complex situational discussion, thus lack of robustness.
Newton's method \cite{Benthin2006, Abert2006, Geimer2005, Martin2000, Toth1985, Barth1993EfficientRT} is a powerful tool that can deliver an accurate solution with a fast convergence speed, but it depends severely on the initial conditions and has no guarantee for a correct solution since it only solves the problem locally. Also, the system would become ill-conditioned if a ray intersects the surface tangentially.



To solve for a global solution robustly, researchers have introduced subdivision methods \cite{Nishita1990, Campagna1997, Rogers1984,  Woodward1989} that recursively prune the parameter domain until convergence.
B\'{e}zier clipping \cite{Sederberg1990, Nishita1990, Tejima2015} is a classical subdivision method that utilizes the convex-hull property of B\'{e}zier patches. It has been widely used for ray-tracing any surfaces that can be converted into rational B\'{e}zier patches, including B-splines \cite{Tejima2015} and NURBS \cite{Efremov2005RobustAN}.
Meanwhile, subdivision-based methods often expend more time and space cost than Newton’s methods.
Therefore, a popular idea is to first conduct subdivision methods to construct a BVH for culling and getting a rough solution as an initial guess and then use Newton's descent to get a final answer \cite{Barth1993EfficientRT, Abert2006, Geimer2005, Martin2000}.
Simple geometric primitives, such as Chebyshev boxing/sphere \cite{Slusallek1995, Barth1993EfficientRT}, Convex hulls \cite{Pabst2006, Martin2000}, and axis aligned bounding box \cite{Geimer2005, Abert2006} are the popular choices for the bounding volume representations. 

\subsection{Collision Detection of Parametric Surfaces}

Continuous collision detection (CCD) for high-order meshes or parametric surfaces is mostly conducted by detecting collisions between two approximated triangle meshes \cite{Panozzo2023}, due to the maturity of traditional CCD techniques for linear FEM. As can be imagined, this would introduce inevitable false positives and negatives where the original surface sinks or extrudes from the linear mesh. 

Conducting DCD/CCD directly on parametric surfaces would result in solving a 4D/5D nonlinear system with a  complex solution manifold, which is usually reduced to point pairs in implementation \cite{Herzen90,Synder93}. Similar to the ray-tracing problem, solutions for CCD are always established on either subdivision methods \cite{Herzen90, Hughes96dcd,Synder93} to recursively solve for a global optimum or Newton’s methods \cite{Lu2014,LU2011,Synder93} to directly solve for a local optimum. Researchers need to get an appropriate initial guess for Newton’s method and cull unnecessary checks via assisting strategies including tessellation \cite{Lu2014}, BVH \cite{Krishnan98}, spatial partition \cite{Teschner05}, etc.
Recently, another promising approach proposed by Zhang et al.\ \cite{Sos2021,Sos2023} modeled the CCD problem between polynomial surfaces as a sum-of-squares programming (SOSP) so bypassed linearizing high-order surfaces.
This method includes an auxiliary hyperparameter in SOSP that directly determines the certificate of the exact solution, which requires a trade-off between efficiency and effectiveness.

\begin{figure}[t]
    \newcommand{\formattedgraphics}[1]{\includegraphics[trim=5cm 2.5cm 5cm 4cm,clip,width=0.494\linewidth]{#1}}
    \centering
    \formattedgraphics{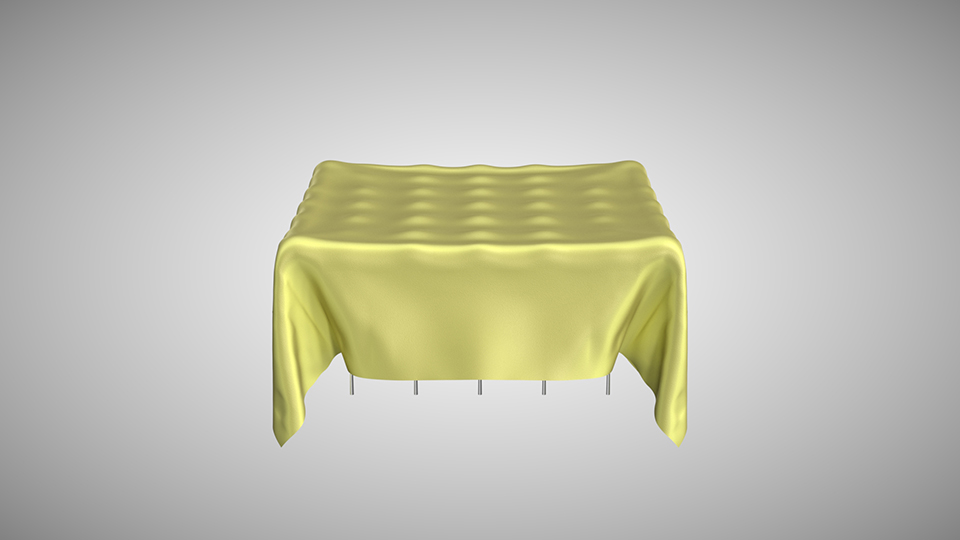}
    \formattedgraphics{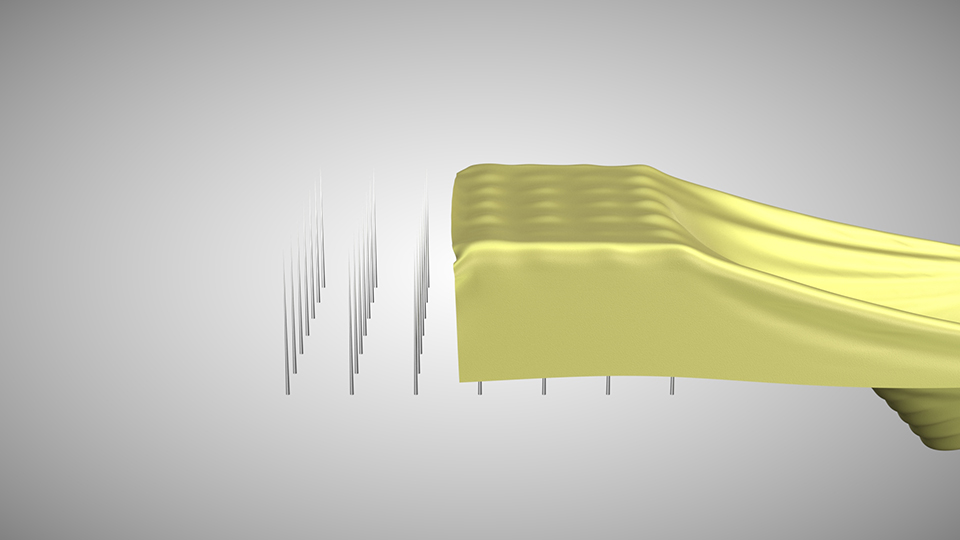}
    \caption{A sheet of cloth ($30\times30$ patches) falls on a needle array and then gets pulled away from aside. The bulges on the cloth surface are pushed by the needle tips. This sharp geometry deformation caused by the tiny contact demonstrates well the fine resolution of the patch interpolation. }
    \label{fig:needles}
\end{figure}

\section{Thin-Shell Theory}
\label{sec:theory}

To make the paper self-contained, in this section, we will briefly review the thin-shell theory based upon the Kirchhoff--Love assumption,
which generally follows the work of \cite{Cirak2000} but is not restricted to linearized kinematics or quasi-static analysis.

\paragraph*{Conventions and notations.}
The Einstein summation convention is followed, where an index variable appears twice in a single term implying summation of that term over all the values of the index.
We further assume that the value of an index denoted by a lowercase Latin letter (e.g., $i$, $j$, and $k$) ranges over the set $\{1,2,3\}$, while that denoted by a lowercase Greek letter (e.g., $\alpha$, $\beta$, and $\gamma$) ranges over the set $\{1,2\}$.
Besides, indices appearing after commas imply partial derivatives.

\subsection{Geometries}
\label{sec:theory_geometry}

We begin by considering a \emph{midsurface} $\varOmega\subset\mathbb{R}^3$.
As a surface, $\varOmega$ is parameterized with curvilinear coordinates $(\xi^1,\xi^2)$. The possible values of these coordinates form a parameter space $\omega$, and the parameterization is then given by a mapping $\bm{x}:\omega\to\varOmega$ such that the following properties hold:
\begin{itemize}
  \item At each point of $\varOmega$, the two partial derivatives ${\partial\bm{x}}/{\partial\xi^1}$ and ${\partial\bm{x}}/{\partial\xi^2}$ exist and are linearly independent;
  \item As a function $\omega\to\mathbb{R}^3$, $\bm{x}(\xi^1,\xi^2)$, as well as its first- and second-order derivatives, is square-integrable.
\end{itemize}
The former property allows the definition of the unit normal vector $\bm{a}_3={\bm{a}_1\times\bm{a}_2}/{\lVert \bm{a}_1 \times \bm{a}_2 \rVert}$, in which $\bm{a}_\alpha$ denotes ${\partial\bm{x}}/{\partial\xi^\alpha}$, while the second property is further required by analysis of elastic energy.

According to the Kirchhoff--Love assumption, the midsurface is extruded by a constant distance $h/2$ both along and opposite to the surface normal direction, which forms the volume of a $h$-thick shell. 
With $\omega^h$ defined as $\omega\times[-h/2,+h/2]$, the extrusion of $\varOmega$ is described by a function $\bm{r}:\omega^h\to\varOmega^h$ as follows:
\begin{equation}
  \bm{r}(\xi^1,\xi^2,\xi^3)=\bm{x}(\xi^1,\xi^2)+\xi^3\bm{a}_3(\xi^1,\xi^2)\text{,}\quad-\frac{h}{2}\le\xi^3\le+\frac{h}{2}\text{,}
\end{equation}
which lays the foundation of a thin-shell geometry.

To analyze shell deformation, the geometric difference between deformed and undeformed (reference) configurations needs evaluating.
We assume that each point of $\omega^h$ always maps to the same material point, and use symbols with overbars to indicate quantities in the reference configuration. A function $\bar{\bm{r}}:\omega^h\to\bar{\varOmega}^h$ can be similarly defined by
\begin{equation}
  \bar{\bm{r}}(\xi^1,\xi^2,\xi^3)=\bar{\bm{x}}(\xi^1,\xi^2) + \xi^3\bar{\bm{a}}_3(\xi^1,\xi^2)\text{,}\quad-\frac{h}{2}\le\xi^3\le+\frac{h}{2}\text{.}
\end{equation}

\subsection{Strains}

Given the parametric description of shell geometry, we acquire the tangent basis vectors of $\varOmega^h$ as follows:
\begin{subnumcases}{\label{eqn:gdef}\bm{g}_i=\frac{\partial\bm{r}}{\partial\xi^i}=}
  \bm{a}_{\alpha}+\xi^3\bm{a}_{3,\alpha}\text{,}&$i=\alpha<3$,\\
  \bm{a}_3\text{,}&$i=3$.
\end{subnumcases}
Dot products of these vectors result in covariant components of the metric tensor. To be specific, $g_{ij}=\bm{g}_i\cdot\bm{g}_j$ is formulated by
\begin{subnumcases}{g_{ij}=}
  a_{\alpha\beta}-2b_{\alpha\beta}\xi^3+c_{\alpha\beta}(\xi^3)^2\text{,}&$i=\alpha<3\land j=\beta<3$,\\
  1\text{,}&$i=j=3$,\\
  0\text{,}&otherwise,
\end{subnumcases}
where $a_{\alpha\beta}=\bm{a}_\alpha\cdot\bm{a}_\beta$, $b_{\alpha\beta}=\bm{a}_{\alpha,\beta}\cdot\bm{a}_3$, and $c_{\alpha\beta}=\bm{a}_{3,\alpha}\cdot\bm{a}_{3,\beta}$
respectively correspond to the \nth{1}, \nth{2}, and \nth{3} fundamental forms of $S$.

With $\{\bar{\bm{g}}_i\}$ being the basis of the tensor space, the \emph{Green--Lagrange strain} is defined as half the difference between the metric tensors in the deformed and undeformed configurations, i.e.,
\begin{equation}
  E_{ij}=\frac{1}{2}(g_{ij}-\bar{g}_{ij})\text{.}
\end{equation}
As will be readily seen, $E_{ij}$ can be expanded as
\begin{subnumcases}{}
  E_{\alpha\beta} = A_{\alpha\beta}-2B_{\alpha\beta}\xi^3+C_{\alpha\beta}(\xi^3)^2\text{,}\label{eqn:eab}\\
  E_{3\alpha} = E_{\alpha 3} = 0\text{,}\\
  E_{33} = 0\text{,}
\end{subnumcases}
followed by definitions $A_{\alpha\beta}=(a_{\alpha\beta}-\bar{a}_{\alpha\beta})/2$, $B_{\alpha\beta}=(b_{\alpha\beta}-\bar{b}_{\alpha\beta})/2$, and $C_{\alpha\beta}=(c_{\alpha\beta}-\bar{c}_{\alpha\beta})/2$.
Note that the \nth{0}-order term $A_{\alpha\beta}$ in Eq.~(\ref{eqn:eab}) represents the \emph{membrane strain}, while the other terms $B_{\alpha\beta}\xi^3+C_{\alpha\beta}(\xi^3)^2$ characterize the \emph{curvature strain}.

\subsection{Energies}

The elastic strain energy $V_\mathrm{e}$ is the most distinctive energy that is stored by a thin shell. 
As suggested by most of the previous studies \cite{Cirak2000, Weischedel2012}, the formula of $V_\mathrm{e}$ is usually derived from the Green--Lagrange strain in the material space, based upon a St.\ Venant--Kirchhoff constitutive model, which is given in the form of areal density $\bar{\mathcal{V}}_\mathrm{e}$ as
\begin{equation}
  \bar{\mathcal{V}}_\mathrm{e}=\frac{\mathrm{d}V_\mathrm{e}}{\mathrm{d}\bar{\varOmega}}=\left(A_{\alpha\beta}A_{\gamma\delta}h+\frac{1}{3}B_{\alpha\beta}B_{\gamma\delta}h^3\right)\bar{H}^{\alpha\beta\gamma\delta}\text{,}
\end{equation}
with $\bar{H}^{\alpha\beta\gamma\delta}$ defined by
\begin{equation}
  \bar{H}^{\alpha\beta\gamma\delta}=\frac{\lambda}{2}\bar{a}^{\alpha\beta}\bar{a}^{\gamma\delta}+\mu\bar{a}^{\beta\gamma}\bar{a}^{\alpha\delta}\text{.}
\end{equation}
Here, $\bar{a}^{\alpha\beta}$ is a contravariant tensor component, which can be calculated by the matrix inversion ${(\bar{a}^{\alpha\beta})}_{2\times2}=(\bar{a}_{\alpha\beta})^{-1}_{2\times2}$.
$\lambda$ and $\mu$, known as the first and the second \emph{Lam\'{e} parameters}, are deduced from Young's modulus $Y$ and Poisson's ratio $\nu$ as $\lambda={Y\nu}/{(1-\nu^2)}$ and $\mu={Y}/{2(1+\nu)}$, respectively. 

In addition to the elastic energy, the motion of a thin shell is also influenced by the kinetic energy $T$. With $\bar{\rho}$ denoting mass density, similar to $V_\mathrm{e}$, $T$ is also written in the form of areal density $\bar{\mathcal{T}}$ as
\begin{equation}
  \bar{\mathcal{T}}=\frac{\mathrm{d}T}{\mathrm{d}\bar{\varOmega}}=\frac{1}{2}\bar{\rho}h(\dot{\bm{x}}\cdot\dot{\bm{x}})\text{,}
\end{equation}
which is derived in the material space by integrating energy density along the thickness and ignoring high-order infinitesimals.

\subsection{Equations of Motion}

Finally, the equations of motion for a thin shell can be constructed by analyzing the interchange of energy.
Since the volumetric shell is replaced by its midsurface, we should also reduce any force that applies to the shell into surface force by integration.
Supposed that the areal density of external force applied on $\bar{\varOmega}$ is $\bar{\bm{f}}$, and the linear density on $\bar{\varGamma}=\partial{\bar{\varOmega}}$ is $\bar{\bm{t}}$, D'Alembert's principle states that for any virtual deformation $\delta\bm{x}$, the following equation holds:
\begin{equation}
  \label{eqn:dalembert1}  \iint_{\bar{\Omega}}\left(\delta\bar{\mathcal{T}}+\delta\bar{\mathcal{V}}_\mathrm{e}\right)\mathrm{d}\bar{\Omega}=\iint_{\bar{\Omega}}\bar{\bm{f}}\cdot\delta\bm{x}\,\mathrm{d}\bar{\Omega}+\int_{\bar{\Gamma}}\bar{\bm{t}}\cdot\delta\bm{x}\,\mathrm{d}\bar{\Gamma}\text{,}
\end{equation}
in which $\delta\bar{\mathcal{T}}=\bar{\rho}h\ddot{\bm{x}}\cdot\delta\bm{x}$ can be interpreted as the virtual work done by the inertia force, and the variation of $\bar{\mathcal{V}}_\mathrm{e}$ can be expanded as
\begin{equation}
  \label{eqn:deltave}
  \delta\bar{\mathcal{V}}_\mathrm{e}=\left(hA_{\gamma\delta}\,\delta a_{\alpha\beta}+\frac{1}{3}h^3B_{\gamma\delta}\,\delta b_{\alpha\beta}\right)\bar{H}^{\alpha\beta\gamma\delta}\text{,}
\end{equation}
owing to the exchange symmetry of indices.

It is noteworthy that the surface area element $\mathrm{d}\bar{\varOmega}$ can be expressed in the parameter space by
\begin{gather}
  \mathrm{d}\bar{\varOmega}=\sqrt{\bar{a}}\,\mathrm{d}\xi^1\mathrm{d}\xi^2\text{,}\\
  \bar{a}=(\bar{\bm{a}}_1\times\bar{\bm{a}}_2)^2=\bar{a}_{11}\bar{a}_{22}-\bar{a}_{12}\bar{a}_{21}\text{.}
\end{gather}

\begin{figure}[t]
    \newcommand{\formattedgraphics}[1]{\includegraphics[trim=5cm 2.5cm 5cm 2.5cm,clip,width=0.23\textwidth]{#1}}
    \centering
    \formattedgraphics{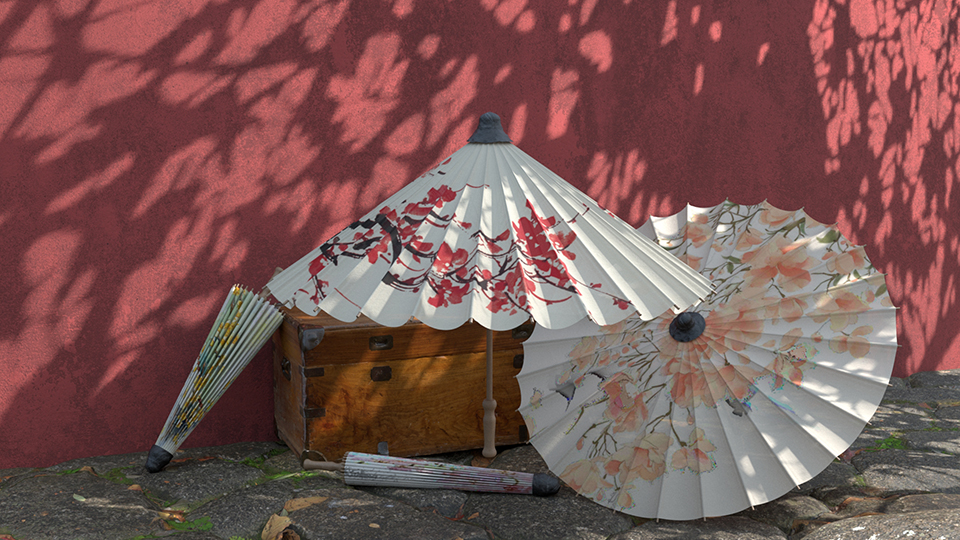}
    \formattedgraphics{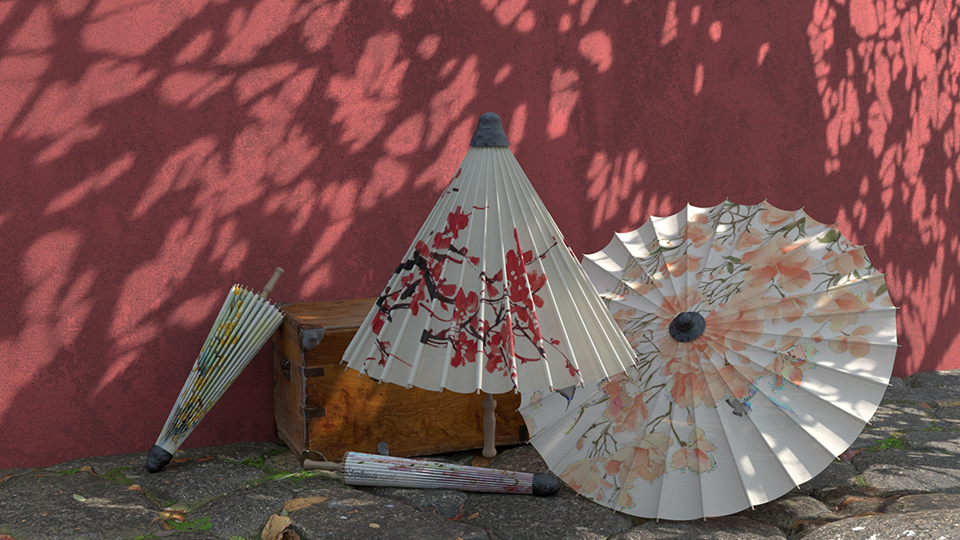}
    \caption{The folding process of an oriental paper parasol ($280\times5$ patches), simulated by jointly controlling nodal positions and their first-order derivatives. Bending and subtle wrinkles along the ribs can be observed.}
    \label{fig:umbrella}
    \vspace{-0.3cm}
\end{figure}

\section{The Bicubic Hermite Element Method}
\label{sec:bhem}

\subsection{Geometric Discretization}

\begin{figure}
\centering
\begin{subfigure}{0.495\linewidth}
\centering
\input{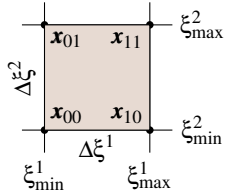}
\end{subfigure}
\begin{subfigure}{0.495\linewidth}
\centering
\begin{tikzpicture}




\draw (4.5,-0.4) -- (6.5,-0.4);
\draw (4.5,1) -- (6.5,1);
\draw (4.8,-0.7) -- (4.8,1.3);
\draw (6.2,-0.7) -- (6.2,1.3);
\fill (4.8,-0.4) circle (0.05);
\fill (4.8,1) circle (0.05);
\fill (6.2,-0.4) circle (0.05);
\fill (6.2,1) circle (0.05);
\definecolor{mycolor}{RGB}{231,218,210}
\draw[color=black, fill = mycolor] (4.8,-0.4) rectangle (6.2,1);
\node at(5.1,-0.2) {$\bm{x}_{00}$};
\node at(5.9,0.8) {$\bm{x}_{11}$};
\node at(5.1,0.8) {$\bm{x}_{01}$};
\node at(5.9,-0.2) {$\bm{x}_{10}$};
\node at(6.9,1) {$\xi_\mathrm{max}^2$};
\node at(6.9,-0.4) {$\xi_\mathrm{min}^2$};
\node at(4.8,-1) {$\xi_\mathrm{min}^1$};
\node at(6.2,-1) {$\xi_\mathrm{max}^1$};
\node at(5.5,-0.65) {$\Delta\xi^1$};
\node[rotate=90] at(4.5,0.3) {$\Delta\xi^2$};
\end{tikzpicture}
\end{subfigure}
\caption{A surface that is homeomorphic to a rectangle is embedded into a Cartesian grid, of which grid cells are taken as patches. Each BH patch corresponds to an axis-aligned rectangle in the parameter space.}
\label{fig:regular}
\end{figure}

\begin{figure}
  \centering
  \begin{subfigure}{0.495\linewidth}
    \centering
    \begin{asy}
    size(100);
    defaultpen(fontsize(10pt));
    import graph3;
    
    pen xarcPen=deepblue+0.5bp;
    pen yarcPen=brown+0.5bp;
    
    currentprojection=perspective(3,3,2);
    
    int m=8;
    int n=16;
    real R=2;
    real a=0.6;
    
    triple fs(pair t) {
    return (R*t.x,a*Cos(t.y),a*Sin(t.y));
    };
    
    pair p,q;
    
    for(int i=0;i<n;++i){
    for(int j=0;j<m;++j){
      p=(j/m,i*360/n);
      q=((j+1)/m,i*360/n);
      draw(fs(p)..fs((p+q)/2)..fs(q),xarcPen);
      q=(j/m,(i+1)*360/n);
      draw(fs(p)..fs((p+q)/2)..fs(q),yarcPen);
    }
    };
    for(int i=0;i<n;++i){
      p=(1,i*360/n);
      q=(1,(i+1)*360/n);
      draw(fs(p)..fs((p+q)/2)..fs(q),yarcPen);
    }
    
    surface s=surface(fs,(0,0),(1,360),1,32,Spline);
    //draw(s,surfacepen=material(lightyellow+opacity(.9), emissivepen=0.2*white),render(compression=Low,merge=true));
    draw(s,surfacepen=material(RGB(190,184,220)+opacity(.9), emissivepen=0.15*white),render(compression=Low,merge=true));

    p=((m-1)/m,0*360/n);
    q=((m-2)/m,0*360/n);
    draw(fs(p)..fs((p+q)/2)..fs(q),xarcPen,Arrow3
    //(TeXHead2(normal=Y))
    //,L=Label("$u$",position=EndPoint,align=N)
    );
    
    p=(m/m,1*360/n);
    q=(m/m,(1+1)*360/n);
    draw(fs(p)..fs((p+q)/2)..fs(q),yarcPen,Arrow3
    //(TeXHead2)
    //,L=Label("$v$",position=EndPoint,align=N+E)
    );
    \end{asy}
    \caption{Cylinders;}
    \label{fig:cylinder}
  \end{subfigure}
  \begin{subfigure}{0.495\linewidth}
    \centering
    \begin{asy}
    size(100);
    defaultpen(fontsize(10pt));
    import graph3;
    
    pen xarcPen=deepblue+0.5bp;
    pen yarcPen=brown+0.5bp;
    
    currentprojection=perspective(5,4,6);
    
    int m=16;
    int n=8;
    real R=2;
    real a=1;
    
    triple fs(pair t) {
    return ((R+a*Cos(t.y))*Cos(t.x),(R+a*Cos(t.y))*Sin(t.x),a*Sin(t.y));
    };
    
    pair p,q,v;
    
    for(int i=0;i<n;++i){
    for(int j=0;j<m;++j){
      p=(j*360/m,i*360/n);
      q=((j+1)*360/m,i*360/n);
      draw(fs(p)..fs((p+q)/2)..fs(q),xarcPen);
      q=(j*360/m,(i+1)*360/n);
      draw(fs(p)..fs((p+q)/2)..fs(q),yarcPen);
    }
    };
    
    surface s=surface(fs,(0,0),(360,360),10,10,Spline);
    draw(s,surfacepen=material(RGB(190,184,220)+opacity(.9), emissivepen=0.15*white),render(compression=Low,merge=true));

    p=(0*360/m,1*360/n);
    q=((0+1)*360/m,1*360/n);
    draw(fs(p)..fs((p+q)/2)..fs(q),xarcPen,Arrow3
    //,L=Label("$u$",position=MidPoint,align=S)
    );
    
    p=(0,1*360/n);
    q=(0,(1+1)*360/n);
    draw(fs(p)..fs((p+q)/2)..fs(q),yarcPen,Arrow3
    //,L=Label("$v$",position=MidPoint,align=N)
    );
    \end{asy}
    \caption{Torus.}
    \label{fig:torus}
  \end{subfigure}
  \caption{Some complex surfaces can be divided into BH patches with the help of periodic boundary condtions.}
\end{figure}

To numerically analyze the statics and dynamics of a thin shell, we divide its midsurface into a collection of bicubic Hermite patches that share information via nodes, where the bicubic Hermite interpolation is used to naturally maintain the $\mathcal{C}^1$ smoothness and square-integrability (\S\ref{sec:theory_geometry}) in surface reconstruction.

As shown in Fig.~\ref{fig:regular}, each patch corresponds to an axis-aligned rectangle in the parameter space, and each rectangle side is entirely shared by two adjacent patches.
In the context of such discretization, with the function value and its derivatives (i.e., $\bm{x}$, $\partial\bm{x}/\partial\xi^1$, $\partial\bm{x}/\partial\xi^2$, and $\partial^2\bm{x}/\partial\xi^1\partial\xi^2$) stored at nodes,
any point on a single patch satisfying $\xi^1\in[\xi_\mathrm{min}^1,\xi_\mathrm{max}^1]$ and $\xi^2\in[\xi_\mathrm{min}^2,\xi_\mathrm{max}^2]$ can be expressed with the piecewise bicubic Hermite interpolation as
\begin{equation}
  \label{eqn:hermite}
  \bm{x}(\xi^1,\xi^2)=\sum_{p,q,r,s\in\{0,1\}}w_{pq,rs}(\theta^1,\theta^2)\bm{x}_{pq,rs}\text{.}
\end{equation}
Here $\bm{x}_{pq,rs}$ represents $4$ generalized coordinates of node $\bm{x}_{pq}$, with $r$ and $s$ denoting the order of partial derivatives w.r.t. $\xi^1$ and $\xi^2$, respectively.
The weight function takes the form of
\begin{equation}
  w_{pq,rs}(\theta^1,\theta^2)=w_{p,r}(\theta^1)\,w_{q,s}(\theta^2)\text{,}
  \label{eqn:hermite_weight}
\end{equation}
with $\theta^\alpha\in[0,1]$ defined as $(\xi^\alpha-\xi_\mathrm{min}^\alpha)/\Delta\xi^\alpha$.
We refer readers to \S\ref{apx:hermite} for the details.

It is easy to formulate the Hermite interpolation in an FEM-like style by converting the weights (\S\ref{apx:hermite}) into \emph{shape functions}.
Concretely speaking, the midsurface can be parameterized by 
\begin{equation}
  \label{eqn:shape}
  \bm{x}(\xi^1,\xi^2)=\sum_{I=1}^{N}\varPhi^I(\xi^1,\xi^2)\,\bm{q}_I\text{,}
\end{equation}
in which $\bm{q}_I\in\mathbb{R}^3$ ($I=1,2,3,\ldots,N$), treated as \emph{generalized coordinates}, denotes a value or derivative that is stored at nodes, and each $\bm{q}_I$ corresponds to a shape function $\varPhi^I(\xi^1,\xi^2)$.
These shape functions have compact supports, so the summand $\varPhi^I\bm{q}_I$ takes nonzero values only if $\bm{q}_I$ is stored at the $4$ nodes of the patch that $(\xi^1,\xi^2)$ lies in. Typically, the number of such $\bm{q}_I$ is $16$.

The bicubic Hermite patches can be readily used to discretize surfaces that are homeomorphic to a rectangle by embedding it in a Cartesian grid and taking grid cells as patches (Fig.~\ref{fig:regular}). More complex surfaces, e.g., cylinders and torus, can also be divided into BH patches with the help of periodic boundary conditions (\S\ref{sec:bhem_bc}), which is illustrated in Figs. \ref{fig:cylinder} and \ref{fig:torus}.

\begin{figure*}[t]
    \newcommand{\formattedgraphics}[1]{\includegraphics[trim=10cm 0cm 10cm 2.5cm,clip,width=0.163\textwidth]{#1}}
    \centering
    \formattedgraphics{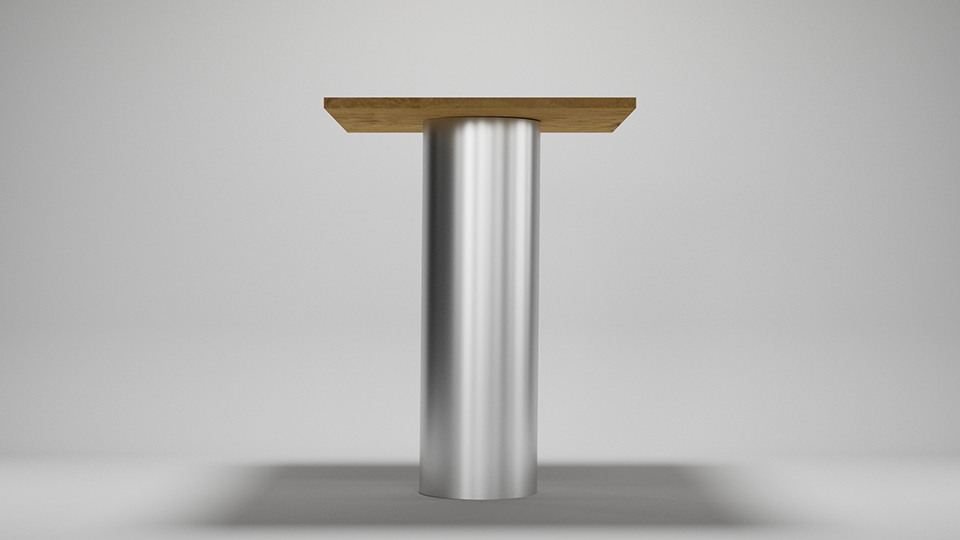}
    \formattedgraphics{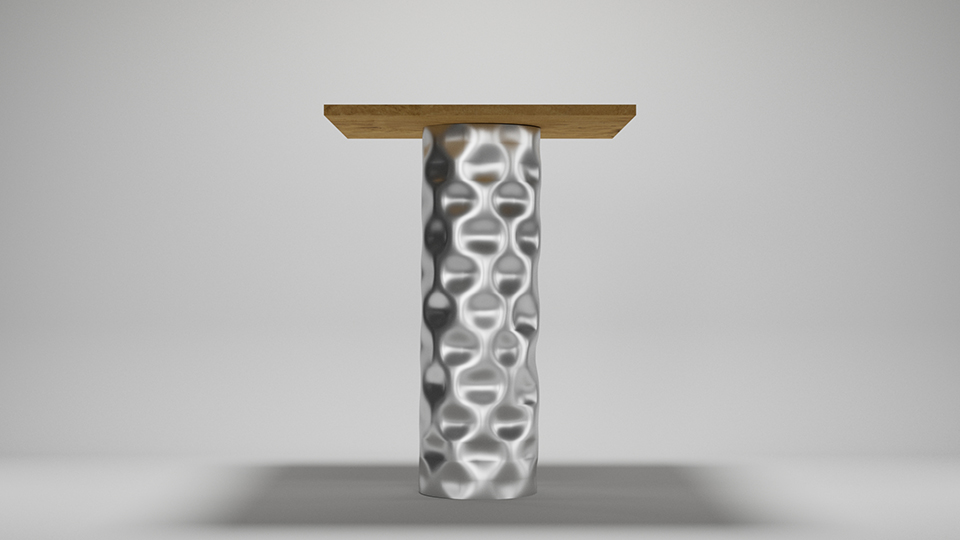}
    \formattedgraphics{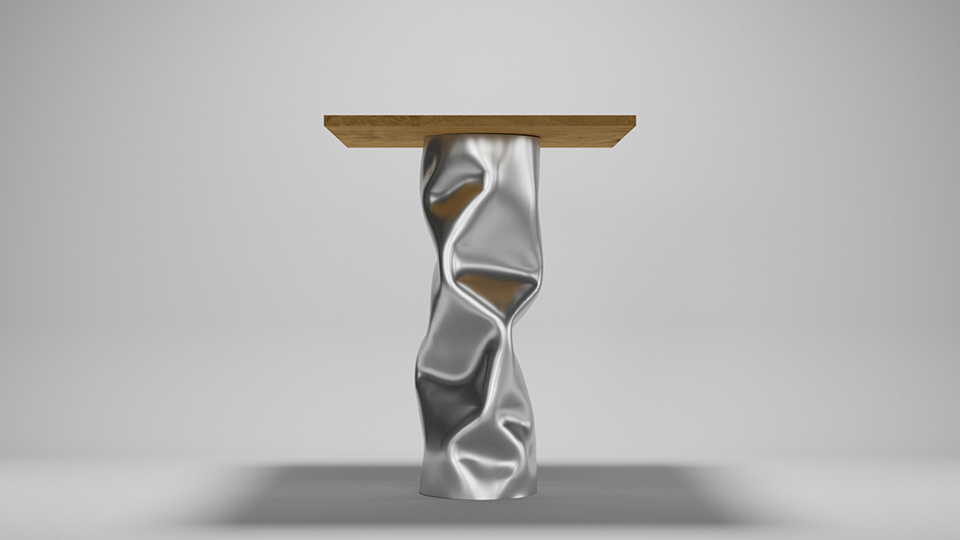}
    \formattedgraphics{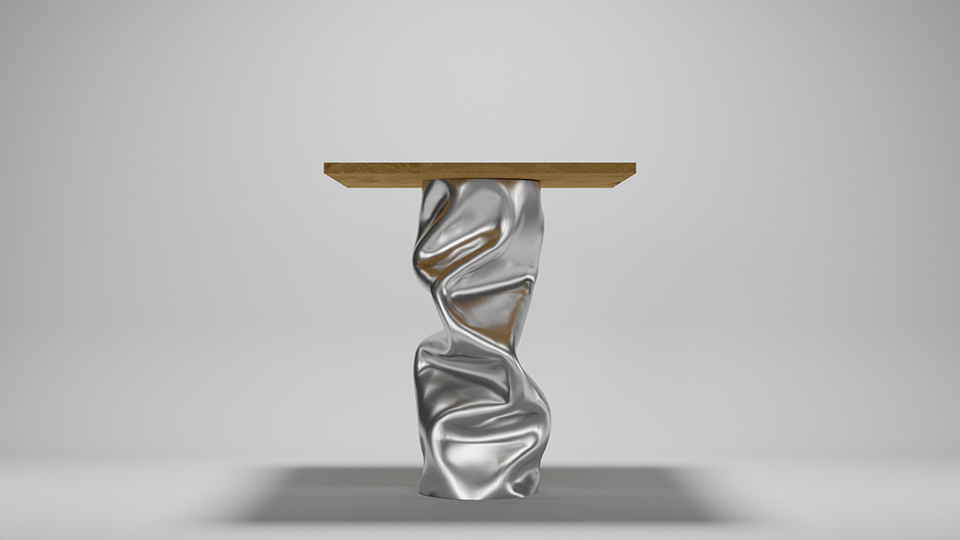}
    \formattedgraphics{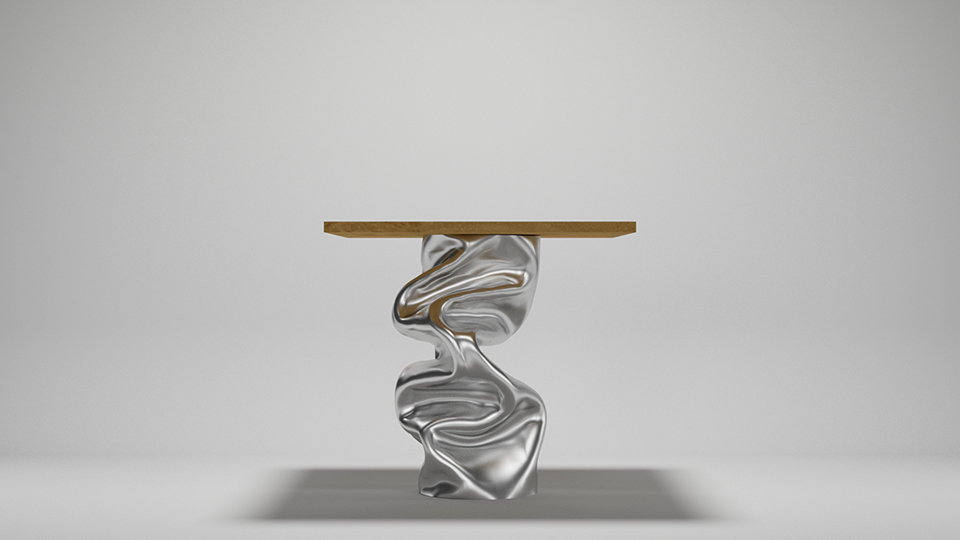}
        \formattedgraphics{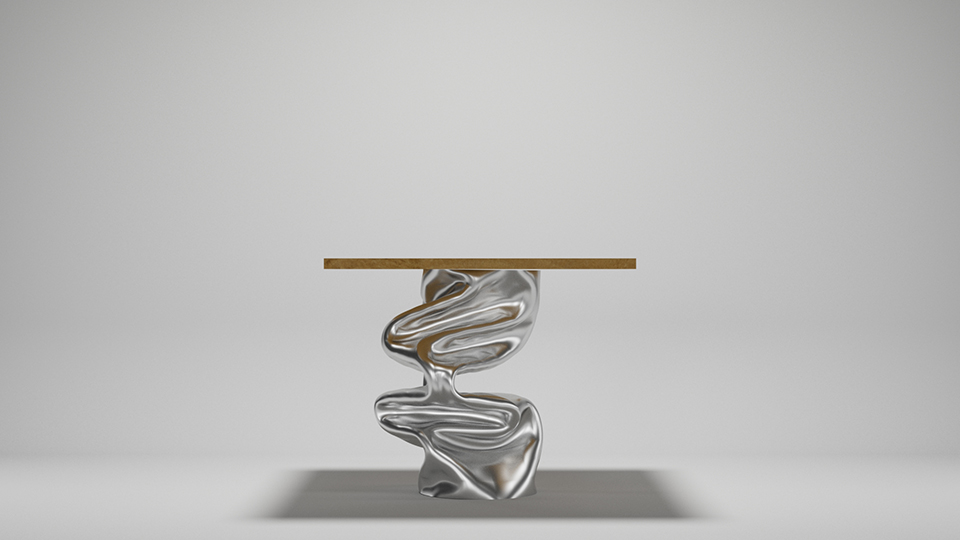}
    \caption{ A hollow cylindrical shell ($20\times20$ patches) deforms severely under the gradually increased compression. Initially, the cylinder is mounted on the ground; as the wood plank gets pushed down, the deformation of the cylinder manifests locally at first; at some point, the  unshaping takes a sudden change; further compression produces severe deformation and high degree of self contact.
}
    \label{fig:can}
\end{figure*}

\subsection{Boundary Conditions}
\label{sec:bhem_bc}

\paragraph*{Periodic boundary conditions.}
For BHEM simulations, it is significant to incorporate the periodic boundary conditions in order to complete geometric discretization.
As an example, in the discretization of a cylinder (Fig.~\ref{fig:cylinder}), a node with an azimuthal coordinate of $2\pi$ is identical to that with an azimuthal coordinate of $0$. Whenever we encounter such a situation, only one of the overlapping nodes needs maintaining. Any calculation that involves these nodes is referred to as the maintained one.

\paragraph*{Positional constraints.}
The BHEM prescribes two categories of positional constraints, namely Dirichlet and Neumann boundary conditions, which respectively constrain $\bm{x}(\xi^1,\xi^2)$ and its first-order derivatives at specific points.
When at nodes, these constraints can be simply realized by removing their corresponding generalized coordinates from the solution variables and taking their influence back to the governing equations as a given term.
Based on this, a constraint imposed on an element's whole boundary curve can be realized through the combination of Dirichlet constraints and Neumann constraints along the boundary tangent direction for every node on the boundary.
Furthermore, as conventional finite element methods, Lagrange multipliers should be introduced to apply the constraints at arbitrary points of the midsurface.

\subsection{Governing Equations}
\label{sec:governing}

Now we consider discretizing Eq.~(\ref{eqn:dalembert1}) using the variational method based on the BH patches.
Taking $\delta\bar{\mathcal{V}}_\mathrm{e}={\partial\bar{\mathcal{V}}_\mathrm{e}}/{\partial\bm{q}_I}\cdot\delta\bm{q}_I$, substituting Eq.~(\ref{eqn:shape}) into Eq.~(\ref{eqn:dalembert1}) yields
\begin{gather}
  \label{eqn:galerkin}
  {\small%
  \sum_{I=1}^{N}\left(\iint_{\bar{\varOmega}}\left(\varPhi^I\bar{\bm{f}}^*
  -\frac{\partial\bar{\mathcal{V}}_\mathrm{e}}{\partial\bm{q}_I}\right)\mathrm{d}\bar{\varOmega}+\int_{\bar{\Gamma}}\varPhi^I\bar{\bm{t}}\,\mathrm{d}\bar{\Gamma}\right)\cdot\delta\bm{q}_I=0}\text{,}\\
\bar{\bm{f}}^*=\bar{\bm{f}}-\bar{\rho}h\sum_{J=1}^{4N}\varPhi^J\ddot{\bm{q}}_J\text{,}
\end{gather}
which indicates that Eq.~(\ref{eqn:dalembert1}) is always true for any virtual deformation interpolated through Eq.~(\ref{eqn:shape}). Due to the arbitrariness of $\delta\bm{q}_I$, every coefficient of $\delta\bm{q}_I$, i.e., the terms in the outermost parentheses of Eq.~(\ref{eqn:galerkin}), must equal to zero. Thus, for any $I$ ($1\le I\le N$), the following equations hold:
\begin{equation}
  \label{eqn:linear}
  \iint_{\bar{\varOmega}}\varPhi^I\bar{\bm{f}}\,\mathrm{d}\bar{\varOmega}+\int_{\bar{\Gamma}}\varPhi^I\bar{\bm{t}}\,\mathrm{d}\bar{\Gamma}-\iint_{\bar{\varOmega}}\frac{\partial\bar{\mathcal{V}}_\mathrm{e}}{\partial\bm{q}_I}\,\mathrm{d}\bar{\varOmega}-\sum_{J=1}^{N}M^{IJ}\ddot{\bm{q}}_J=\bm{0}
  \text{,}
\end{equation}
where each coefficient of the mass matrix is defined as
\begin{equation}
  \label{eqn:mass}
  M^{IJ}=\iint_{\bar{\varOmega}}\bar{\rho}h\varPhi^I\varPhi^J\,\mathrm{d}\bar{\varOmega}=\iint_{\varOmega}\bar{\rho}h\varPhi^I\varPhi^J\sqrt{\bar{a}}\,\mathrm{d}\xi^1\mathrm{d}\xi^2\text{.}
\end{equation}
These constitute the governing equations of a dynamic BHEM shell.
For static analysis, the last term in the left-hand side of Eq.~(\ref{eqn:linear}) is omitted.

\paragraph*{Properties of the mass matrix.}
Due to the compact support of the shape functions $\varPhi$, a coefficient $M^{IJ}$ is nonzero only when the affected scopes of $\varPhi^I$ and $\varPhi^J$ overlap, i.e., the corresponding generalized coordinates of $\varPhi^I$ and $\varPhi^J$ belong to the nodes of the same patch.
This implies the sparsity of the mass matrix.
Furthermore, with $\varPhi^I$ being a polynomial of no more than degree 3, $\varPhi^I\varPhi^J$ reaches a sixth-order at most in each parametric dimension. Thus we conclude that $M^{IJ}$ can be calculated precisely without much effort, given that $\sqrt{\bar{a}}$ is also a polynomial concerning the position.

\paragraph*{External Forces.}
We briefly introduce how to calculate the external force term in Eq.~(\ref{eqn:linear}), taking $\iint_{\bar{\varOmega}}\varPhi^I\bar{\bm{f}}\,\mathrm{d}\bar{\varOmega}$ as an example.
For an areal force, the integration cannot be avoided. Typical examples are gravity force (given by $\bar{\bm{f}}_\mathrm{grav}=\bar{\rho}h\bm{g}$ with $\bm{g}$ standing for the gravity acceleration) and pressure force (given by $\bar{\bm{f}}_\mathrm{press}=p\bm{n}$ with $p$ denoting the magnitude of pressure).
On the other hand, a point force $\bar{\bm{f}}_\mathrm{pt}$ exerted at an arbitrary point can be reformulated as an areal force multiplied by a Dirac $\delta$ function. Given the point $\bm{x}_\mathrm{pt}(\xi^1_\mathrm{pt},\xi^2_\mathrm{pt})$, the closed integral form of $\bar{\bm{f}}_\mathrm{pt}$ can be calculated by
\begin{equation}
\iint_{\bar{\varOmega}}\varPhi^I\bar{\bm{f}}_\mathrm{pt}\delta(\bm{x}_\mathrm{pt})\,\mathrm{d}\bar{\varOmega}=\sum_I\varPhi^I(\xi^1_\mathrm{pt},\xi^2_\mathrm{pt})\bar{\bm{f}}_\mathrm{pt}\text{,}
\label{eqn:point_force}
\end{equation}
which implies that a point force only influences the 16 generalized coordinates of its nearest 4 nodes.

\begin{figure*}[t]
    \newcommand{\formattedgraphics}[2]{\begin{subfigure}{0.196\linewidth}\centering\includegraphics[trim=0.75cm 1cm 2.5cm 0.5cm,clip,width=\textwidth]{#1}\caption{#2}\vspace{-0.3cm}\end{subfigure}}
    \centering
    \formattedgraphics{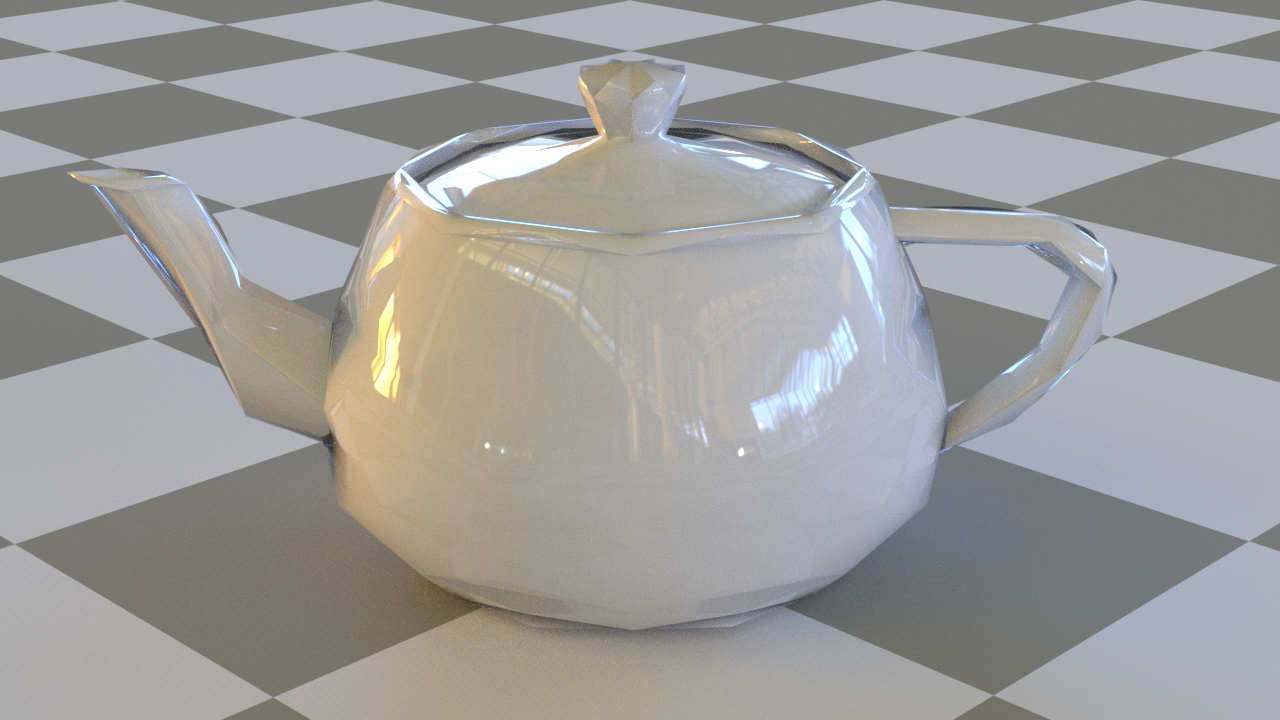}{265 vertices;}
    \formattedgraphics{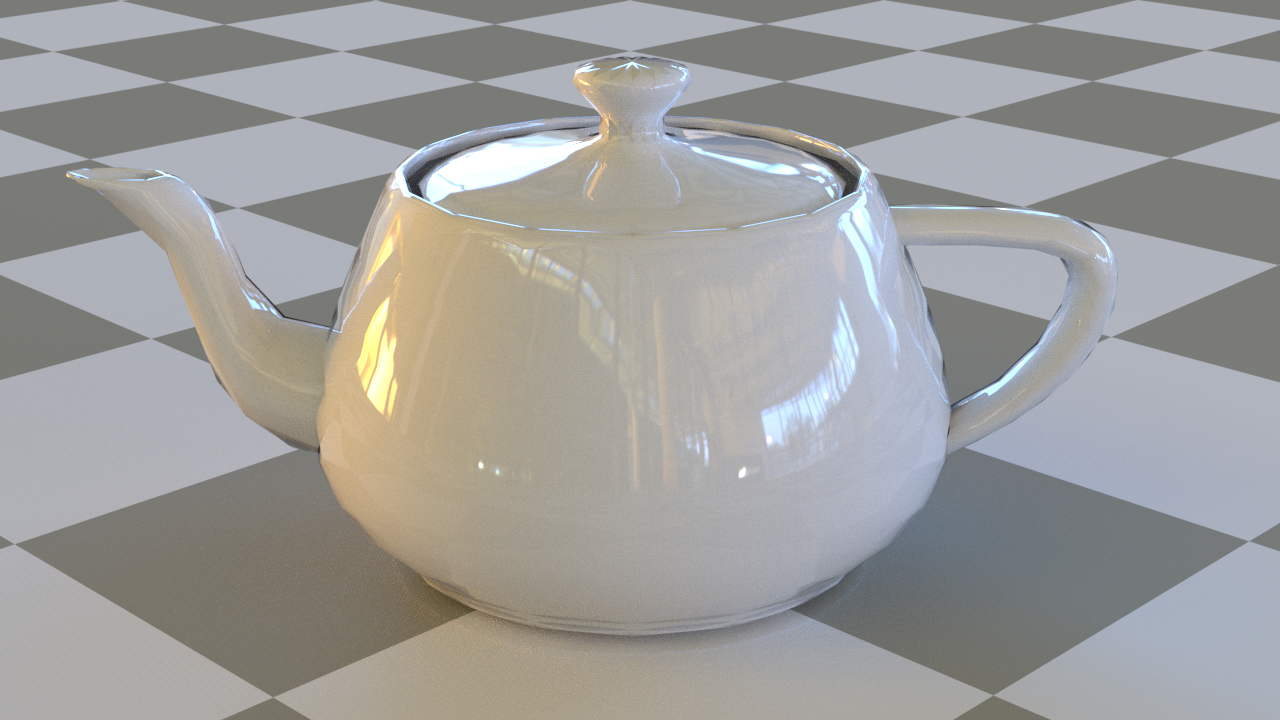}{1,041 vertices;}
    \formattedgraphics{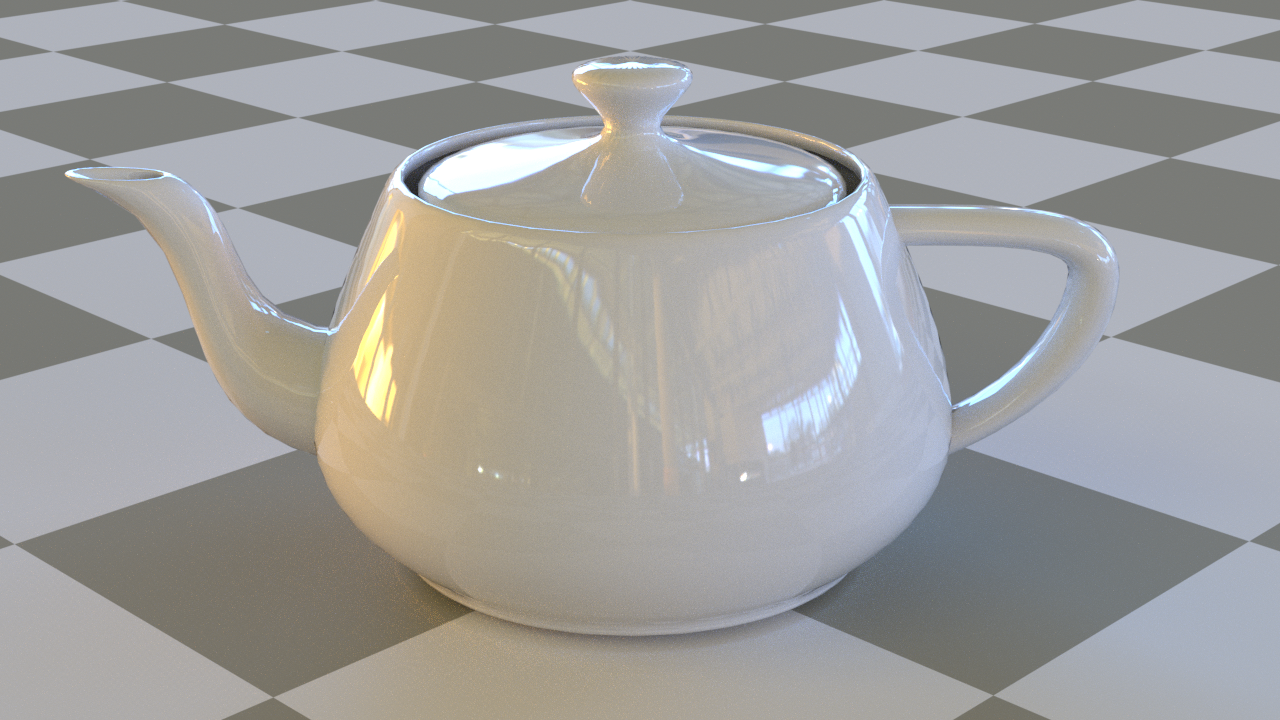}{4,129 vertices;}
    \formattedgraphics{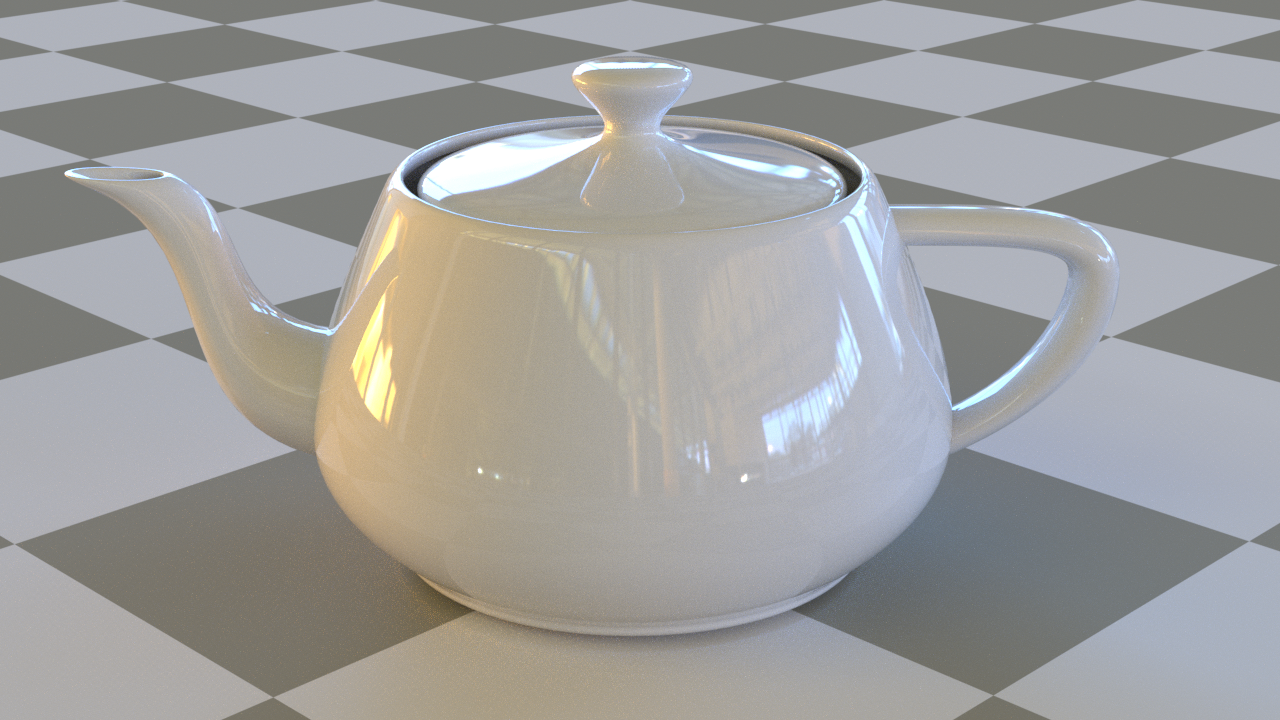}{16,449 vertices;}
    \formattedgraphics{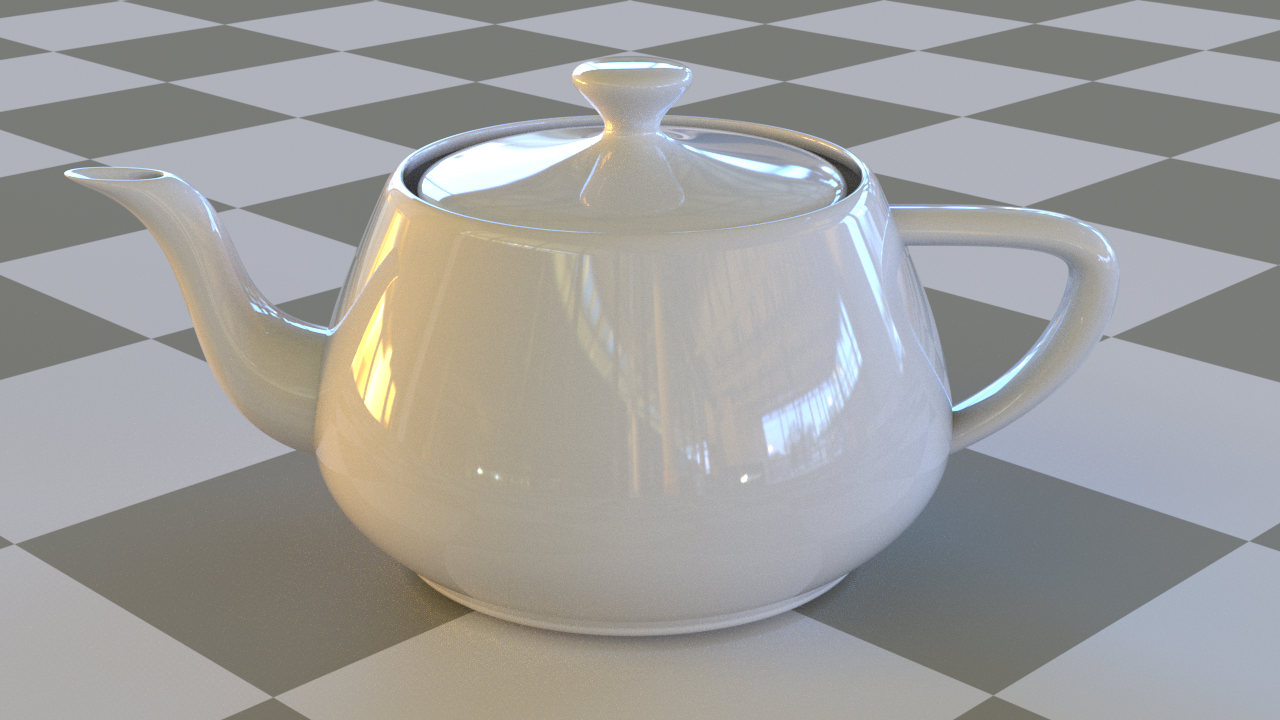}{32 patches.}
    \caption{Parametric surface rendering. A Utah teapot represented in bicubic patches is rendered by our scheme, shown in (e). The teapots rendered by traditional methods with each patch triangulated into 4/16/64/256 facets are shown in (a)/(b)/(c)/(d), where artifacts can be seen at the mouth and silhouettes of the teapots, especially for coarse mesh decimation. }
    \label{fig:teapot}
    \vspace{-0.3cm}
\end{figure*}

\section{Ray--Patch Intersection Detection}
\label{sec:detection}

We can equivalently transform a BH surface into a B\'{e}zier surface by regarding the cubic polynomials as linear combinations of \emph{Bernstein basis polynomials} of degree $3$: 
\begin{equation}
  \bm{x}(\xi^1,\xi^2)=\sum_{i=0}^3\sum_{j=0}^3B^i(\xi^1)B^j(\xi^2)\bm{p}_{ij}\text{,}
\end{equation}
where $B^i(x)$ is defined as $\tbinom{3}{i}x^i(1-x)^{3-i}$ and $\bm{p}_{ij}\in \mathbb{R}^3$ denote the control points.
A B\'{e}zier form provides us with a strong property that a surface lies completely within the convex hull of its control points, and thus also completely within the bounding box of them in any given Cartesian coordinate system, which lays the foundation of our ray-surface intersection detection algorithms.

The surface intersection is an important geometric operation in CAGD. We mainly focus on finding the intersection point between a ray and a surface in this paper. According to whether the surface moves during ray propagation, two kinds of intersections are detected.

\subsection{Static Ray--Patch Intersection}
\label{sec:render}
Given a ray defined as
\begin{equation}
  \bm{x}_\mathrm{ray}(\tau) = \bm{x}_\mathrm{0} + \tau\bm{d}\text{,}
  \label{eqn:ray}
\end{equation}
where $\bm{d}$ represents the direction and $\tau$ denotes the (pseudo) time, the goal of intersection tests is to find $(\xi^1,\xi^2)$ and minimum $\tau>0$ such that $\bm{x}_\mathrm{ray}(\tau)=\bm{x}(\xi^1,\xi^2)$ holds.
The convex-hull property of B\'{e}zier surfaces implies that for any $\xi^1$, $\xi^2$, and $\tau$ satisfying this intersection equation, the following inequalities hold:
\begin{equation}
  \min_{i,j\in\{0,1,2,3\}}\left\{\bm{p}_{ij}\cdot\hat{\bm{e}}_k\right\}\le(\bm{x}_0+\tau\bm{d})\cdot\hat{\bm{e}}_k\le\max_{i,j\in\{0,1,2,3\}}\left\{\bm{p}_{ij}\cdot\hat{\bm{e}}_k\right\}\text{.}
  \label{eqn:ineqray}
\end{equation}
Here, $\hat{\bm{e}}_k$ ($k\in\{1,2,3\}$) denotes the $k$-th vector of the standard basis.
These inequalities correspond to an intersection test between the ray and an axis-aligned bounding box (AABB). By solving Eq.~(\ref{eqn:ineqray}) for $\tau$, we can obtain an interval $[\tau_\mathrm{min},\tau_\mathrm{max}]$ or determine that no $\tau>0$ satisfies all the inequalities, indicating that the ray does not intersect the current surface.

When the former case is encountered after solving Eq.~(\ref{eqn:ineqray}), a divide-and-conquer strategy is employed.
The B\'{e}zier surface is split into four smaller B\'{e}zier surfaces using De Casteljau's algorithm \cite{BOEHM1999587} at specific parameter coordinates. This recursive process allows for ray-AABB tests to be performed, enabling the elimination of surfaces that cannot be intersected by the ray.
This continues until either all surfaces are discarded or an arbitrarily small interval of $\tau$ is obtained.

The above subdivision-based algorithm is trivial but effective and robust, converging linearly to the accurate intersection point.
To minimize the number of surfaces requiring collision checks, the selection of the next surface is improved by using a min-heap of candidate surfaces. 
Every time a new surface is generated, it is inserted into the heap with $\tau_\mathrm{min}$ as the key. When moving to another surface, the top of the heap is chosen.
The first surface that satisfies the termination condition provides an accurate approximation of the minimum $\tau$ among all the intersection points.

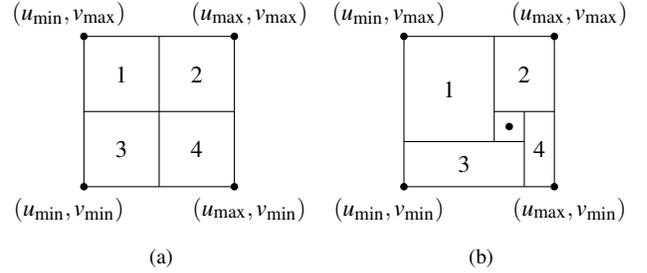
\begin{figure}
\begin{subfigure}{0.495\linewidth}
\begin{tikzpicture}
	\draw[step=2] (0,0) grid (2,2);
	\draw (0,1)--(2,1);
	\draw (1,0)--(1,2);
	\fill (0,0) circle (0.05);
    \fill (0,2) circle (0.05);
    \fill (2,0) circle (0.05);
    \fill (2,2) circle (0.05);
	\node at(-0.2,-0.3) {$(u_\mathrm{min},v_\mathrm{min})$};
    \node at(2.2,2.3) {$(u_\mathrm{max},v_\mathrm{max})$};
    \node at(-0.2,2.3) {$(u_\mathrm{min},v_\mathrm{max})$};
    \node at(2.2,-0.3) {$(u_\mathrm{max},v_\mathrm{min})$};
    \node at(0.5,1.5) {1};
    \node at(1.5,1.5) {2};
    \node at(0.5,0.5) {3};
    \node at(1.5,0.5) {4};
\end{tikzpicture}
\caption{}
\label{fig:submid}
\end{subfigure}
\begin{subfigure}{0.495\linewidth}
\begin{tikzpicture}
	\draw[step=2] (0,0) grid (2,2);
	\fill (1.4,0.8) circle (0.05);
	\draw (1.2,0.6)--(1.2,2);
	\draw (1.6,0.6)--(0,0.6);
	\draw (1.6,1)--(1.6,0);
	\draw (1.2,1)--(2,1);
	\fill (0,0) circle (0.05);
    \fill (0,2) circle (0.05);
    \fill (2,0) circle (0.05);
    \fill (2,2) circle (0.05);
	\node at(-0.2,-0.3) {$(u_\mathrm{min},v_\mathrm{min})$};
    \node at(2.2,2.3) {$(u_\mathrm{max},v_\mathrm{max})$};
    \node at(-0.2,2.3) {$(u_\mathrm{min},v_\mathrm{max})$};
    \node at(2.2,-0.3) {$(u_\mathrm{max},v_\mathrm{min})$};
    \node at(0.6,1.3) {1};
    \node at(1.6,1.5) {2};
    \node at(0.8,0.3) {3};
    \node at(1.8,0.5) {4};
\end{tikzpicture}
\caption{}
\label{fig:subany}
\end{subfigure}
\caption{Ray surface intersection subdivide schemes. Left: splitting at midpoint; Right: abandoning the neighborhood of intersection point and splitting the rest region.}
\end{figure}

Finally, Newton's method is employed to determine where to subdivide surfaces by directly solving $\bm{x}(\xi^1,\xi^2)=\bm{x}_\mathrm{ray}(\tau)$. 
The central parameter coordinates of a surface serve as the initial guess for Newton's method during the subdivision process. 
Within a fixed number of iterations, if Newton's method converges to $\xi^1\in[u_\mathrm{min},u_\mathrm{max}]$ and $\xi^2\in[v_\mathrm{min},v_\mathrm{max}]$, we obtain an intersection point and use the corresponding $\tau$ to update the current optimal solution $\tau_\mathrm{opt}$ if $\tau$ is smaller and positive. 
Note that this intersection point may not be the closest one. We remove it from the search space by abandoning 
the neighborhood of $(\xi^1,\xi^2)$ in the parameter space and subdividing the other region as illustrated in Fig.~\ref{fig:subany}. In the case Newton's method fails to converge, the surface is split from its midpoint as shown in (Fig.~\ref{fig:submid}).

\subsection{Dynamic Ray--Patch Intersection}
\label{sec:ccd}


Assuming constant velocity within a single time step $\Delta t$, a time-varying B\'{e}zier surface is given by
\begin{equation}
	\bm{x}(\xi^1,\xi^2,t)=\sum_{i=0}^3\sum_{j=0}^3B_i^3(\xi^1)B_j^3(\xi^2)(\bm{p}_{ij}+t\dot{\bm{p}}_{ij})\text{,}
\end{equation}
where $\bm{p}_{ij}$ and $\dot{\bm{p}}_{ij}$ are the initial position and velocity of the $(i,j)$-th control point, and $t\in[0,\Delta t]$ indicates the time. The trajectory of a moving point can be depicted 
as a parameterized ray as
\begin{equation}
    \bm{x}_\mathrm{p}(t)=\bm{x}_\mathrm{p0}+t\dot{\bm{x}}_\mathrm{p}\text{,}
\end{equation}
where $\bm{x}_\mathrm{p0}$ and $\dot{\bm{x}}_\mathrm{p}$ denote the initial position and velocity.

According to the convex-hull property, a necessary condition for the intersection to occur at the moment $t$ is that the point lies inside the axis-aligned bounding box of the surface at that moment, which can be written as
{\footnotesize%
\begin{equation}
    \min_{i,j}
    \left\{(\bm{p}_{ij}+t\dot{\bm{p}}_{ij})\cdot\hat{\bm{e}}_k\right\}
    \le (\bm{x}_\mathrm{p0}+t\dot{\bm{x}}_\mathrm{p})\cdot\hat{\bm{e}}_k
    \le \max_{i,j}\left\{(\bm{p}_{ij}+t\dot{\bm{p}}_{ij})\cdot\hat{\bm{e}}_k\right\}\text{,}
    \label{eqn:ineqccd}
\end{equation}}%
with $i,j\in\{0,1,2,3\}$. 
By solving Eq.~(\ref{eqn:ineqccd}) for $t$ similar to the approach described in \S\ref{sec:render} for $\tau$, we can employ a similar subdivision-based algorithm.

For each inequality on the left-hand side, the result is equivalent to the union of the intervals solved from the inequalities
\begin{equation}
  (\bm{p}_{ij}+t\dot{\bm{p}}_{ij})\cdot\hat{\bm{e}}_k\le(\bm{x}_\mathrm{p0}+t\dot{\bm{x}}_\mathrm{p})\cdot\hat{\bm{e}}_k\text{,}
\end{equation}
where $i$ and $j$ ranges in $\{0,1,2,3\}$. The same interpretation holds for the right-hand side of Eq.~(\ref{eqn:ineqccd}). The solution of Eq.~(\ref{eqn:ineqccd}) is the intersection of the two unions, which can be found using segment-tree or greedy algorithms. 
If no feasible solution exists within $[0,\Delta t]$, then no intersection occurs during the time step.
Otherwise, we can utilize the divide-and-conquer framework described in \S\ref{sec:render} to recursively detect possible intersections.
Similarly, Newton's method can be employed to 
accelerate convergence.

\begin{figure}
    \centering
    \begin{minipage}[h]{\linewidth}
        \centering
        \captionsetup*{type = table}
        \caption{Rendering performance. The three algorithms from left to right are: our subdivision-based algorithm that splits only at the midpoint, our algorithm with Newton's method, and B\'{e}zier clipping.}
        \begin{tabular}{|c|c|c|c|} \hline
                &  Subdivision &  Subdivision (Opt.) &B\'ezier clipping\\ \hline
        Plane    & $\SI{58.3}{\second}$ & $\SI{4.5}{\second}$  & $\SI{6.0}{\second}$\\ \hline
        Cylinder & $\SI{91.6}{\second}$ & $\SI{9.7}{\second}$  & $\SI{14.2}{\second}$\\ \hline
        Drape    & $\SI{86.0}{\second}$ & $\SI{14.8}{\second}$ & $\SI{33.9}{\second}$\\ \hline
        \end{tabular}
        \label{tab:performance}
    \end{minipage}
    \begin{minipage}[h]{\linewidth}
        \centering
    \begin{annotationimage}{trim=0cm 0cm 0cm 0cm,clip,width = 0.32\linewidth}{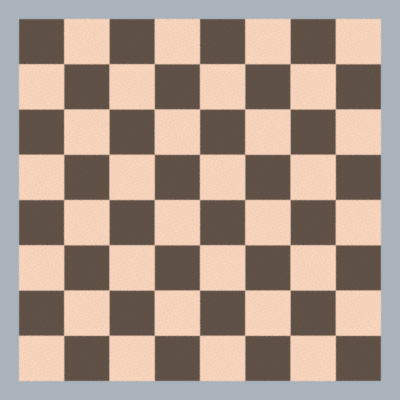}
        \draw[coordinate label = {Plane at (0.138,0.07)}];
    \end{annotationimage}
    \begin{annotationimage}{trim=3.5cm 0cm 3.5cm 0cm,clip,width = 0.32\linewidth}{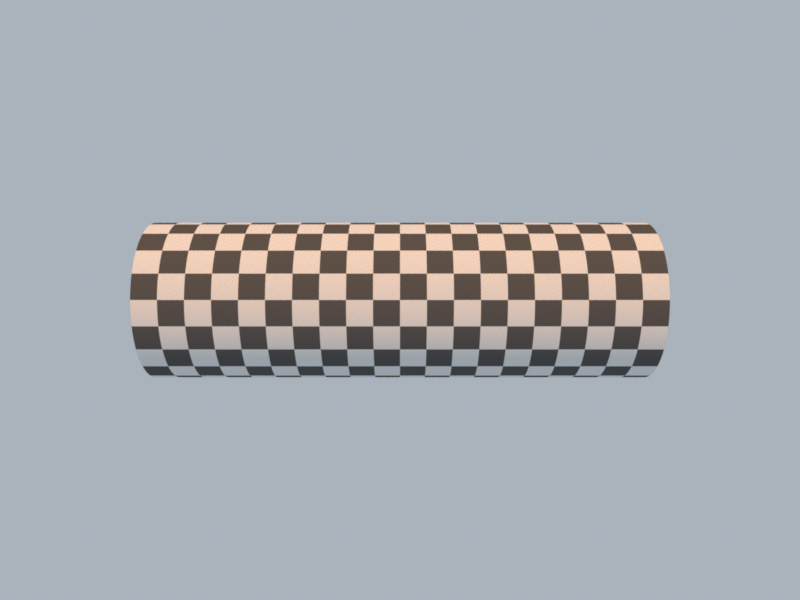}
        \draw[coordinate label = {Cylinder at (0.195,0.08)}];
    \end{annotationimage}
    \begin{annotationimage}{trim=3.5cm 0cm 3.5cm 0cm,clip,width = 0.32\linewidth}{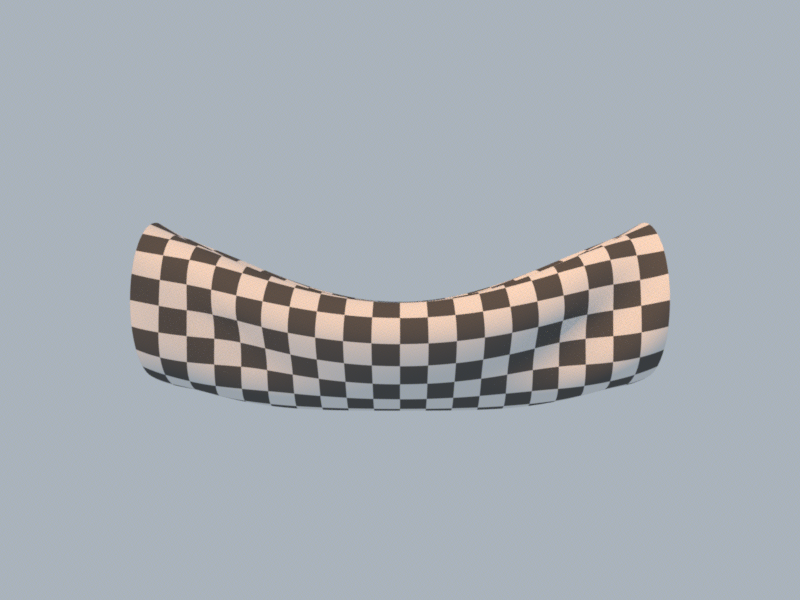}
        \draw[coordinate label = {Drape at (0.15,0.08)}];
    \end{annotationimage}
    \caption{Three test cases with only 1 patch for Plane and $10\times10$ patches for Cylinder and Drape each. The scenes all have the same setting of lights and 16 samples per pixel.}
    \label{fig:performance}
    \end{minipage}
\end{figure}

\section{Implementation}
\label{sec:implementation}

\subsection{Implicit BHEM Solver}
\label{sec:solver}

Having established spatially discretized equations of motion (\S\ref{sec:governing}), we here discuss how to discretize Eq.~(\ref{eqn:linear}) in time.

As a preparation, we assemble the mass matrix $\bm{M}\in\mathbb{R}^{3N\times3N}$ by a $16$-point Gauss--Legendre quadrature method (\S\ref{apx:quadrature}) according to the component definition in Eq.~(\ref{eqn:mass}).
The generalized coordinates and the corresponding generalized forces exerted on them are also numerically integrated and stacked into a $3N$-dimensional $\bm{q}$ and $\bm{F}$, respectively. 
Then Eq.~(\ref{eqn:linear}) is reformulated as 
\begin{equation}
    \bm{M}\ddot{\bm{q}}=\bm{F}\text{,}
\end{equation}
which is further discretized by the implicit Euler scheme in time:
\begin{subnumcases}{\label{eqn:implicit}}
\dot{\bm{q}}^{n+1}=\dot{\bm{q}}^n+\Delta t\,\bm{M}^{-1}\bm{F}(\bm{q}^{n+1},\dot{\bm{q}}^{n+1},t^{n+1})\text{,}\\
\bm{q}^{n+1}=\bm{q}^n+\Delta t\,\dot{\bm{q}}^{n+1}\text{.}
\end{subnumcases}
By convention, we adopt Newton's method to solve the above equations iteratively, where quadratic and cubic line searchers referring to the widely used library \emph{ArcSim} \cite{Narain2012,Narain2013} are integrated.
During the per Newton step, the linear system is solved by a direct sparse LDLT Cholesky factorizations \cite{eigenweb}.

Besides, for the aforementioned positional constraints at arbitrary points, we use the augmented Lagrangian method \cite{Hestenes1969MultiplierAG} to solve the optimization problem, which acquires a better convergence rate than a pure method of Lagrange multipliers.

\paragraph*{The Hessian matrix.}
In a Newton's method, the analytical form of $\partial\bm{F}/\partial\bm{q}_I$ ($I=1,2,3,\ldots,N$) is required. The most tricky part of the derivatives is the contribution of the elastic force, namely the Hessian matrix of $V_\mathrm{e}$ w.r.t. the generalized coordinates.
We have carefully derived the elastic energy's first- and second-order derivatives for the BHEM solver and provide the concrete formulations in \S\ref{apx:derivative}.
In some situations, an inexact Hessian matrix may reduce the time cost of the solver's convergence, due to the high overhead of assembling the exact one. This alternative, which we called the \emph{pseudo} Hessian matrix, is also given in the appendix.
Moreover, when facing with ill-conditioned Hessian matrix, a diagonal regularizer is added to make the matrix positive definite.

\subsection{Collision Handling}
We utilize the point-to-surface intersection detection scheme for CCD through sampling strategies adjusted to specific scenarios.
For collision detection between shells and colliders with simple geometry shapes, such as spheres, cylinders, or planes, the midsurface is uniformly sampled. 
As for the collision detection with colliders possessing complex shapes, we sample the collider surface, or directly take the surface vertices as the sample points, if the collider has a triangular mesh.
For the thin-shell self-collision, we iteratively sample each Hermite patch and perform CCD on sampling points against all the other surfaces. Penetrations occurring within a single surface can be handled by subdividing the current surface with the scheme illustrated in Fig.~\ref{fig:subany} and treating the $4$ subdivided regions recursively.

After we have examined all the collision primitive pairs and got the list of earliest simultaneous collision pairs, we roll back to the collision moment. This prohibits penetration throughout the simulation. 
Then we follow the impulse-based method~\cite{Vouga08} to compute the simultaneous collision responses according to the conservation of momentum. Using a zero restitution coefficient~\cite{Bridson2002}, we model each collision as the momentum change at the collision point, which is formulated as a linear constraint of generalized velocities. 
Then we handle the multiple collisions in one batch by solving the linear system of constraints for the impulses in a least-square sense and update velocities.
When friction exists, we compute the applied frictional impulses on the tangent direction of collision points according to the Column cone and solve for the normal and tangent updates together.
To avoid the resolved collision being detected at the beginning of the next round of CCD,
we additionally update the displacement by pushing out each collision point a subtle distance ($10^{-4}$ in practice) opposite the normal direction.
This is also modeled as a constraint linear to the generalized positions.
Then the position constraints are resolved in a similar way after the velocity updates.


\begin{figure*}[t]
    \newcommand{\formattedgraphics}[2]{\begin{subfigure}{0.196\linewidth}\centering\includegraphics[trim=0cm 0cm 0cm 0cm,clip,width=\textwidth]{#1}\caption{#2}\vspace{-0.3cm}\end{subfigure}}
    \centering
    \formattedgraphics{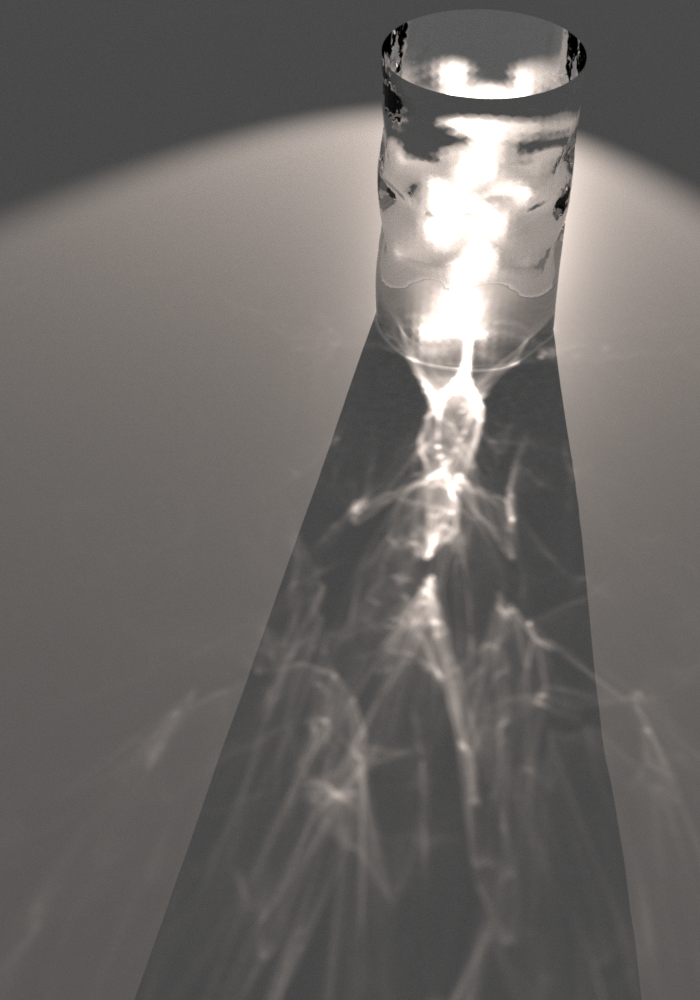}{3,160 vertices;}
    \formattedgraphics{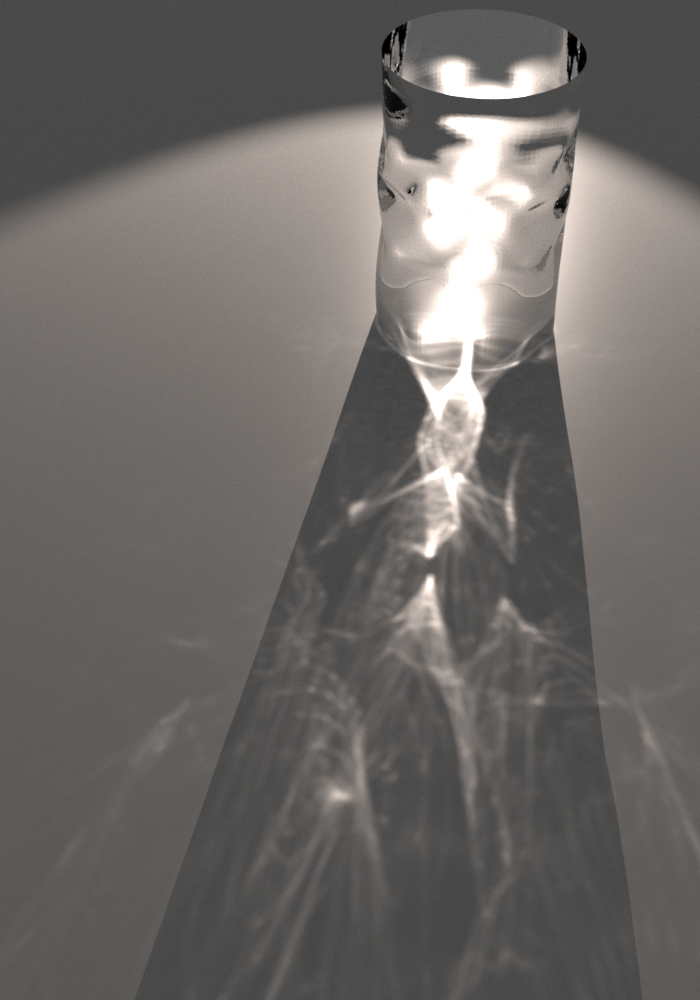}{14,520 vertices;}
    \formattedgraphics{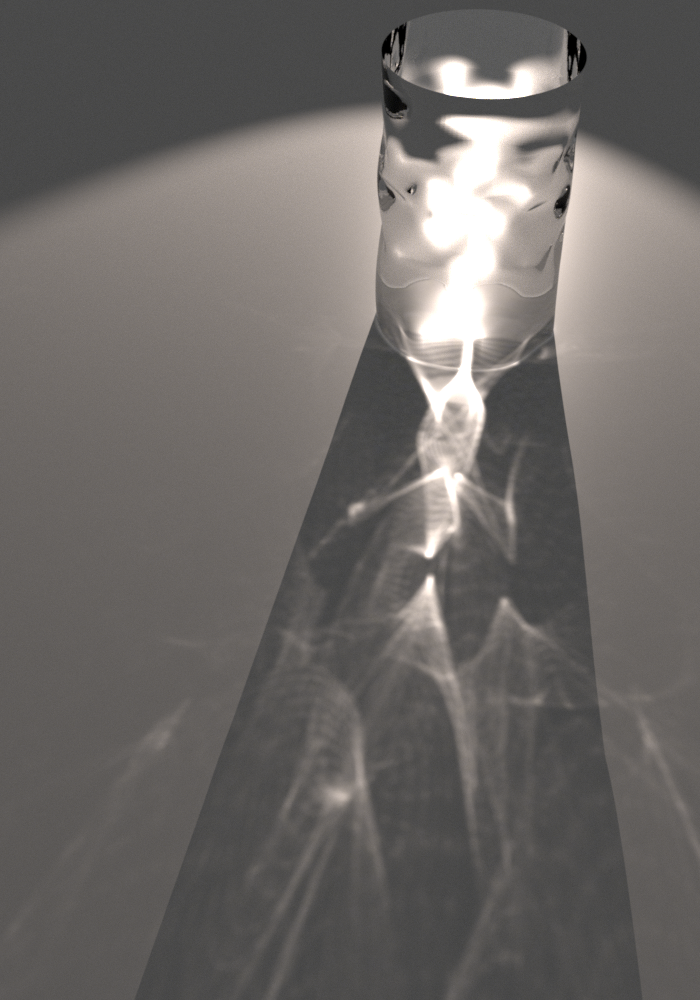}{57,840 vertices;}
    \formattedgraphics{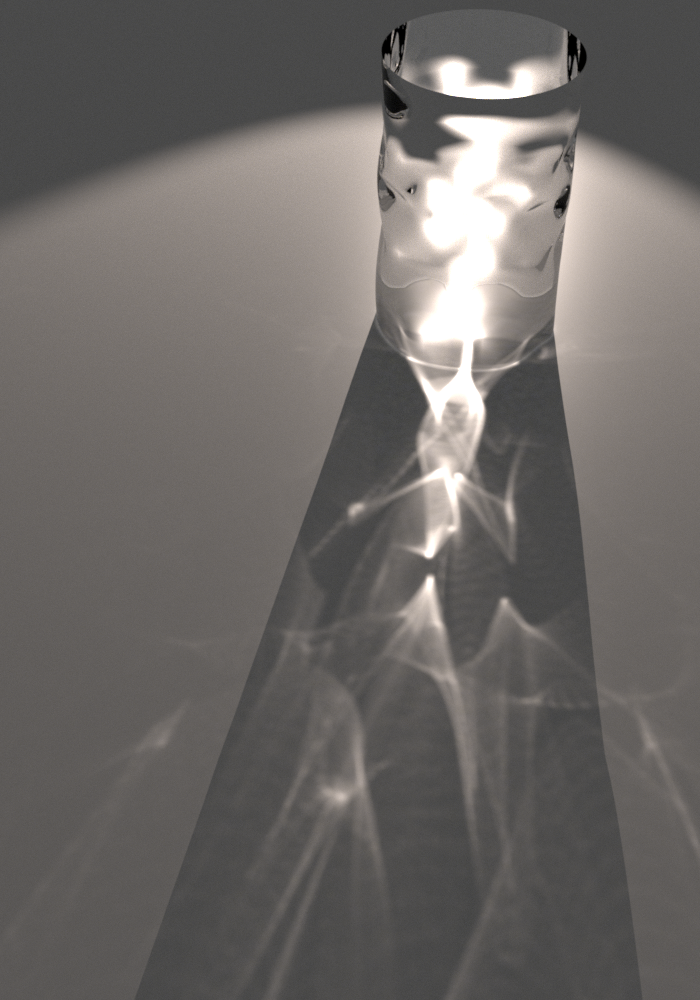}{230,880 vertices;}
    \formattedgraphics{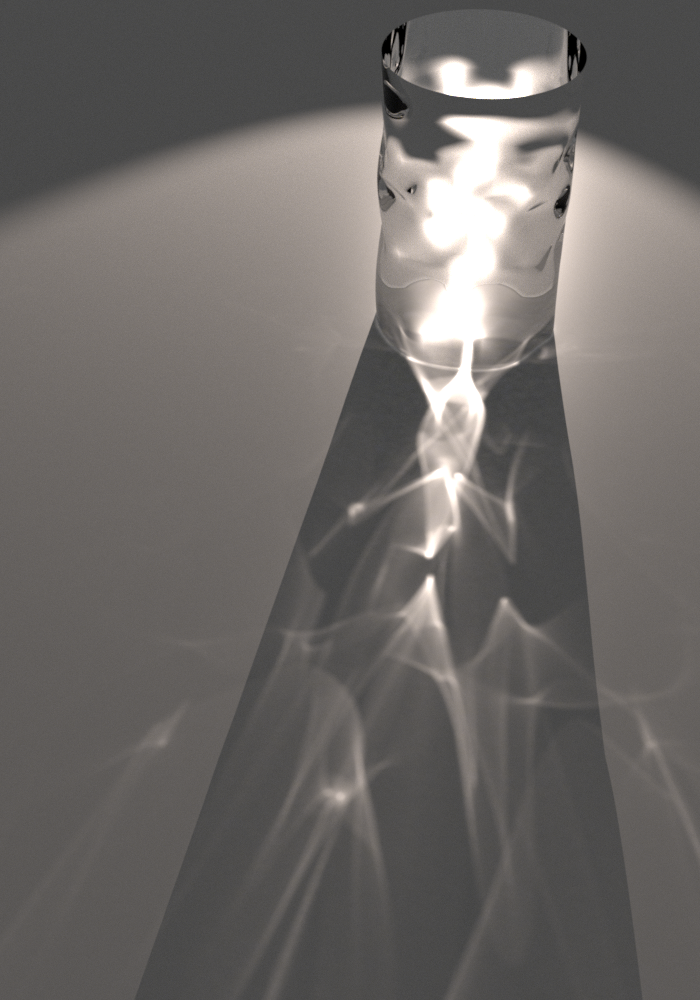}{450 patches.}
    \caption{Parametric surface rendering. The caustic lighting effect amplifies the imperfection of the surface, such as the un-smooth surface normal distribution. Fine-grained caustic rays clutter the ground for the scenarios with low-resolution mesh objects. 
    }
    \label{fig:caustic}
\end{figure*}

\subsection{Rendering}
In a ray-tracing rendering framework, the most expensive part is to test whether, where, and when a ray intersects the object. Traditional subdivision-based methods own their advantage in robustness but often fall in efficiency.
We implement our ray-tracing algorithm in the rendering system of pbrt-v4 \cite{Pharr2023} using our algorithm of static ray-patch intersection detection.
Our improvements of the subdivision-based algorithm lead to a great acceleration, resulting in even faster performance than a theoretically quadratic-convergent algorithm, such as Bézier clipping, as shown in Tab. 1 and Fig. 10. The algorithm optimized by Newton’s method shows the best in efficiency over three scenes.


\section{Results}
We design a wide range of validation tests and simulation experiments to evaluate the validity, fidelity, and effectiveness of our framework from various aspects.
All of the experiments are run on 
a $3.50 GHz$ 13th Gen Intel(R) Core(TM) i5-13600KF desktop with $32$ GB RAM.

\subsection{Validation}
\begin{figure}[t]
    \centering
    \newcommand{\formattedgraphics}[1]{\includegraphics[trim=2cm 4cm 2cm 4cm,clip,width=\textwidth]{#1}}
    \begin{subfigure}{0.495\linewidth}
        \centering
        \formattedgraphics{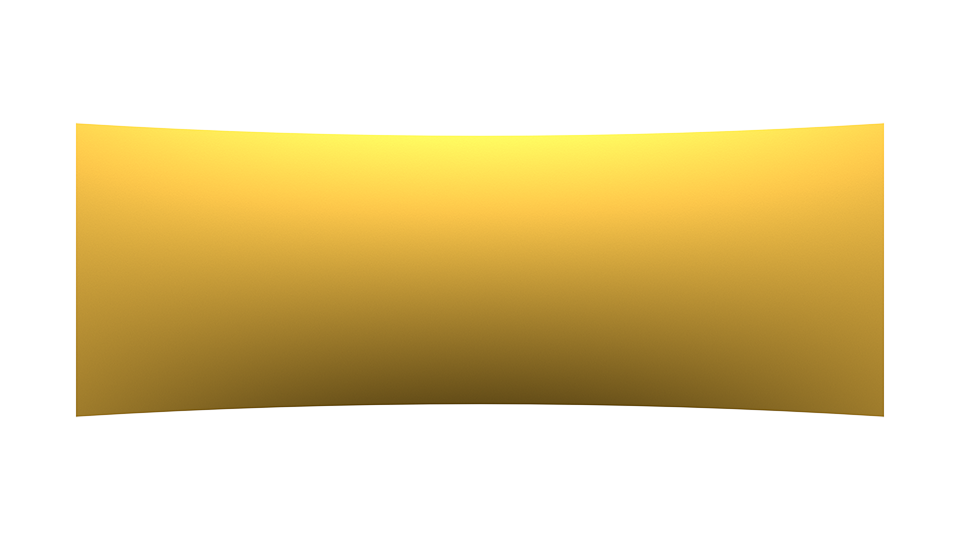}
        \caption{$1\times1$ patch;}
    \end{subfigure}
    \begin{subfigure}{0.495\linewidth}
        \centering
        \formattedgraphics{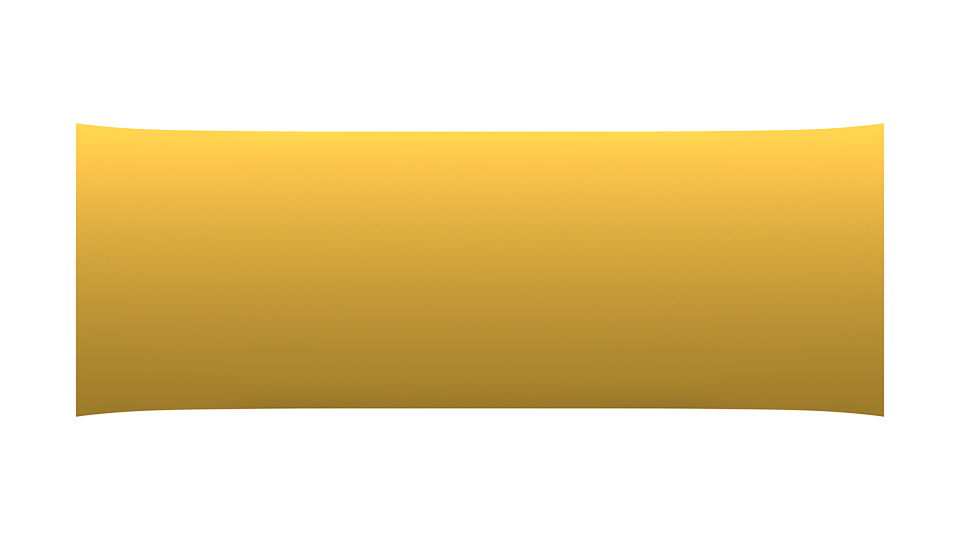}
        \caption{$5\times5$ patches;}
    \end{subfigure}
    \begin{subfigure}{0.495\linewidth}
        \centering
        \formattedgraphics{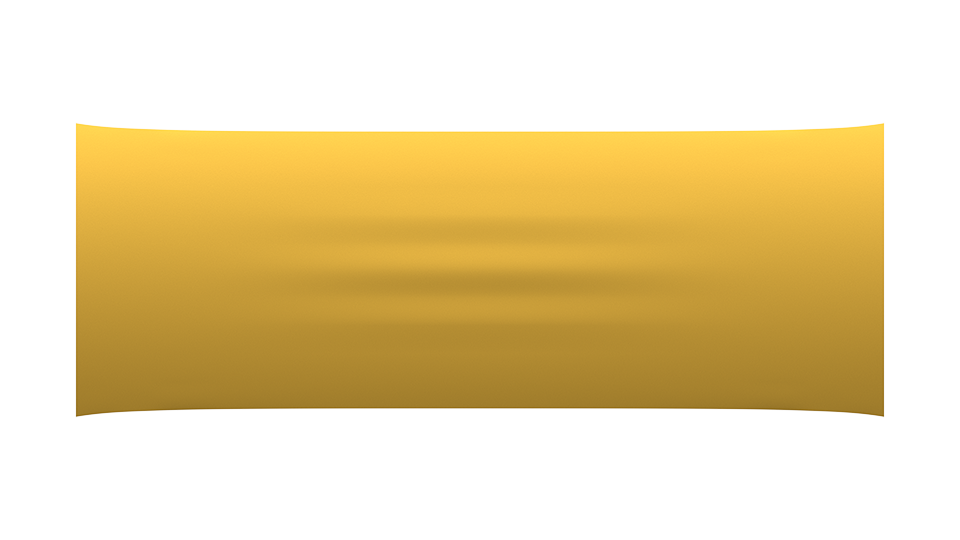}
        \caption{$10\times10$ patches;}
    \end{subfigure}
    \begin{subfigure}{0.495\linewidth}
        \centering
        \formattedgraphics{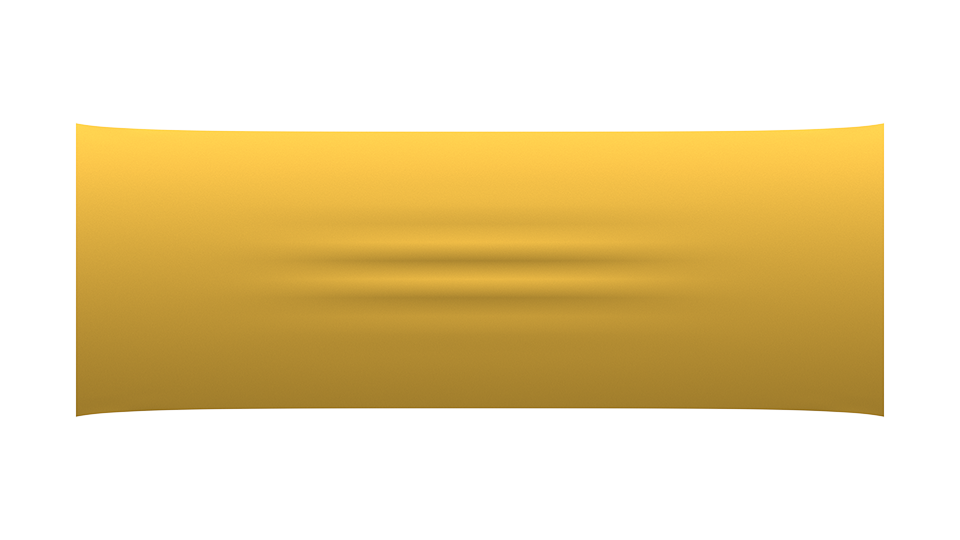}
        \caption{$15\times15$ patches;}
    \end{subfigure}
    \begin{subfigure}{0.495\linewidth}
        \centering
        \formattedgraphics{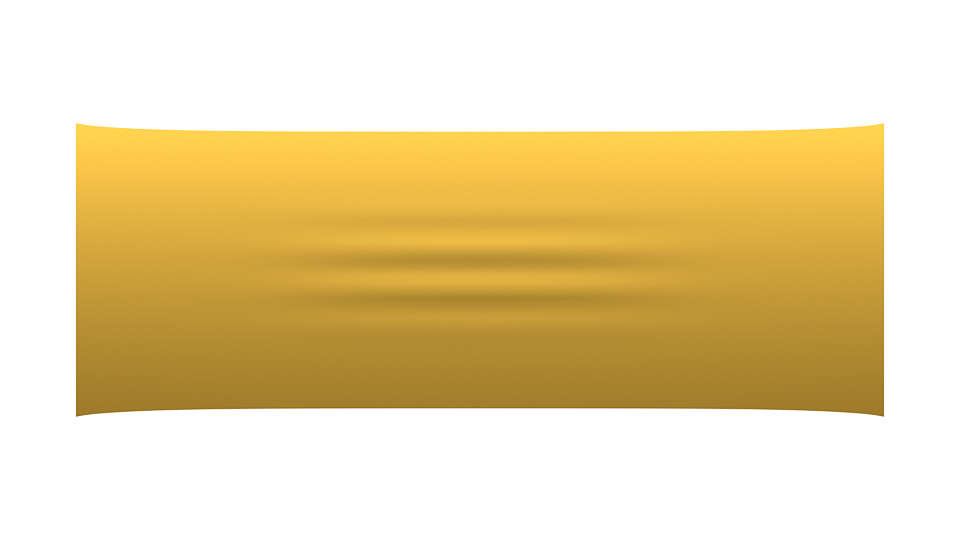}
        \caption{$20\times20$ patches;}
    \end{subfigure}
    \begin{subfigure}{0.495\linewidth}
        \centering
        \formattedgraphics{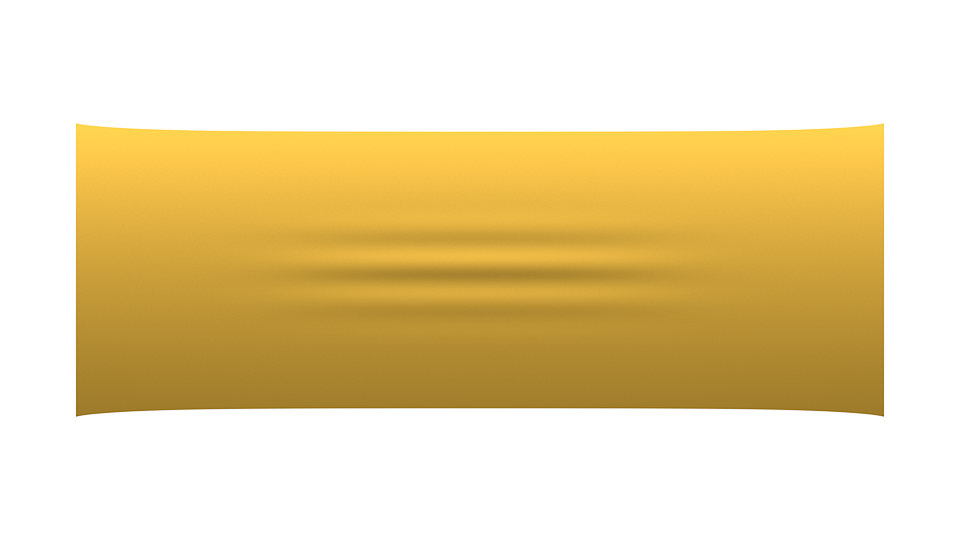}
        \caption{$30\times30$ patches.}
    \end{subfigure}
    \caption{Wrinkled sheets. Sheets of different resolutions buckle under uniaxial stretching. (a) There appears only one artificial bump on a single patch. (b) No apparent wrinkle on a shell composed of $5\times5$ patches. (c) A $10\times10$-patch sheet manages to produce several shallow wrinkles. (d) Clearer wrinkles are produced by a $15\times15$-patch sheet. (e) Though wrinkles are finer, the difference is subtle compared with the previous. (f) The wrinkle pattern converges when it is finer than $30\times30$ patches.}
    \label{fig:wrinkle}
\end{figure}

\begin{figure}[t]
    \centering
    \newcommand{\formattedgraphics}[1]{\includegraphics[width=0.99\textwidth]{#1}}
    \begin{subfigure}{\linewidth}
        \centering
        \formattedgraphics{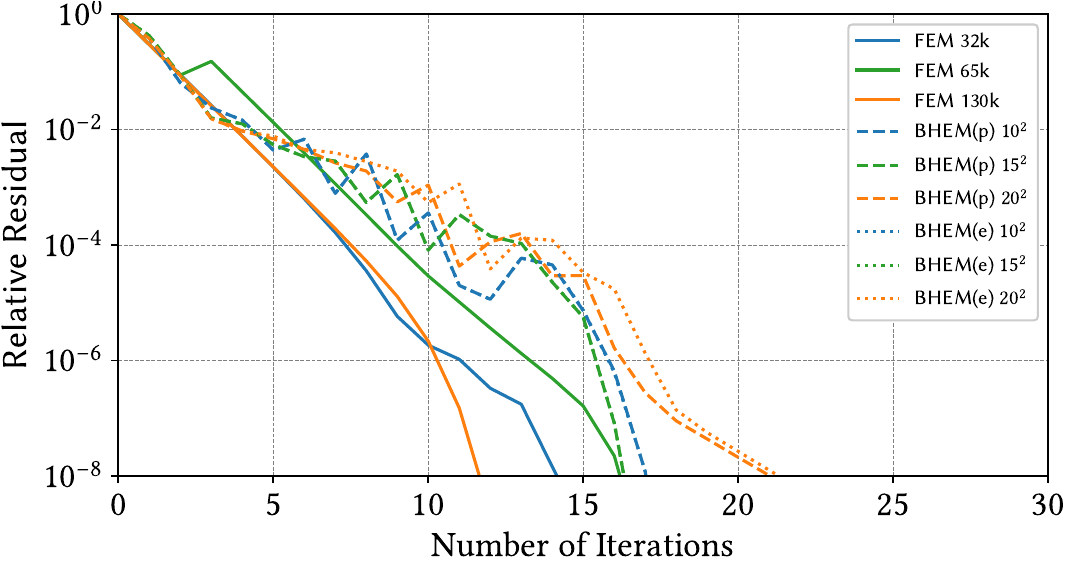}
    \end{subfigure}
    \begin{subfigure}{\linewidth}
        \centering
        \formattedgraphics{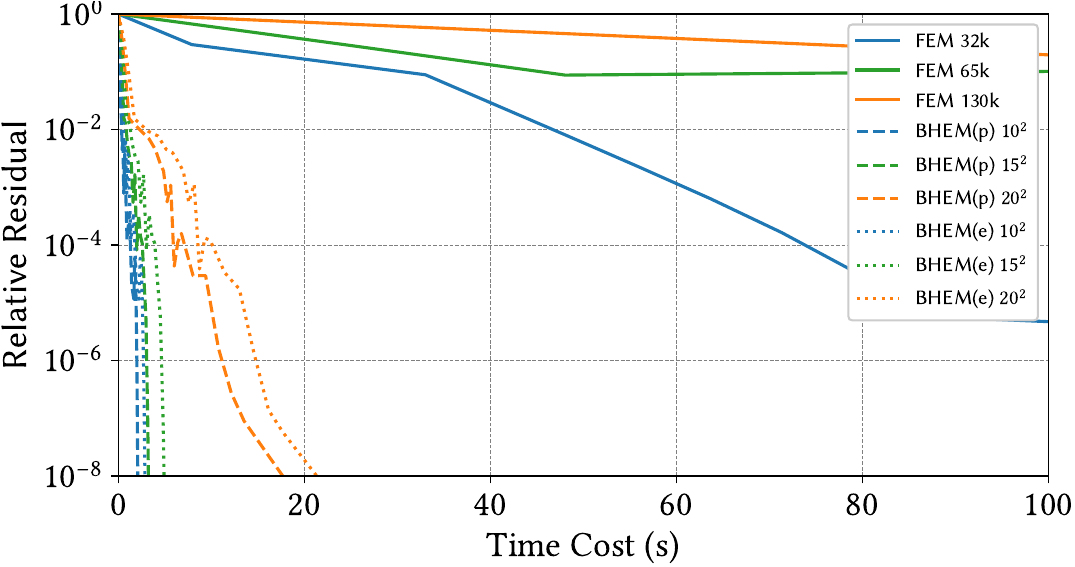}
    \end{subfigure}
    \caption{Performance curves of examples in Fig.~\ref{fig:wrinkle}. "BHEM(p)" and "BHEM(e)" denote BHEM solvers using pseudo-Hessian and exact Hessian matrices respectively. 
    }
    \label{fig:wrinkle_performance}
\end{figure}

\begin{figure*}[t]
    \centering
    \newcommand{\formattedgraphics}[1]{\includegraphics[trim=8cm 2cm 8cm 2cm,clip,width=\textwidth]{#1}}
    \begin{subfigure}{0.163\linewidth}
        \centering
        \formattedgraphics{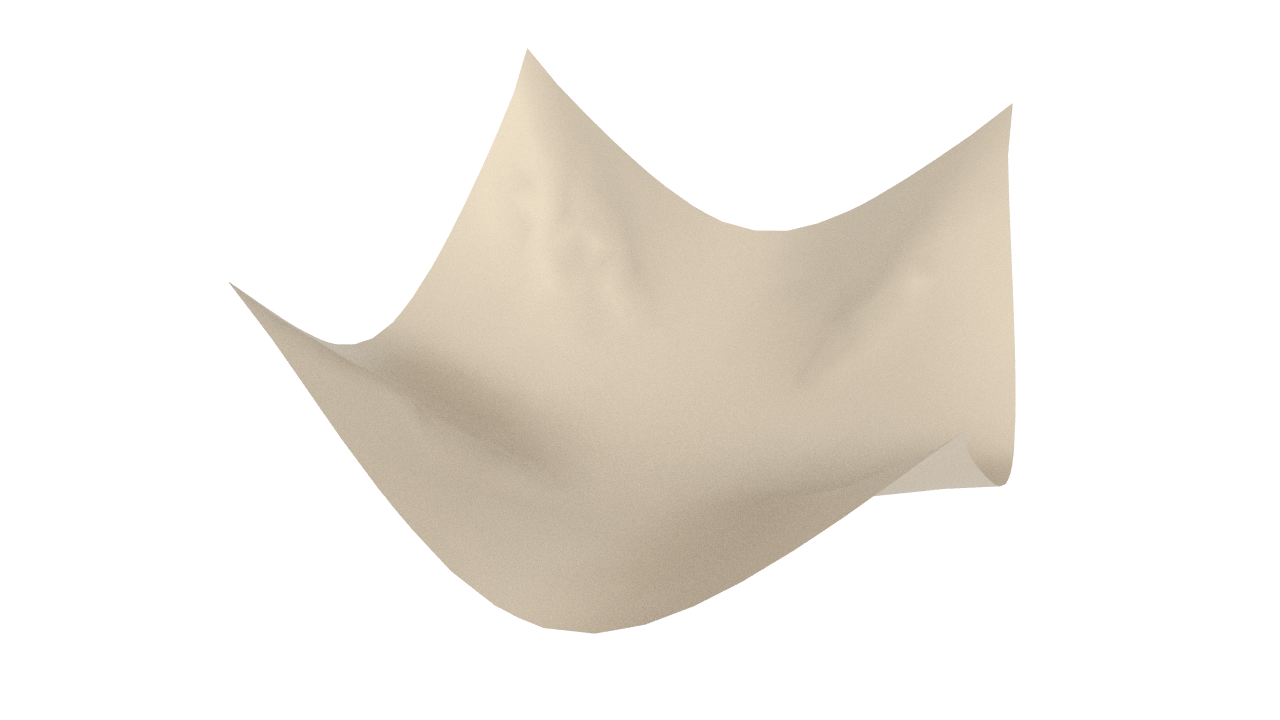}
        \caption{500 vertices;}
    \end{subfigure}
    \begin{subfigure}{0.163\linewidth}
        \centering
        \formattedgraphics{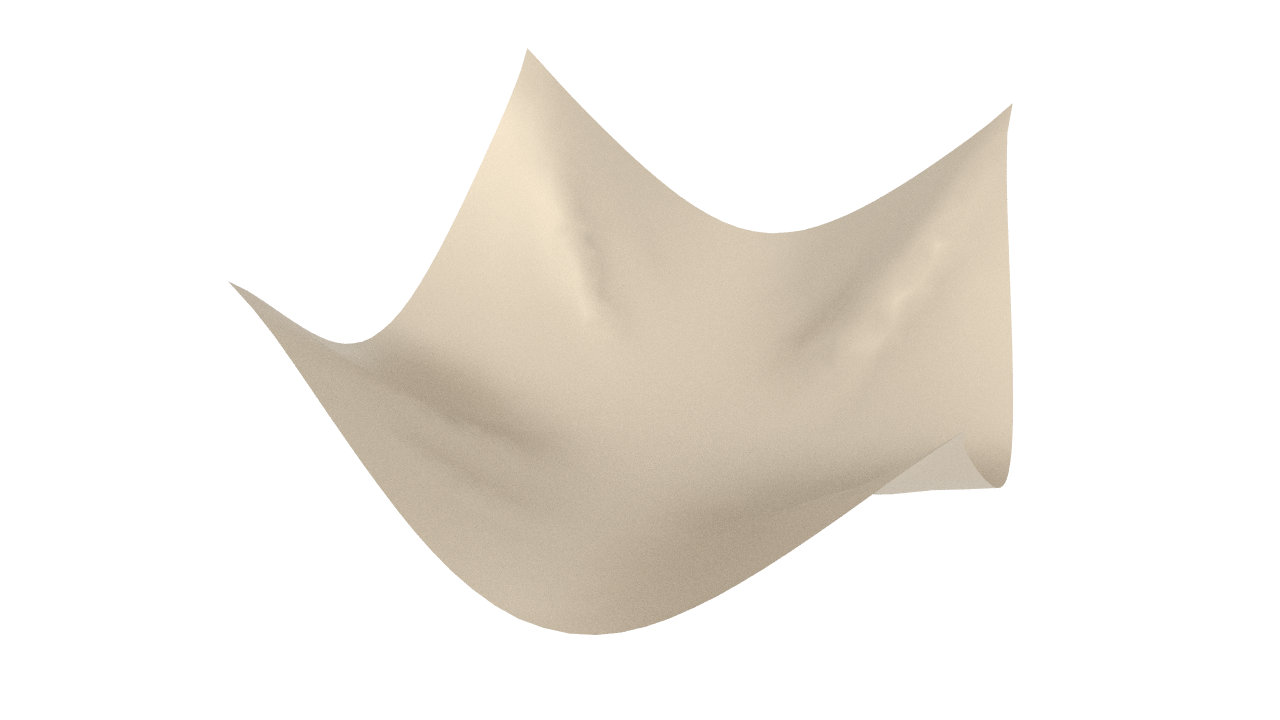}
        \caption{1k vertices;}
    \end{subfigure}
    \begin{subfigure}{0.163\linewidth}
        \centering
        \formattedgraphics{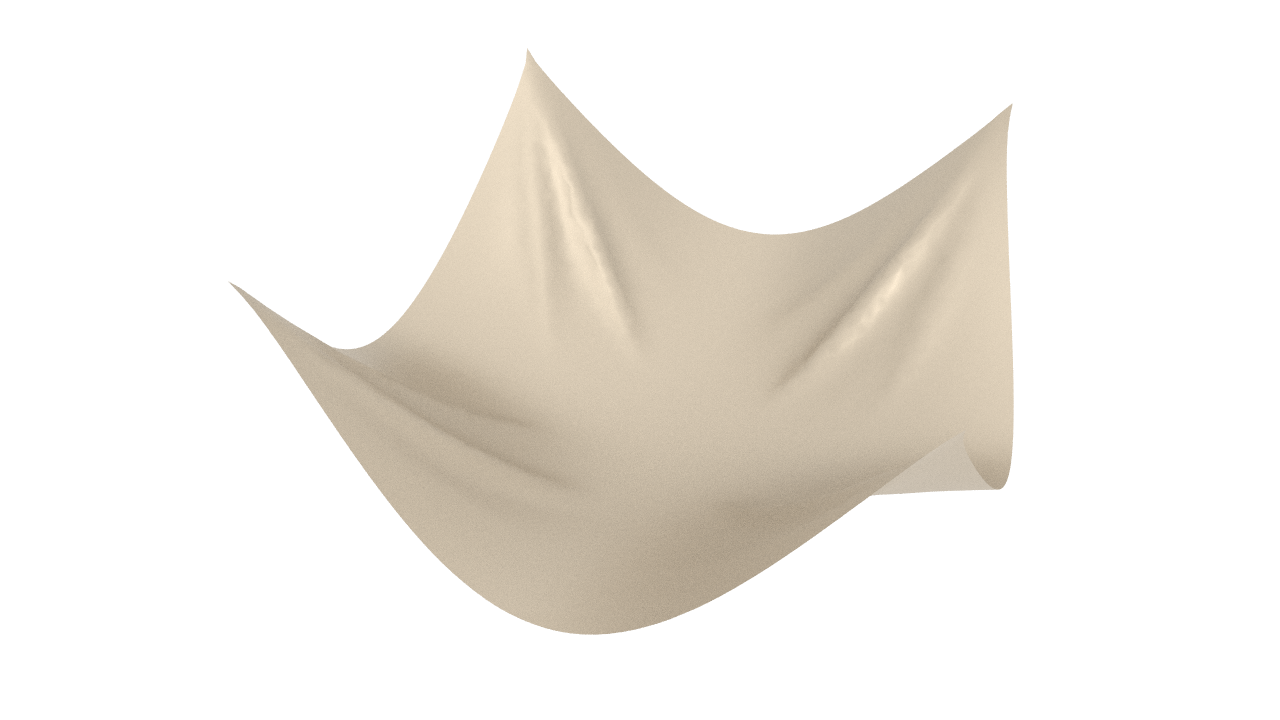}
        \caption{4k vertices;}
    \end{subfigure}
    \begin{subfigure}{0.163\linewidth}
        \centering
        \formattedgraphics{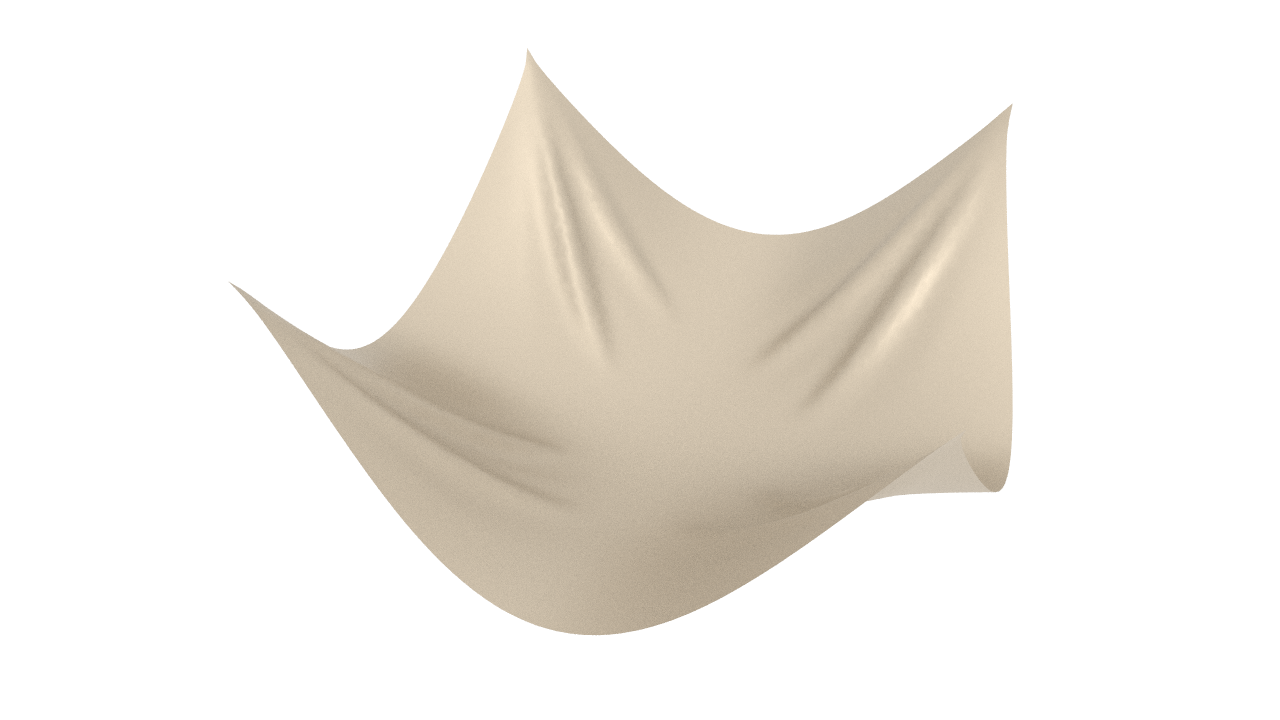}
        \caption{8k vertices;}
    \end{subfigure}
    \begin{subfigure}{0.163\linewidth}
        \centering
        \formattedgraphics{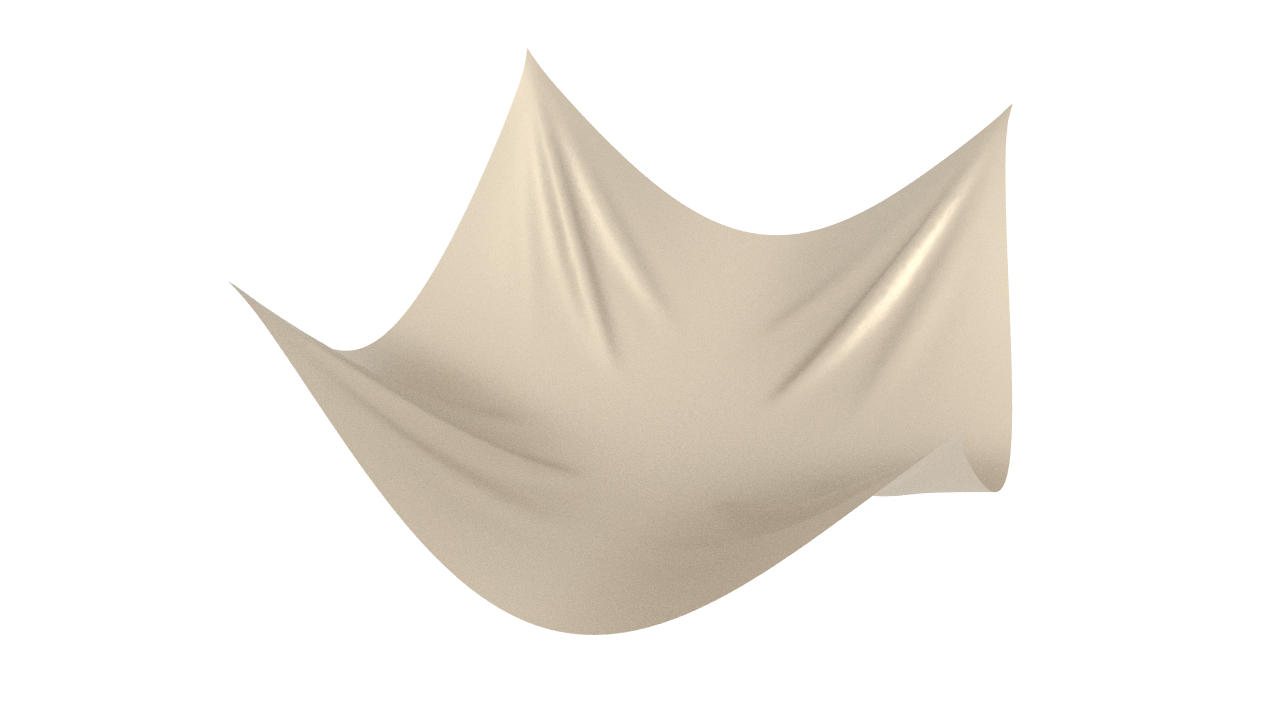}
        \caption{16k vertices;}
    \end{subfigure}
    \begin{subfigure}{0.163\linewidth}
        \centering
        \formattedgraphics{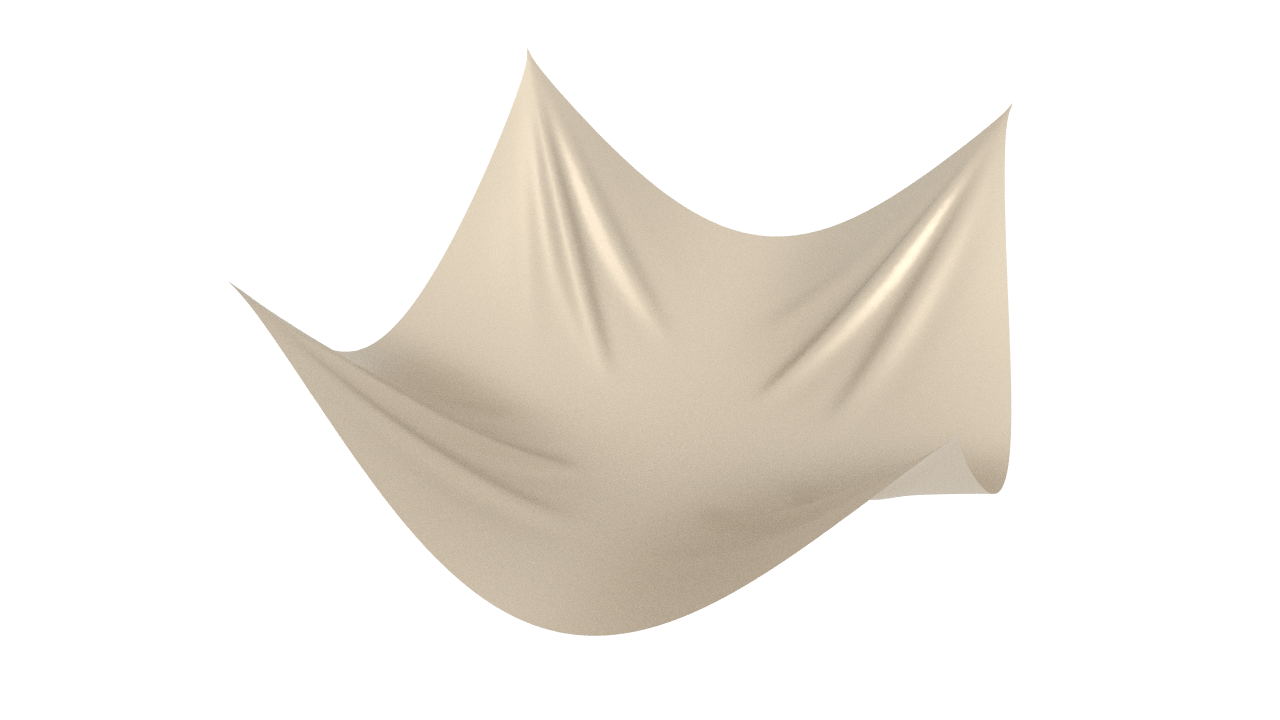}
        \caption{32k vertices;}
    \end{subfigure}
    \begin{subfigure}{0.163\linewidth}
        \centering
        \formattedgraphics{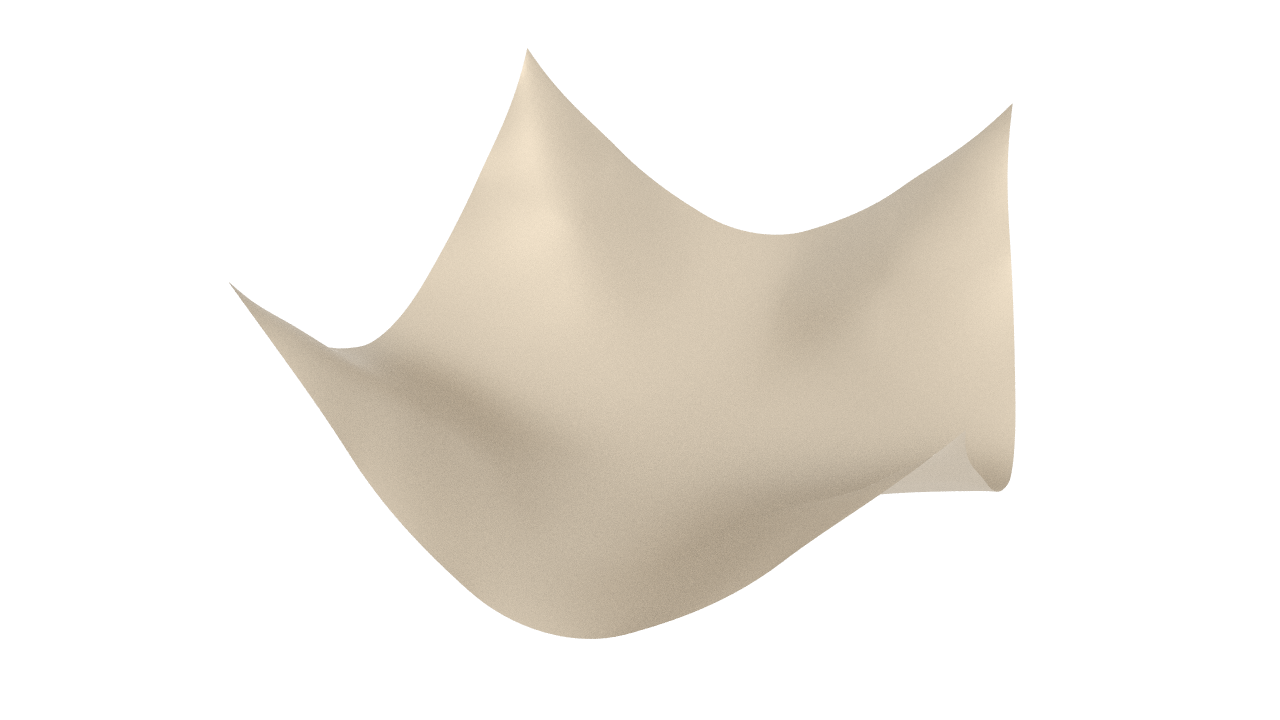}
        \caption{$5\times5$ patches;}
    \end{subfigure}
    \begin{subfigure}{0.163\linewidth}
        \centering
        \formattedgraphics{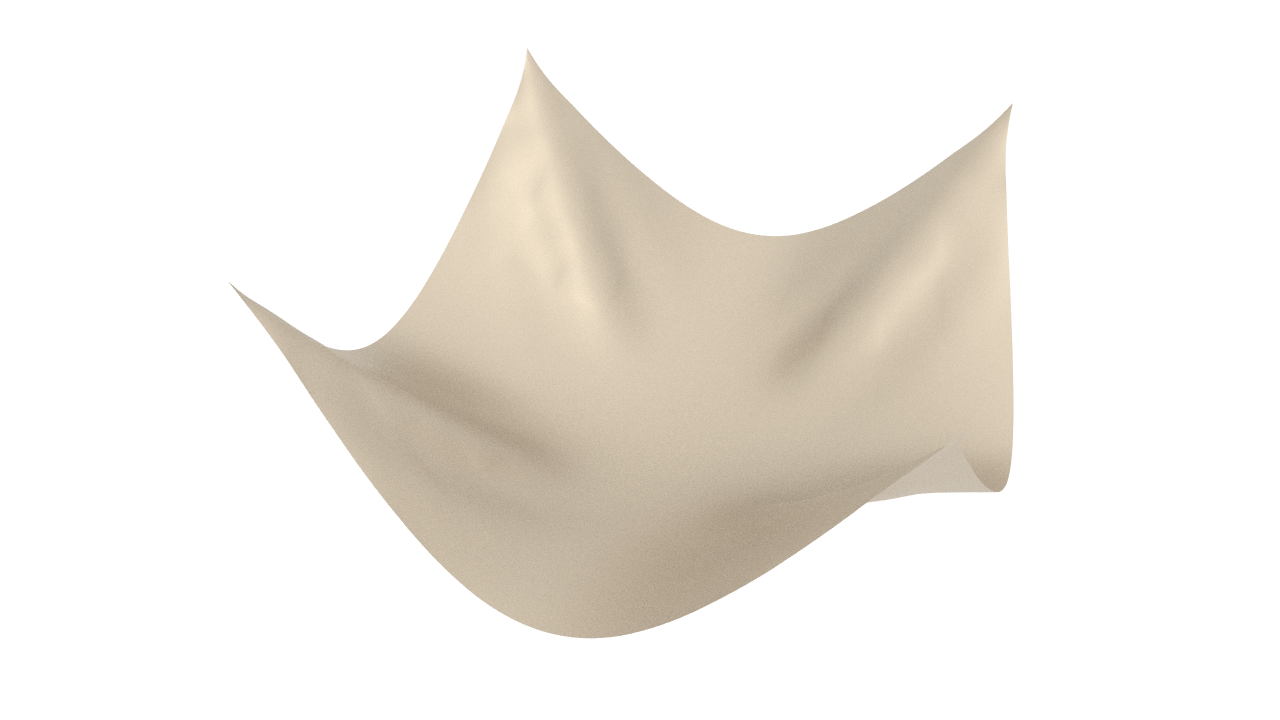}
        \caption{$10\times10$ patches;}
    \end{subfigure}
    \begin{subfigure}{0.163\linewidth}
        \centering
        \formattedgraphics{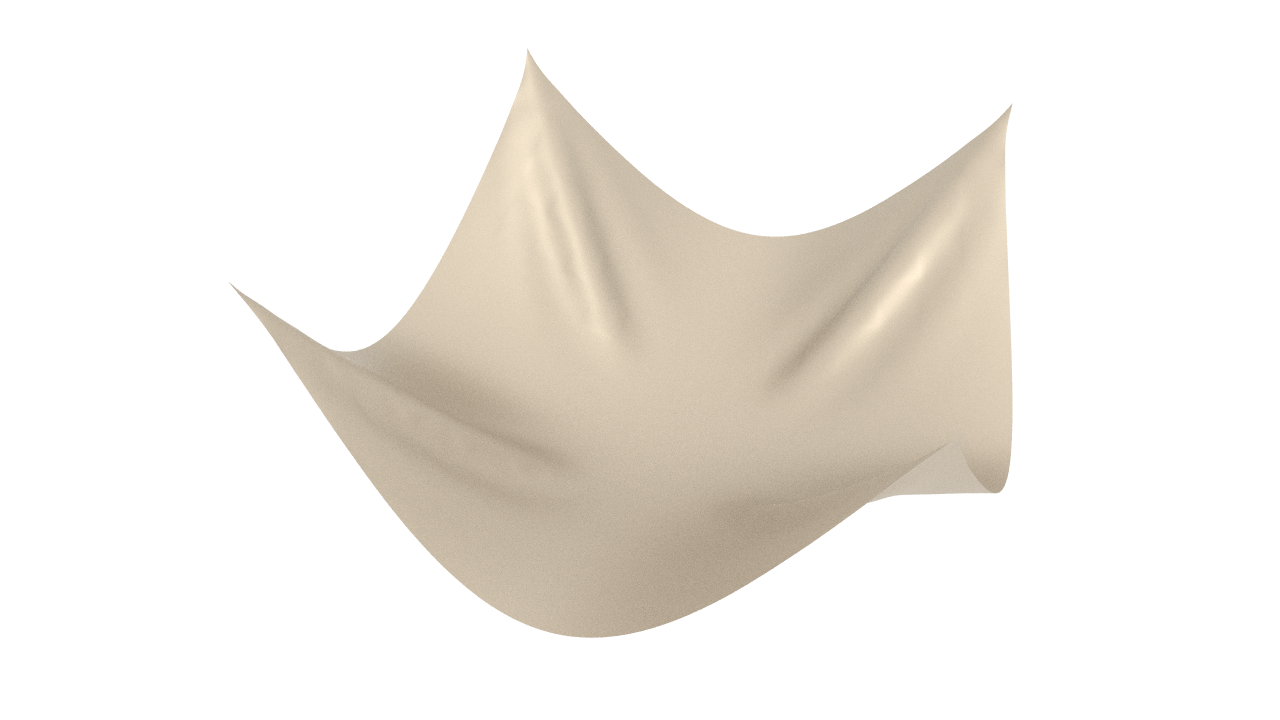}
        \caption{$15\times15$ patches;}
    \end{subfigure}
    \begin{subfigure}{0.163\linewidth}
        \centering
        \formattedgraphics{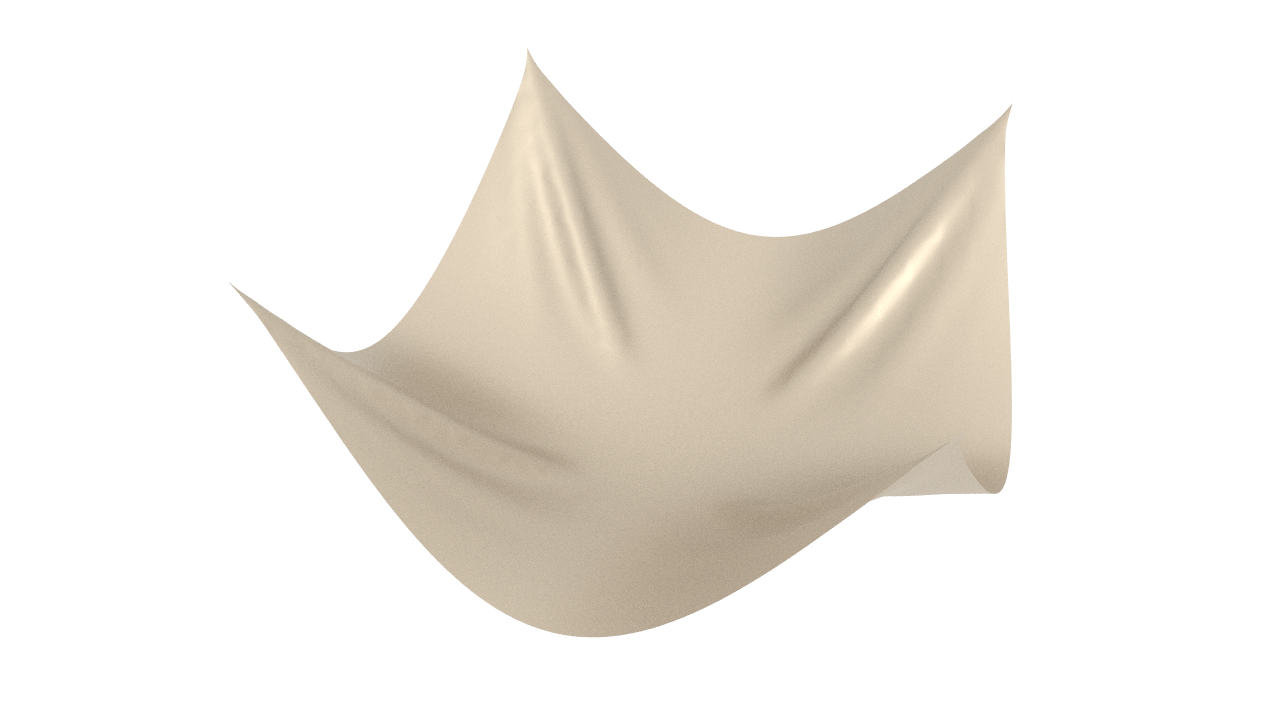}
        \caption{$20\times20$ patches;}
    \end{subfigure}
    \begin{subfigure}{0.163\linewidth}
        \centering
        \formattedgraphics{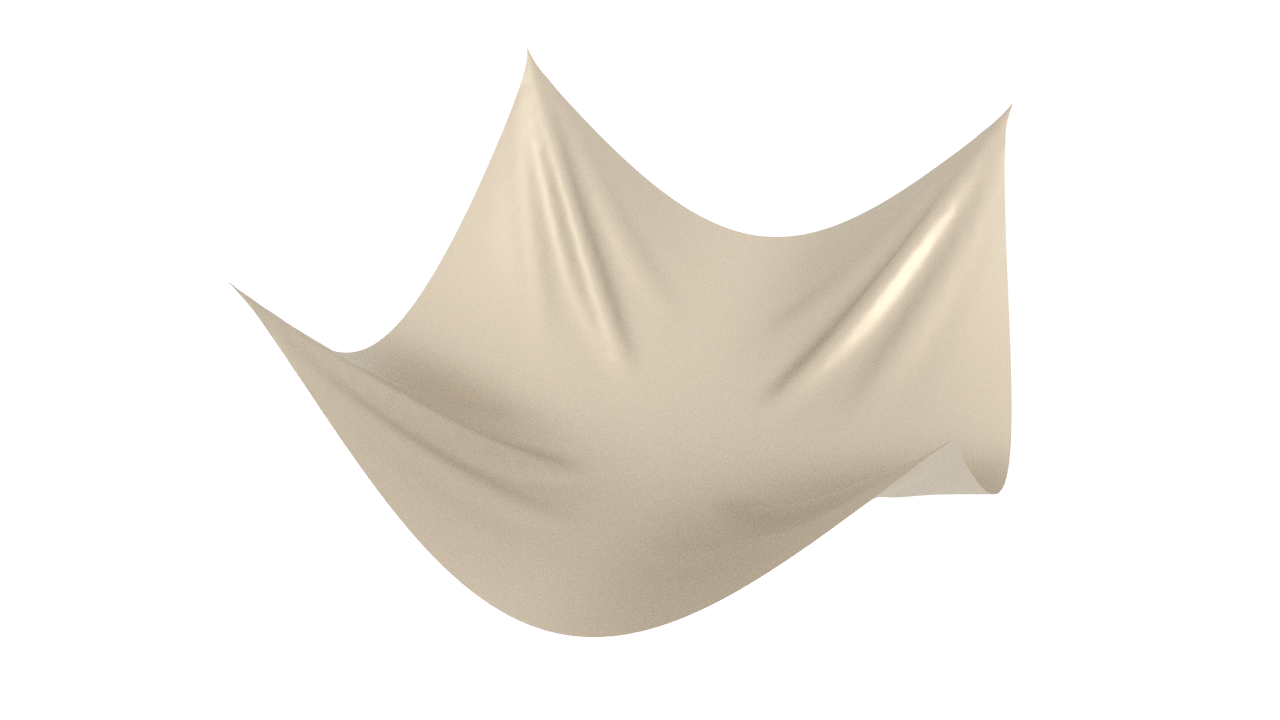}
        \caption{$25\times25$ patches;}
    \end{subfigure}
    \begin{subfigure}{0.163\linewidth}
        \centering
        \formattedgraphics{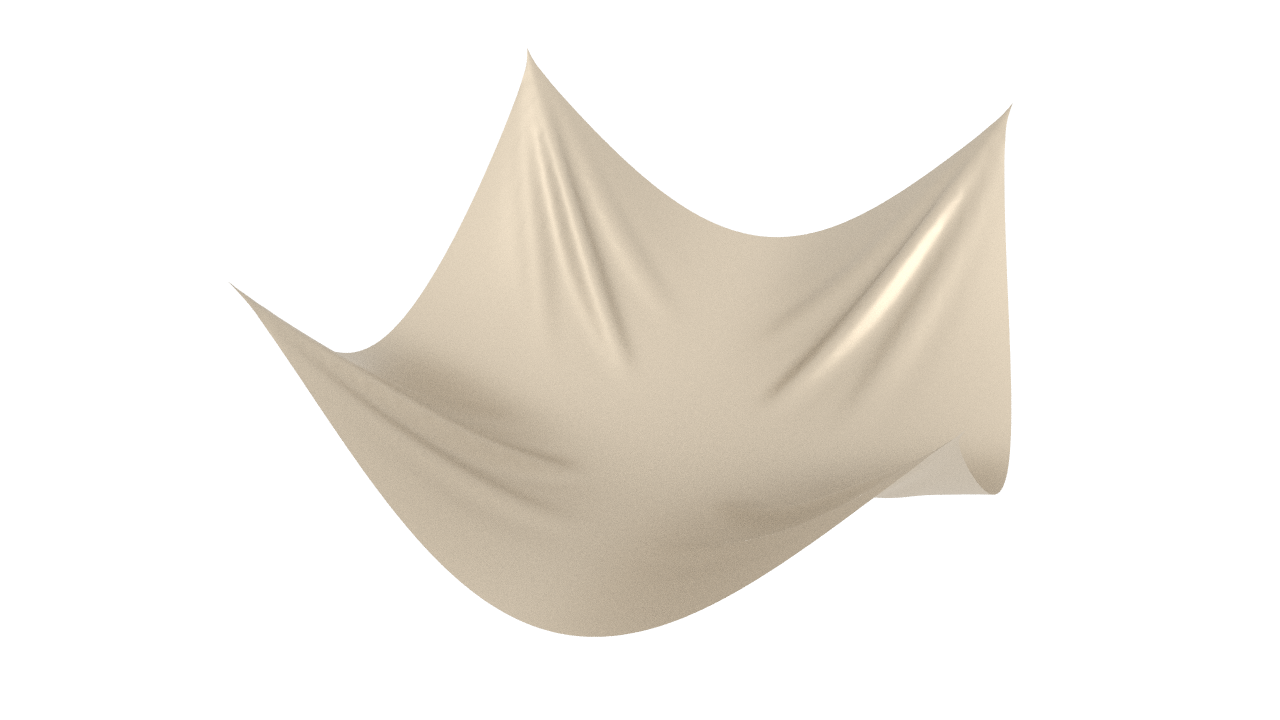}
        \caption{$30\times30$ patches.}
    \end{subfigure}
    \caption{Draped cloth. The top row and bottom row show the results obtained by FEM and BHEM, respectively. The resolution of the discretized surfaces increases from left to right. We choose the examples with the highest resolutions and similar plausible wrinkles and test the performance of the solvers under these resolutions. The performance curves are shown in Fig.~\ref{fig:drape_performance}}
    \label{fig:drape}
\end{figure*}
\begin{figure}[t]
    \centering
    \newcommand{\formattedgraphics}[1]{\includegraphics[width=0.99\textwidth]{#1}}
    \begin{subfigure}{\linewidth}
        \centering
        \formattedgraphics{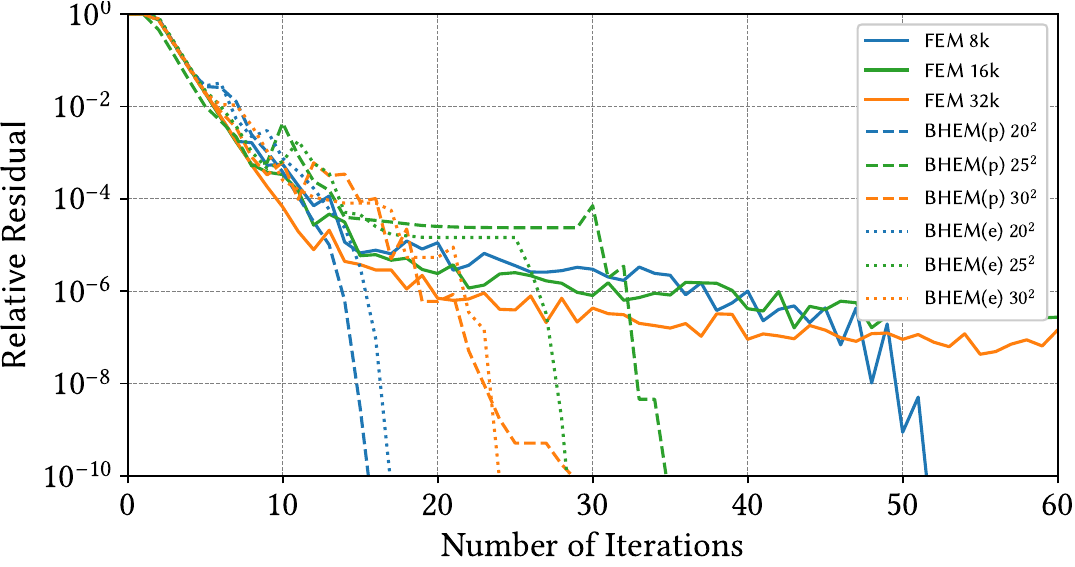}
    \end{subfigure}
    \begin{subfigure}{\linewidth}
        \centering
        \formattedgraphics{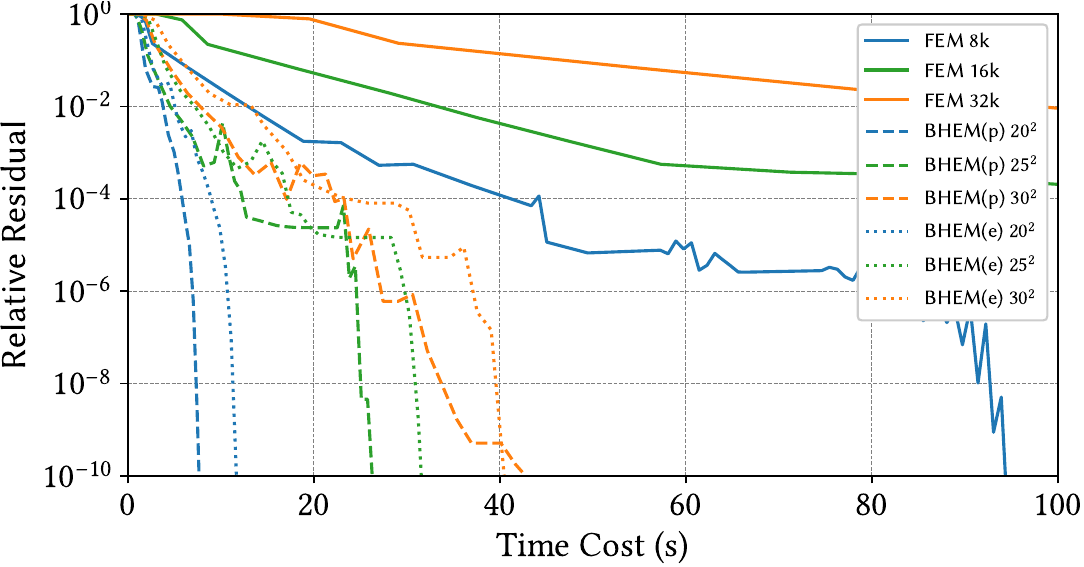}
    \end{subfigure}
    \caption{Performance curves of the example in Fig.~\ref{fig:drape}.}
    \label{fig:drape_performance}
\end{figure}

\paragraph*{Wrinkled sheets.}
We first validate the accuracy and convergence of our BHEM method with a standard stretched sheet experiment, which was first proposed by the pioneering work of Cerda et al.\ \cite{Cerda2003} and later used in both physical engineering \cite{WANG2018} and computer graphics \cite{Chen2021} community as well. In this experiment, a rectangular thin sheet is pulled apart from its two ends. Due to the high Poisson ratio, the sheet compresses in the perpendicular (vertical) and generates horizontal wrinkles. 

We simulate this problem with the same physical parameter settings as in the work of \cite{Chen2021}, which are $\SI{0.25}{\meter} \times \SI{0.1}{\meter}$ for size, $\SI{0.1}{\milli\meter}$ for thickness, Poisson ratio  $\nu= 0.5$ and
Young’s modulus $Y=\SI{1}{\mega\pascal}$.
According to the conclusion of Chen et al.\ \cite{Chen2021}, the traditional triangular finite-element-based method needs a high resolution of up to tens of thousands of vertices to generate correct wrinkle patterns in this example.  
As shown in Fig.~\ref{fig:wrinkle}, the BHEM starts to give apparent wrinkles at a resolution of merely $10^2$ patches (1,452 DoFs in total). As the resolution increases over $15^2$, the difference in the shape of wrinkles is already imperceptible. 

\begin{figure}[t]
\centering
\includegraphics[width=\linewidth]{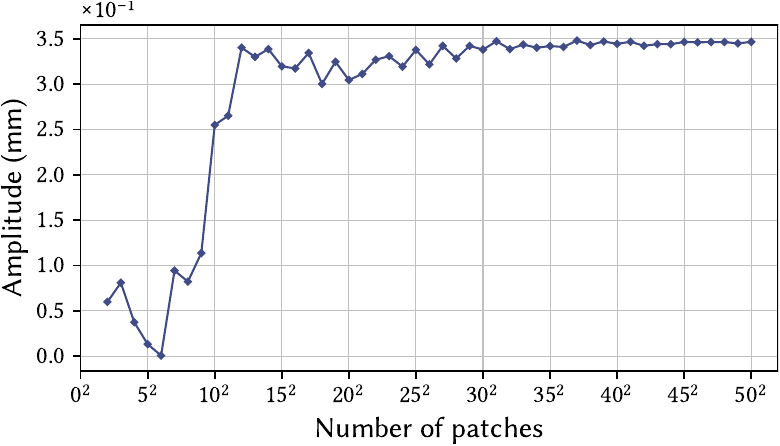}
\caption{Statistic of the wrinkle's peak value. The simulated amplitude increases with the refinement of the sheet and converges rapidly to a physical real value ($\SI{3.5e-1}{\milli\meter}$). 
}.
\label{fig:wrinkle_curve}
\end{figure}

We further verify this observation through a quantitative experiment that measures the peak amplitude of wrinkles produced with increased resolution. 
According to the physical experiment results reported by Wang et al.\ \cite{WANG2018}, the sheet is expected to produce wrinkles with a peak amplitude of $\SI{0.35}{\milli\meter}$. 
The curve illustrated in Fig.~\ref{fig:wrinkle_curve} shows that our BHEM can stably yield $\SI{0.34}{\milli\meter}$ peak amplitude when the simulation resolution is greater than $30^2$ patches (11.5k DoFs in total). While the peak amplitude can only reach $\SI{0.31}{\milli\meter}$ for the traditional FEM simulated on a triangular mesh with 130k vertices (390k DoFs in total).

In Fig.~\ref{fig:wrinkle_performance}, we also demonstrate the efficiency of BHEM by comparing the convergence curves with those of randomly triangulated thin shells using the standard FEM (\emph{Libshell} \cite{libshell}) under different resolutions. FEM solver takes fewer iterations but much more time to get a converged solution. This is probably because the BH system mixes DoFs of different orders, which, in the meanwhile, allows the BH surface to present similar high-frequency visual effects with fewer DoFs. 

\paragraph*{Draped cloth.}
We conduct a set of comparative experiments to further demonstrate the superiority of our geometric discretization format. In this scenario, a square piece of cloth, with its four corners moved inward a bit and clamped,
drapes from a flattened configuration under the influence of gravity. 
In Fig.~\ref{fig:drape}, we demonstrate the results simulated with our BHEM and standard FEM (\emph{Libshell} \cite{libshell}) under different discretization resolutions respectively. The comparison shows that BHEM produces more vivid wrinkles when the number of DoFs is comparable (11.5k DoFs for $30^2$-patch BH surface, and 12k DoFs for $4$k-vertex FE surface). 
Additionally, when using a low discretization resolution, BHEM (432 DoFs for $5^2$ patches) presents an over-smoothed surface but FEM (1.5k DoFs for 500 vertices) produces fake wrinkles, as shown in the first column.
When using a high discretization resolution, the quality of standard FEM improved, however, BHEM can achieve a similar quality requiring only 11.5k DoFs ($30^2$ BH patches), and thus takes much less iterations and time to converge, as shown in Fig.~\ref{fig:drape_performance}.
We compare the convergence speed of the examples in the three columns on the right in Fig.~\ref{fig:drape_performance}. It also shows that a BHEM solver with exact Hessian generally requires more time to compute but fewer iterations to converge, compared to a BHEM solver using pseudo-Hessian.


\begin{figure}[t]
    \centering
    \newcommand{\formattedgraphics}[1]{\includegraphics[trim=0cm 0cm 0cm 0cm,clip,width=0.324\linewidth]{#1}}
    \formattedgraphics{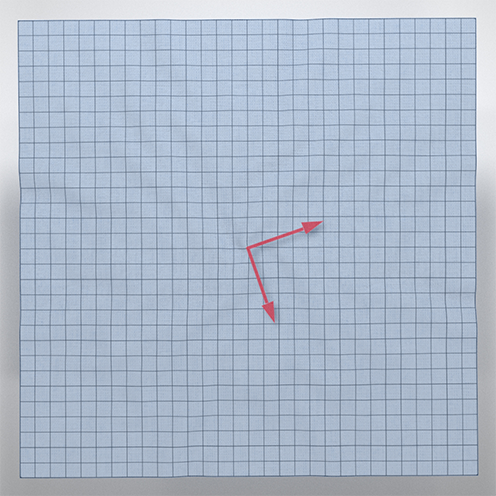}
    \formattedgraphics{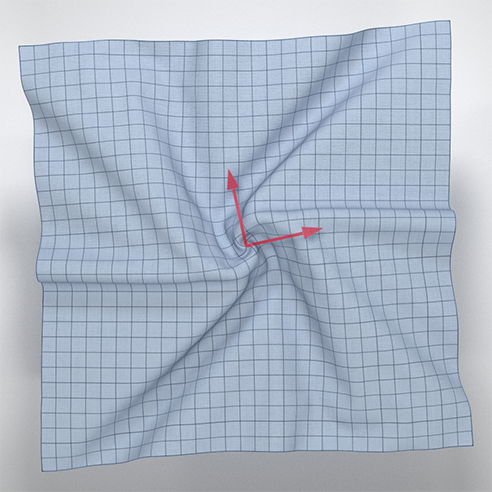}
    \formattedgraphics{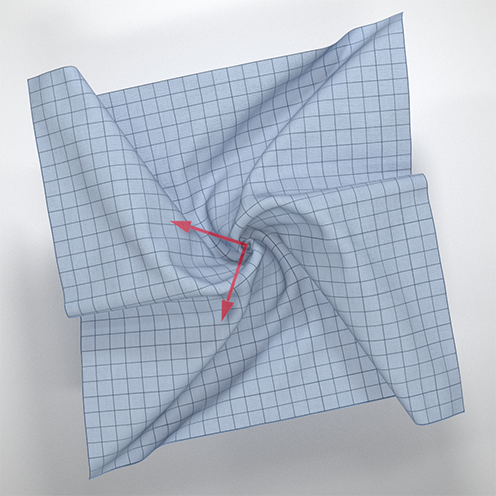}
    \caption{Twisting cloth ($30\times30$ patches). A buckling swirl around the center as the first-order derivatives (indicated by the red arrows) of the center point rotate horizontally with a constant speed.}
    \label{fig:drill}
\end{figure}


\paragraph*{Parametric surface rendering.}
Our proposed ray-surface intersection algorithm would be regarded as a substantial enrichment for the current off-the-shell ray-tracing rendering engine. By simply substituting the current ray-polygon intersection detection module in pbrt-v4 \cite{Pharr2023} with our algorithm, pbrt-v4 can realize parametric surface ray-tracing rendering with good visual effects, as shown in the rightmost columns of Fig.~\ref{fig:teapot} and Fig.~\ref{fig:caustic}. The other columns are the rendering results for the polygon meshes generated from the parametric surface subdivision. From left to right, the subdivision level gradually increases. A parametric surface naturally processes continuous normal vectors on its surface. Fairly fine mesh is required to achieve a similar rendering result. 



\begin{figure}[t]
\includegraphics[trim = 0cm 0cm 0cm 0cm, clip,width=\linewidth]{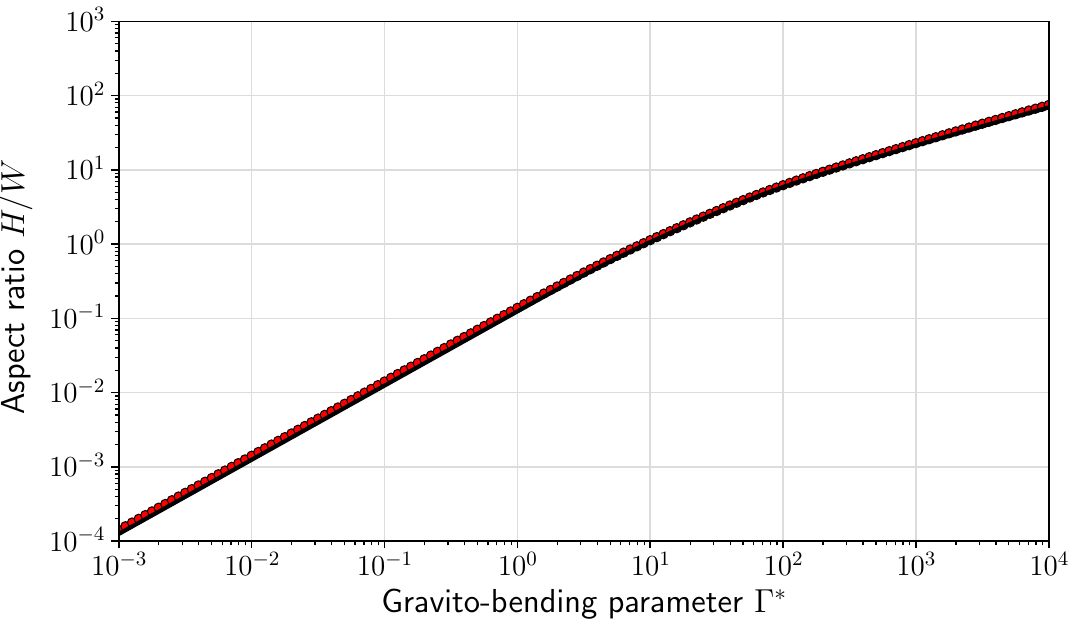}
\caption{Comparisons with theoretical solutions on the cantilever test. We simulate 140 $\Gamma^*$ values and superimpose the data (red dots) onto the master curve (black line). Our results perfectly match the master curve.}
\label{fig:cantilever}
\end{figure}

\begin{figure}[t]
\includegraphics[trim = 0cm 0cm 0cm 0cm, clip,width=\linewidth]{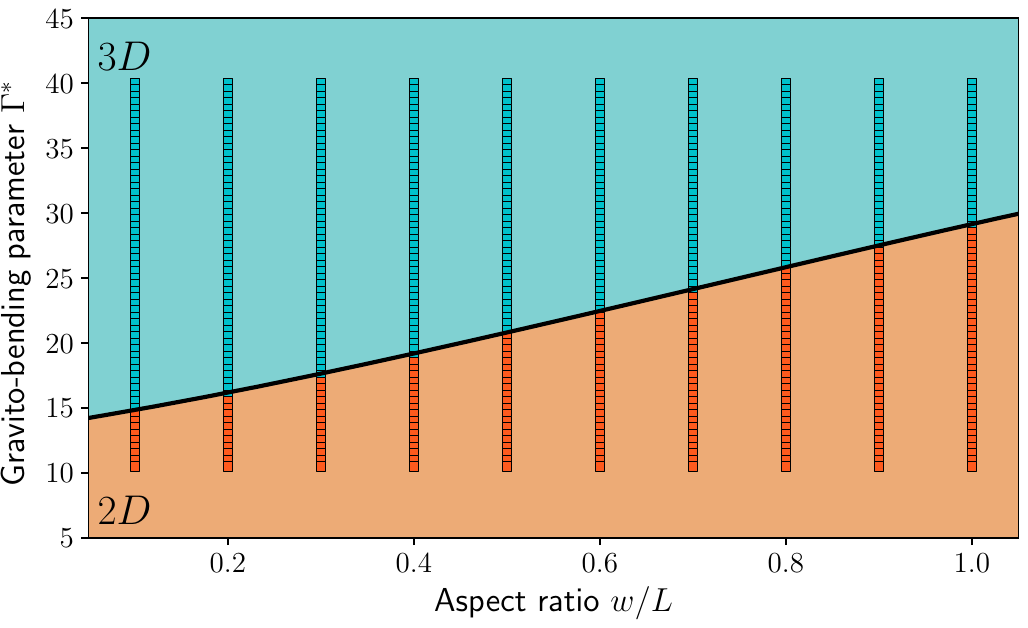}
\caption{Comparisons with theoretical solutions on the lateral buckling test. The master curve separates two areas colored to indicate whether the plate has buckled in 3D (turquoise) or lies in 2D (orange). Our results perfectly match the theoretical solution.}
\label{fig:lateralbuckling}
\end{figure}

\paragraph*{2D cantilever beam and 3D lateral buckling.}
We further validate our method by comparing two experiments with their theoretical solutions proposed by Romero et al.\ \cite{Romero2021}. 
The first is the cantilever beam experiment. 
We fix one end of a beam and let it bend under gravity. Its master curve uniquely determines the aspect ratio $H/W$ of the cantilever beam under equilibrium as a function of the dimensionless parameter $\Gamma^* = 12(1-\nu^2)\rho gL^3/Yh^2$. 
The second is the lateral buckling experiment with the plate lying vertically in the $(x,z)$ plane. We let the plate hang and sag under its weight, waiting for a bifurcation to occur. Its master curve determines when the plate buckle in the third direction as $\Gamma^*$ increases under a given aspect ratio $w/L$.
We conduct our experiments using methods and physical parameters provided by Romero et al.\ \cite{Romero2021}. The beam in the cantilever test is discretized in $2\times 100$ patches. And the plate is discretized in 20 patches per meter. Our results perfectly match the master curves as shown in ~\ref{fig:cantilever} and ~\ref{fig:lateralbuckling}.

\subsection{Experiments}

\begin{figure}[t]
    \newcommand{\formattedgraphics}[1]{\includegraphics[trim=5.5cm 2cm 5.75cm 1.5cm,clip,width=0.324\linewidth]{#1}}
    \centering
    \formattedgraphics{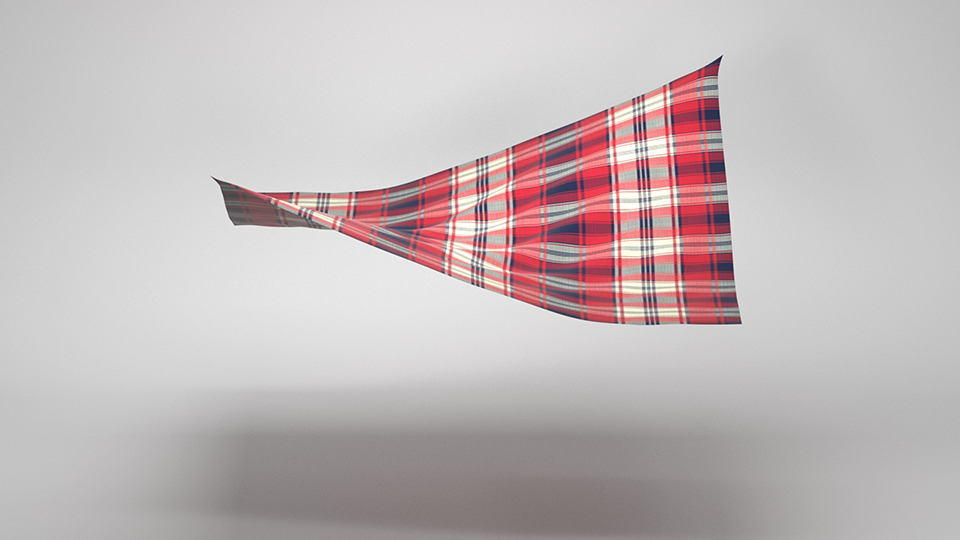}
    \formattedgraphics{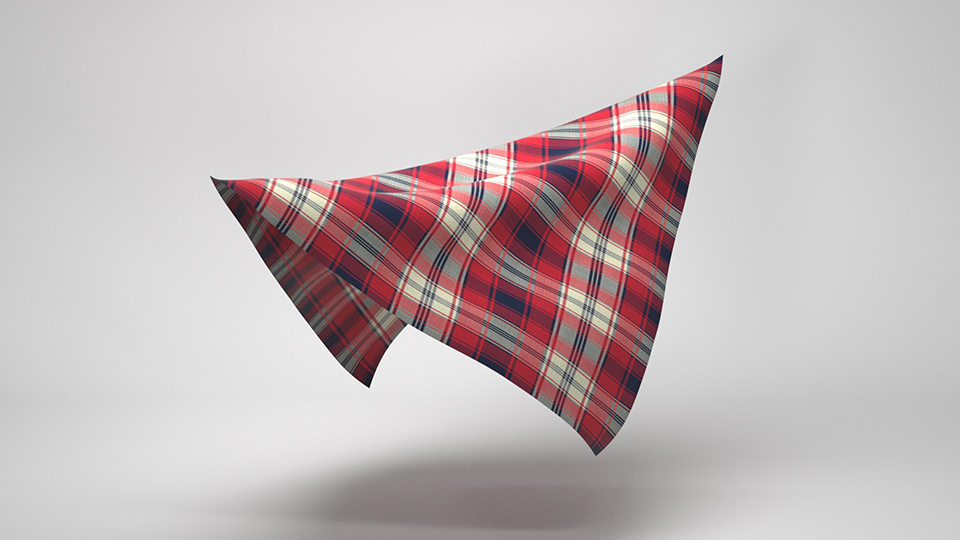}
    \formattedgraphics{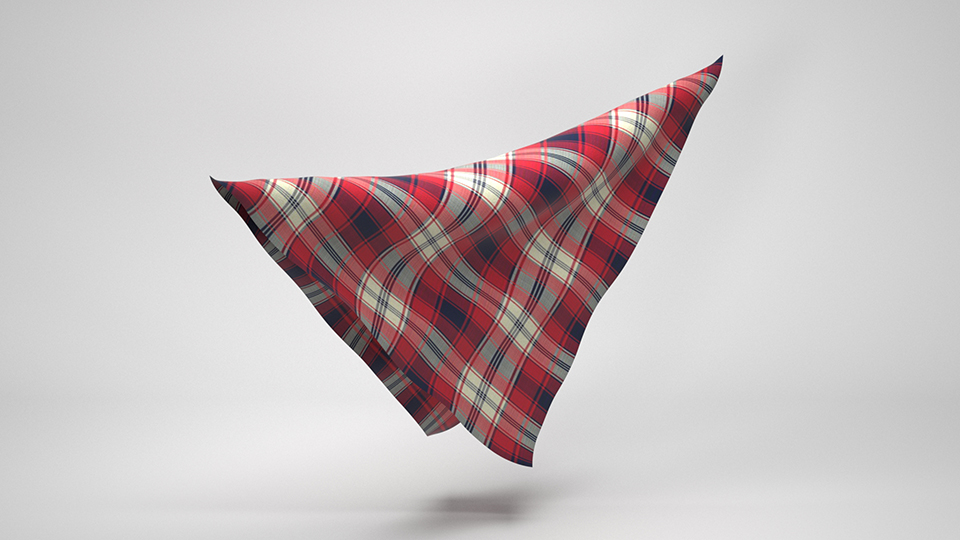}\\
    \formattedgraphics{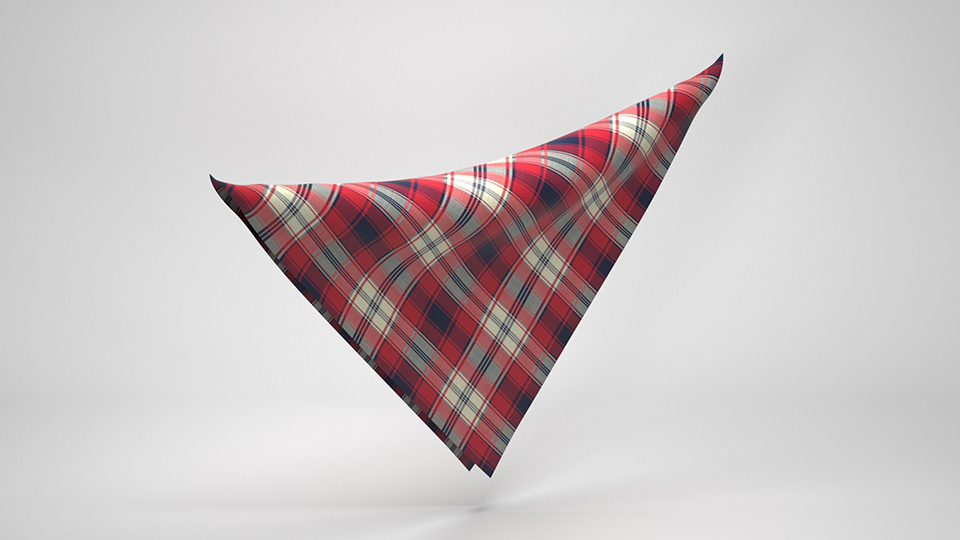}
    \formattedgraphics{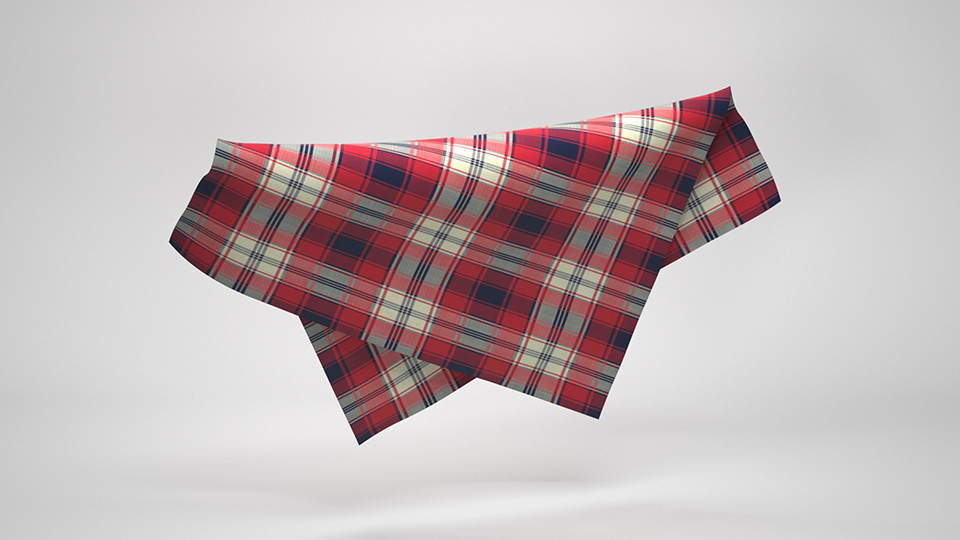}
    \formattedgraphics{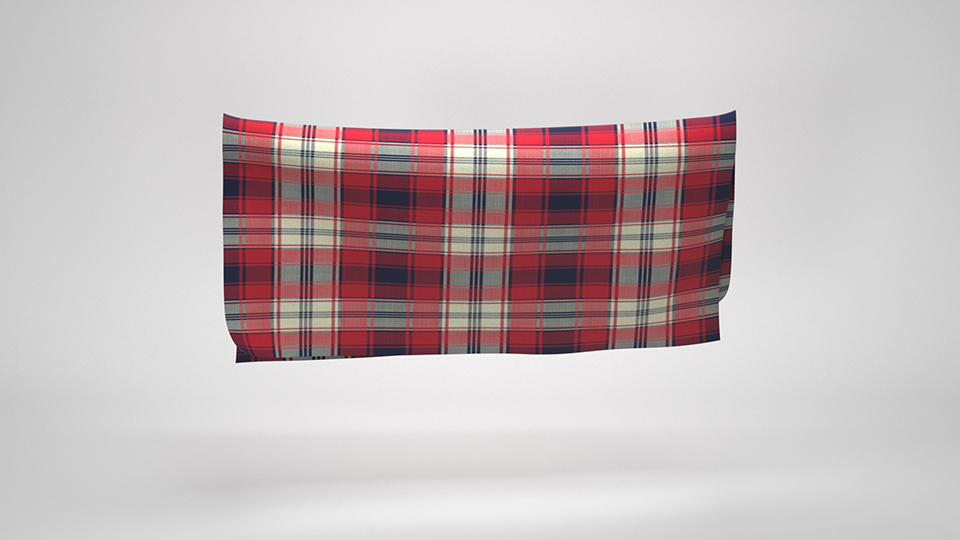}
    \caption{A thin sheet ($10\times10$ patches) with its two diagonal ends fixed bends by gravity. After reaching a steady state, the fixed points slide smoothly along the opposite edges without any locking artifacts from discretization. We emphasize that only $10\times10$ patches are used here.}
    \label{fig:locking}
\end{figure}

\paragraph*{Locking.}
As a high-order method, the BHEM formulation alleviates locking issues remarkably. It doesn't need any special treatment to get plausible effects, such as dynamic remeshing, even with very few degrees of freedom.
As shown in the top row of Fig.~\ref{fig:locking}, a squared piece of cloth, which is composed of $10^2$ patches, is initially pinned at its two diagonal corners. As the pinning points slide along the boundary, the cloth can naturally fold down along any direction as illustrated in the bottom row of Fig.~\ref{fig:locking}. 

\paragraph*{Twisting cloth.}
Our method allows precise and intuitive control over the first-order derivatives of an arbitrary point on the surface. 
In this example, the center point of a squared piece of cloth is fixed. Its two first-order partial derivatives (indicated by the red arrows in Fig.~\ref{fig:drill} are rotated with a constant speed in the horizontal plane. 
This prescript motion results in a persistent wrinkling perpendicular to the first derivative directions and eventually invokes the buckling swirl around the center.
In a pure displacement-based method, a similar result can only be achieved through prescript at least three nodes' motion. 

\paragraph*{Cloth draping on objects.}
Both resting and dynamic contact can be faithfully detected and resolved within our framework. 
Two square sheets of cloth, both of which are composed of $30^2$ patches, drape on a parametric sphere and a triangular meshed armadillo are demonstrated in Fig.~\ref{fig:ball} and Fig.~\ref{fig:armadillo} respectively. Rich deformation details due to the interference from external objects can be observed in the results. Please see the supplemental video for more visual evidence.


\paragraph*{Cloth sliding over needles.}

In this example, we drop a sheet of cloth on a needle array and then pull it away from aside. The subtle bulges on the cloth surface, which are pushed out by the needle tips, can be clearly observed in Fig.~\ref{fig:needles}. 
In the supplemental video, we can easily notice that when pulling the cloth over the needle array, the cloth exhibits natural choppy movement around the needle tips due to the contact interaction. 
These two points verify that our BHEM formulation and collision detection algorithm can handle sharp geometry features robustly.
 

\paragraph*{Folding an oriental paper parasol. }
By jointly controlling node positions and their first-order derivatives, we can mimic the folding process of an oriental paper parasol driven by the motion of its rib as displayed in Fig.~\ref{fig:umbrella}.

\begin{figure}[t]
    \newcommand{\formattedgraphics}[2]{\begin{subfigure}{0.324\linewidth}\centering\includegraphics[trim=11cm 0cm 11cm 2.5cm,clip,width=\textwidth]{#1}\caption{#2}\vspace{-0.3cm}\end{subfigure}}
    \centering
    \formattedgraphics{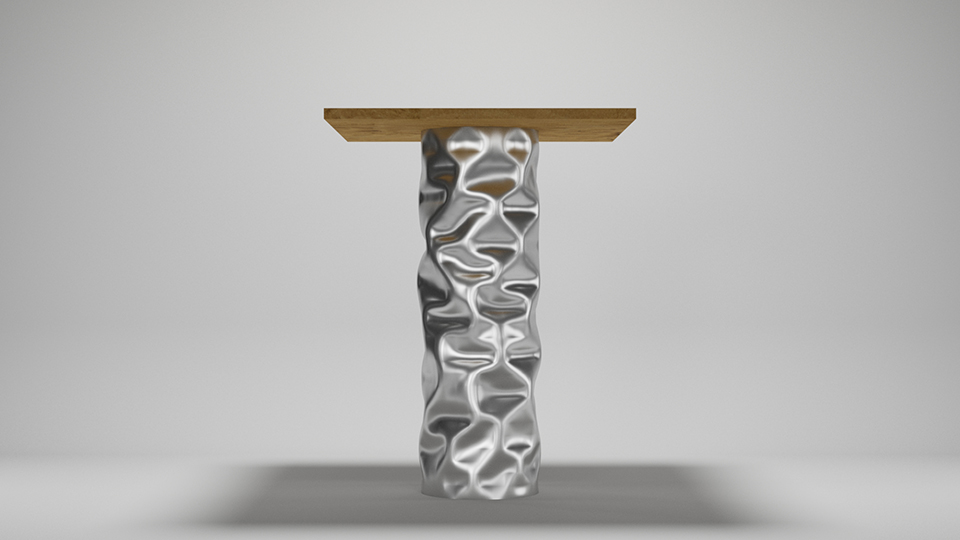}{$h=0.5$mm}
    \formattedgraphics{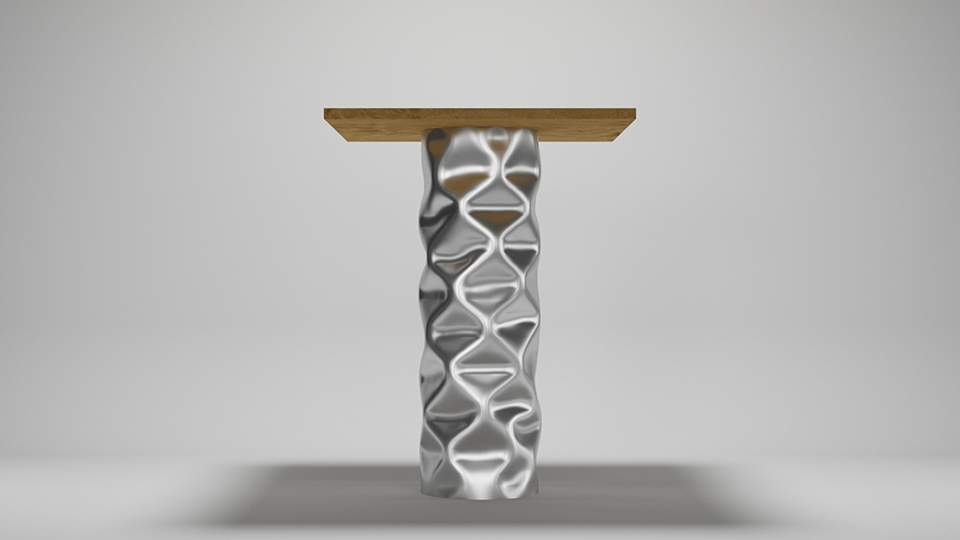}{$h=1$mm}
    \formattedgraphics{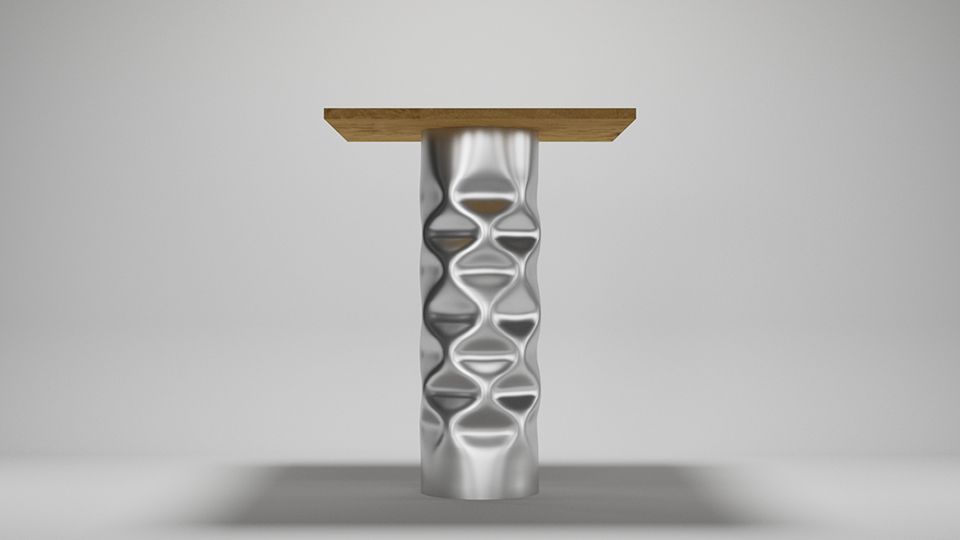}{$h=2$mm}
    \caption{ Hollow cylindrical shells ($30\times30$ patches) with different thickness buckles under the gradually increased axial compression. Inside each shell, there is a rigid cylindrical mandrel. Its radius differs from that of the shell by $10\%$.
    }
    \label{fig:cancomp}
\end{figure}

\paragraph*{Cylindrical shell buckling.}
In Fig.~\ref{fig:can}, a hollow cylindrical shell severely buckles under the gradually increased compression is simulated with our method.  
When there is an inner mandrel, local buckles can no longer grow unrestrained because their radial displacements are arrested by the mandrel. Thereby, more buckles form and eventually accumulate into a diamond-like pattern over the entire cylinder as shown in Fig.~\ref{fig:cancomp}. The deformation mode is consistent with real-world experiments~\cite{Seffen2014}.

\begin{table*}[t]
  \centering
  \caption{Here we list the parameters and the time consumed in each example, including 
  the degrees of freedom, the total number of nonzero elements in the BHEM system matrix, 
  the total number of sampling points on the external collider, 
  the properties of the cloth (Young's Modulus $Y\,[\text{Pa}]$, Poisson's ratio $\nu$ and thickness $h\,[\text{mm}]$, mass density $\rho\,[10^3\text{kg}/\text{m}^3]$), 
  Rayleigh damping coefficient $\alpha$, 
  time step $\Delta t\,[\text{ms}]$,
  time consumed in one integration step $t_\mathrm{int}\,[\text{s}]$, in one CCD step $t_\mathrm{obj}\,[\text{s}]$, and whether we detect self-collision (SC, '--' means that CCD has not been performed in this example).}
  \label{tab:experiments}
  \begin{threeparttable}
  \begin{tabular}{c|c|c|c|c|c|c|c|c|c|c|c|c|c}
      \hline
      Figure& Example& \#DoFs & \#NNZs & \#SPs 
      & $Y$ & $\nu$ & $h$ & $\rho$ 
      & $\alpha$ & $\Delta t$ 
      & $t_\mathrm{int}$ & $t_\mathrm{obj}$ & SC\\
      \hline
      \ref{fig:wrinkle}& Wrinkled & -- & -- & -- & $1\times10^6$ & $0.5$ & $0.1$ & $0.93$ & -- & -- & $63.0$\tnote{\dag} & -- & --\\
      \ref{fig:drape}& Draped  & --& -- & -- & $1\times10^5$ & $0.3$ & $1$ & $0.93$ & $0$ & $10$ & $6.01$\tnote{\dag} & -- & --\\
      \ref{fig:drill}& Twisting & $1.15\times10^4$ & $1.19\times10^6$ & -- & $1\times10^4$ & $0.3$ & $1$ & $0.2$ & $50$ & $2$ & $0.20$ & -- & --\\
      \ref{fig:locking}& Locking & -- & -- & -- & $1\times10^4$ & $0.3$ & $1$ & $0.2$ & $2$ & $2$ & $3.38$\tnote{\ddag} & $4.37$ & yes\\
      \ref{fig:ball}& Ball & $1.15\times10^4$ & $1.17\times10^6$ & -- & $1\times10^5$ & $0.3$ & $0.1$ & $0.2$ & $2.5$ & $2$ & $4.80$ & $0.0023$ & no\\
      \ref{fig:armadillo}& Armadillo  & $1.15\times10^4$ & $1.19\times10^6$ & $2.8$k & $8.21\times10^5$ & $0.243$ & $0.32$ & $0.4726$ & $2.5$ & $1$ & $2.90$ & $1.84$ & no\\
      \ref{fig:needles}& Needles  & $1.15\times10^4$ & $1.19\times10^6$ & $3.6$k & $1\times10^4$ & $0.3$ & $0.1$ & $0.2$ & $5$ & $2$ & $5.05$ & $26.66$ & yes\\
      \ref{fig:umbrella}& Umbrella & $2.02\times10^4$ & $1.76\times10^6$ & -- & $1\times10^6$ & $0.5$ & $0.1$ & $0.93$ & $5$ & $2$ & $12.47$ & -- & --\\
      \ref{fig:can}& Can & $5.3\times10^3$ & $4.97\times10^5$ & -- & $1\times10^8$ & $0.3$ & $2$ & $2.7$ & $5$ & $2$ & $4.55$ & -- & --\\
      \ref{fig:cancomp}& Cans  & $1.15\times10^4$ & $1.13\times10^6$ & -- & $1\times10^9$ & $0.47$ & 
      -- & $1.4$ & $5$ & $0.04$ & $5.29$\tnote{$\ast$} & $0.0024$ & no\\
      \hline
    \end{tabular}
    \begin{tablenotes}
      \footnotesize
      \item[\dag] The given value is obtained by a $30\times30$-patch shell.
      \item[\ddag] The given value is obtained by a $20\times20$-patch shell.
      \item[$\ast$] The given value is obtained by a $\SI{1}{\milli\meter}$-thick shell.
    \end{tablenotes}
  \end{threeparttable}
\end{table*}

\section{Conclusions and Discussions}


In this study, we propose a new computational framework designed for elastodynamic simulation of parametric thin-shell structures. 
The central piece of our framework is a high-order finite element formulation equipped with an implicit Euler solver. This formulation, based on bicubic Hermite interpolation, naturally ensures conforming $\mathcal{C}^1$ continuity.
Capitalizing on the advancements in parametric surface modeling and rendering, we have crafted an intersection detection paradigm that is custom-tailored for bicubic Hermite surfaces. This unified approach empowers us to achieve high-fidelity CCD and rendering without resorting to any form of auxiliary tessellation mesh. 


Nevertheless, several challenges remain unresolved and are open to deeper investigation. We are going to list in the following text. 

\paragraph*{Loss functions.}
When solving the equation of motion by Newton's method, we have yet to account for the unit difference between displacement and its high-order derivatives in the setting of loss functions. We believe that considering this difference when determining the descent direction could potentially yield improved convergence.

\paragraph*{Plasticity.}
While Hermite interpolation excels in depicting complex characteristics like wrinkles or folds in cloth motion, it encounters difficulties when addressing plastic deformation and impact dynamics. The high-order formulation of Hermite interpolation exacerbates the inherent non-linearity and non-smoothness of these tasks, which need sophisticated modeling and analysis.

\paragraph*{Complex geometries.}
The BHEM proficiently manages globally parameterized surfaces but faces challenges with complex geometries. We are considering enhancements such as integrating triangular patches or implementing specialized data structures to bolster its geometric flexibility.
A mixture of BHEM and FEM patches with well-designed shape functions \cite{Schneider2019} may be also helpful.

\paragraph*{Collision handling.}
This paper only presents the algorithm of ray--patch intersection detection, which is insufficient to perform completely non-penetration CCD.
To address this issue, we intend to generalize our algorithm to patch--patch intersection detection.
Moreover, as our current collision-resolving scheme struggles with intricate contact scenarios like tying ribbons into a reef knot, we view the integration of Incremental Potential Contact (IPC) as a natural and significant follow-up step. 
\newcommand{\etalchar}[1]{$^{#1}$}


\appendix

\section{Mathematical tools}

\subsection{Bicubic Hermite Interpolation}
\label{apx:hermite}

As written in Eq.~(\ref{eqn:hermite_weight}), bicubic Hermite interpolation indicates using a cubic basis function in each dimension of the parameter space.
To be specific, $w_{p,r}$ is defined as follows:
\begin{subnumcases}{w_{p,r}(\theta)=}
  f(\theta)\text{,}&$p=0\land r=0$,\\
  f(1-\theta)\text{,}&$p=1\land r=0$,\\
  g(\theta)\text{,}&$p=0\land r=1$,\\
  -g(1-\theta)\text{,}&$p=1\land r=1$,
\end{subnumcases}
where $f(\theta)=2\theta^3-3\theta^2+1$ and $g(\theta)=\theta^3-2\theta^2+\theta$ hold. $w_{q,s}$ is defined similarly by replacing indices.

Furthermore, taking the derivative of Eq.~(\ref{eqn:hermite}) yields
\begin{align}
  \bm{a}_1(\xi^1,\xi^2)&=\sum_{p,q,r,s\in\{0,1\}}\frac{1}{\Delta\xi^1}w'_{p,r}(\theta^1)\,w_{q,s}(\theta^2)\,\bm{x}_{pq,rs}\text{,}\\
  \bm{a}_2(\xi^1,\xi^2)&=\sum_{p,q,r,s\in\{0,1\}}\frac{1}{\Delta\xi^2}w_{p,r}(\theta^1)\,w'_{q,s}(\theta^2)\,\bm{x}_{pq,rs}\text{,}
\end{align}
where
\begin{subnumcases}{w'_{p,r}(\theta)=}
  f'(\theta)\text{,}&$p=0\land r=0$,\\
  -f'(1-\theta)\text{,}&$p=1\land r=0$,\\
  g'(\theta)\text{,}&$p=0\land r=1$,\\
  g'(1-\theta)\text{,}&$p=1\land r=1$,
\end{subnumcases}
with $f'(\theta)=6\theta^2-6\theta$ and $g'(\theta)=3\theta^2-4\theta+1$ holding. $w'_{q,s}(\theta)$ can be still formalized by replacing indices.

It is clear that all the following equations hold: $f(0)=1$, $f(1)=0$, $g(0)=0$, $g(1)=0$, $f'(0)=0$, $f'(1)=0$, $g'(0)=1$, and $g'(1)=0$, which means that the values of $\bm{x}$, $\bm{a}_1$, and $\bm{a}_2$ on a common edge are merely related to the sampled values at the two end nodes of the edge. This implies $\mathcal{C}^1$-smoothness on the whole surface --- the interpolation function itself and its first-order partial derivatives remain continuous across cells, while the second-order partial derivatives come to discontinuity of first kind only at the common edges of the cells.

\subsection{Gauss--Legendre Quadrature}
\label{apx:quadrature}

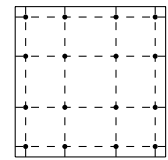
\begin{figure}[ht]
\centering
\begin{tikzpicture}[scale=0.5]
    \draw[step=4] (0,0) grid (4,4);
    \tikzmath{\xa=0.339981;};
    \tikzmath{\xb=0.861136;};
    \tikzmath{\ta=2-\xb*2;};
    \tikzmath{\tb=2-\xa*2;};
    \tikzmath{\tc=2+\xa*2;};
    \tikzmath{\td=2+\xb*2;};
    \foreach \x in {\ta,\tb,\tc,\td}{
      \foreach \y in {\ta,\tb,\tc,\td}{
        \fill (\x,\y) circle (0.06667);
      }
      \draw[dashed] (\x,0)--(\x,4);
      \draw[dashed] (0,\x)--(4,\x);
    }
\end{tikzpicture}
\caption{Quadrature points in a cell.}
\label{fig:quadrature}
\vspace{-0.3cm}
\end{figure}

Any two-dimensional integral that takes the form of
\begin{equation}
  I=
  \int_{\xi_\mathrm{min}^2}^{\xi_\mathrm{max}^2}
  \int_{\xi_\mathrm{min}^1}^{\xi_\mathrm{max}^1}
  f(\xi^1,\xi^2)\,\mathrm{d}\xi^1\mathrm{d}\xi^2\text{,}
\end{equation}
can be computed approximately using the summation of $n$ sampled points that
\begin{equation}
  I\approx\sum_{i=1}^{n}w_if(\xi^1_i,\xi^2_i)\text{.}
\end{equation}
We pick $n=16$ in our framework, with the quadrature points arranged by $4\times4$ in a single cell. As shown in Fig.~\ref{fig:quadrature}, the vertical dashed lines correspond to
\begin{equation}
  \theta^1=\frac{\xi_\mathrm{min}^1+\xi_\mathrm{max}^1}{2}\pm\frac{\Delta\xi^1}{2}\sqrt{\frac{3}{7}\pm\frac{2}{7}\sqrt{\frac{6}{5}}}\text{,}
\end{equation}
respectively, and the horizontal dashed lines correspond to 
\begin{equation}
  \theta^2=\frac{\xi_\mathrm{min}^2+\xi_\mathrm{max}^2}{2}\pm\frac{\Delta\xi^2}{2}\sqrt{\frac{3}{7}\pm\frac{2}{7}\sqrt{\frac{6}{5}}}\text{,}
\end{equation}
respectively. The inside vertical lines equip a weight factor of $(18+\sqrt{30})\Delta\xi^1/72$, while the outside ones equip a weight factor of $(18-\sqrt{30})\Delta\xi^1/72$. The inside horizontal lines equip a weight factor of $(18+\sqrt{30})\Delta\xi^2/72$, while the outside ones equip a weight factor of $(18-\sqrt{30})\Delta\xi^2/72$. The final weight $w_i$ of each point is the product of the two weights of the vertical and horizontal lines it lies on.



\section{Derivatives of the Elastic Energy}
\label{apx:derivative}

We bypass the medium of deformation gradients and compute the first- and second-order partial derivatives of $V_\mathrm{e}$ w.r.t. nodal degrees of freedom $\{\bm{q}_I\}$ directly.

\subsection{The First-Order Derivatives (Force)}

With the total elastic potential energy defined by $V_\mathrm{e}=\iint_{\bar{\Omega}}\bar{\mathcal{V}_\mathrm{e}}\,\mathrm{d}\bar{\Omega}$,
according to Eq.~(\ref{eqn:deltave}), the first-order partial derivatives of the elastic energy are given by
\begin{align}
  \frac{\partial V_\mathrm{e}}{\partial \bm{q}_I}
  =&\iint_{\bar{\Omega}}\frac{\partial\bar{\mathcal{V}}_\mathrm{e}}{\partial\bm{q}_I}\,\mathrm{d}\bar{\Omega}\notag\\
  =&\iint_{\omega}\frac{\partial\bar{\mathcal{V}}_\mathrm{e}}{\partial\bm{q}_I}\sqrt{\bar{a}}\,\mathrm{d}\xi^1\mathrm{d}\xi^2\notag\\
  =&\iint_{\omega}\left(\tau\frac{\partial a_{\alpha\beta}}{\partial\bm{q}_I}A_{\gamma\delta}+\frac{1}{3}\tau^3\frac{\partial b_{\alpha\beta}}{\partial\bm{q}_I}B_{\gamma\delta}\right)\bar{H}^{\alpha\beta\gamma\delta}\sqrt{\bar{a}}\,\mathrm{d}\xi^1\mathrm{d}\xi^2\text{,}\label{eqn:intdeltave}
\end{align}
with partial derivatives of $a_{\alpha\beta}$ and $b_{\alpha\beta}$ are calculated by
\begin{align}
   \label{eqn:paabpqi}
      \frac{\partial a_{\alpha\beta}}{\partial \bm{q}_I}= &\varPhi^I_{,\alpha}\bm{a}_\beta+\varPhi^I_{,\beta}\bm{a}_\alpha\text{,}\\
      \label{eqn:pbabpqi}
      \frac{\partial b_{\alpha\beta}}{\partial \bm{q}_I}= &\varPhi^I_{,\alpha\beta}\bm{a}_3+\frac{1}{\sqrt{a}}\left(\varPhi^I_{,1}\bm{a}_2\times\bm{a}_{\alpha,\beta}+\varPhi^I_{,2}\bm{a}_{\alpha,\beta}\times\bm{a}_1\right. \notag\\ 
       &-\left.\bm{a}_{\alpha,\beta}\cdot\bm{a}_3(\varPhi^I_{,1}\bm{a}_2\times\bm{a}_3+\varPhi^I_{,2}\bm{a}_3\times\bm{a}_1)\right)\text{.}
\end{align}

In order to facilitate numerical calculation, we rewrite Eq.~(\ref{eqn:intdeltave}) in a matrix form using \emph{Voigt notation} as
\begin{equation}
  \label{eqn:matrix}
  \frac{\partial V_\mathrm{e}}{\partial \bm{q}_I}=
  \iint_{\omega}\left(\tau\left(\frac{\partial\bm{\alpha}}{\partial\bm{q}_I}\right)^\mathrm{T}\bm{H}\bm{\alpha}+\frac{1}{12}\tau^3\left(\frac{\partial\bm{\beta}}{\partial\bm{q}_I}\right)^\mathrm{T}\bm{H}\bm{\beta}\right)\sqrt{\bar{a}}\,\mathrm{d}\xi^1\mathrm{d}\xi^2\text{,}
\end{equation}
where $\bm{H}=(H^{ij})_{3\times3}$ is a square matrix, and $\bm{\alpha}=(\alpha_i)_{3\times1}$ and $\bm{\beta}=(\beta_i)_{3\times1}$ are column vectors.
It should be noted that a mixed layout is used here --- first-order partial derivatives are always written as column vectors. Considering the symmetry of the quantities, the matrix and vectors in this integral can be written as
\begin{gather}
  \bm{H}=
  {\small\begin{pmatrix}
    (\lambda+2\mu)(\bar{a}^{11})^2 & \lambda\bar{a}^{11}\bar{a}^{22}+2\mu(\bar{a}^{12})^2 & (\lambda+2\mu)\bar{a}^{11}\bar{a}^{12}\\
    & (\lambda+2\mu)(\bar{a}^{22})^2 & (\lambda+2\mu)\bar{a}^{12}\bar{a}^{22}\\
    \text{sym.} & & (\lambda+\mu)(\bar{a}^{12})^2+\mu\bar{a}^{11}\bar{a}^{22}
  \end{pmatrix}}%
  \text{,}\\
  \bm{\alpha}=
  \begin{pmatrix}
    A_{11} & A_{22} & 2A_{12}
  \end{pmatrix}^\mathrm{T}
  \text{,}\\
  \bm{\beta}=2
  \begin{pmatrix}
    B_{11} & B_{22}& 2B_{12}
  \end{pmatrix}^\mathrm{T}
  \text{,}\\
  \frac{\partial\bm{\alpha}}{\partial\bm{q}_I}=
  \frac{1}{2}
  \begin{pmatrix}
    \frac{\partial a_{11}}{\partial \bm{q}_I}
    & \frac{\partial a_{22}}{\partial \bm{q}_I}
    & 2\frac{\partial a_{12}}{\partial \bm{q}_I}
  \end{pmatrix}^\mathrm{T}
  \text{,}\\
  \frac{\partial\bm{\beta}}{\partial\bm{q}_I}=
  \begin{pmatrix}
    \frac{\partial b_{11}}{\partial \bm{q}_I}
    & \frac{\partial b_{22}}{\partial \bm{q}_I}
    & 2\frac{\partial b_{12}}{\partial \bm{q}_I}
  \end{pmatrix}^\mathrm{T}
  \text{.}
\end{gather}

\subsection{The Second-Order Derivatives (Hessian)}

The Hessian matrix is given by
\begin{align}
    \frac{\partial^2V_\mathrm{e}}{\partial\bm{q}_J\partial\bm{q}_I}&=\iint_{\bar{\Omega}}\frac{\partial^2\bar{\mathcal{V}}_\mathrm{e}}{\partial\bm{q}_J\partial\bm{q}_I}\,\mathrm{d}\bar{\Omega}\notag\\
  &=\iint_{\omega}\left(\tau\,\bm{G}_1+\frac{1}{12}\tau^3\,\bm{G}_2\right)\sqrt{\bar{a}}\,\mathrm{d}\xi^1\mathrm{d}\xi^2\text{,}
\end{align}
in which the following equations hold:
\begin{align}
  \bm{G}_1&=\left(\frac{\partial\bm{\alpha}}{\partial\bm{q}_I}\right)^\mathrm{T}\bm{H}\left(\frac{\partial\bm{\alpha}}{\partial\bm{q}_J}\right)+\sum_{i=1}^3\sum_{j=1}^3\frac{\partial^2\alpha_i}{\partial\bm{q}_J\partial\bm{q}_I}H^{ij}\alpha_j\text{,}\\
  \bm{G}_2&=\left(\frac{\partial\bm{\beta}}{\partial\bm{q}_I}\right)^\mathrm{T}\bm{H}\left(\frac{\partial\bm{\beta}}{\partial\bm{q}_J}\right)+\sum_{i=1}^3\sum_{j=1}^3\frac{\partial^2\beta_i}{\partial\bm{q}_J\partial\bm{q}_I}H^{ij}\beta_j\text{.}
\end{align}
Therefore, the remaining work is to calculate 
$\partial^2\alpha_i/\partial\bm{q}_J\partial\bm{q}_I$ and $\partial^2\beta_i/\partial\bm{q}_J\partial\bm{q}_I$.
The concrete form of the former can be easily deduced as
\begin{align}
  \frac{\partial^2\alpha_1}{\partial\bm{q}_J\partial\bm{q}_I}&=\varPhi^I_{,1}\varPhi^J_{,1}\bm{I}\text{,}\\
  \frac{\partial^2\alpha_2}{\partial\bm{q}_J\partial\bm{q}_I}&=\varPhi^I_{,2}\varPhi^J_{,2}\bm{I}\text{,}\\
  \frac{\partial^2\alpha_3}{\partial\bm{q}_J\partial\bm{q}_I}&=\left(\varPhi^I_{,1}\varPhi^J_{,2}+\varPhi^I_{,2}\varPhi^J_{,1}\right)\bm{I}\text{,}
\end{align}
while it takes some time to compute the latter, which satisfies
\begin{align}
  2\frac{\partial^2B_{\alpha\beta}}{\partial\bm{q}_J\partial\bm{q}_I}=&\frac{\partial^2b_{\alpha\beta}}{\partial\bm{q}_J\partial\bm{q}_I}\notag\\
  =&\varPhi^I_{,\alpha\beta}\frac{\partial\bm{a}_3}{\partial\bm{q}_J}+\varPhi^J_{,\alpha\beta}\left(\frac{\partial\bm{a}_3}{\partial\bm{q}_I}\right)^\mathrm{T}+\varPhi^I_{,1}\varPhi^J_{,1}\bm{D}^{11}\notag\\
  &+\varPhi^I_{,2}\varPhi^J_{,2}\bm{D}^{22}+\varPhi^I_{,1}\varPhi^J_{,2}\bm{D}^{12}+\varPhi^I_{,2}\varPhi^J_{,1}\bm{D}^{21}\text{,}
\end{align}
in which coefficients of the first-order terms are
\begin{align}
  \frac{\partial\bm{a}_3}{\partial\bm{q}_J}&=\frac{1}{\sqrt{a}}\left(-\varPhi^J_{,1}[\bm{a}_2]+\varPhi^J_{,2}[\bm{a}_1]-\bm{a}_3\otimes(\varPhi^J_{,1}\bm{t}_1+\varPhi^J_{,2}\bm{t}_2)\right)\text{,}\\
  \frac{\partial\bm{a}_3}{\partial\bm{q}_I}&=\frac{1}{\sqrt{a}}\left(-\varPhi^I_{,1}[\bm{a}_2]+\varPhi^I_{,2}[\bm{a}_1]-\bm{a}_3\otimes(\varPhi^I_{,1}\bm{t}_1+\varPhi^I_{,2}\bm{t}_2)\right)\text{,}
\end{align}
and coefficients of the second-order terms are respectively calculated by dot products of $\bm{a}_{\alpha,\beta}$ with $\partial^2\bm{a}_3/{\partial\bm{a}_1}^2$, $\partial^2\bm{a}_3/{\partial\bm{a}_2}^2$, ${\partial^2\bm{a}_3}/{\partial\bm{a}_2\partial\bm{a}_1}$, and ${\partial^2\bm{a}_3}/{\partial\bm{a}_1\partial\bm{a}_2}$.
For calculation, we provide relatively simple formulae as follows:
\begin{align}
  \bm{D}^{11}=&\frac{1}{a}\left(\beta_{\alpha\beta}(3\bm{t}_1\otimes\bm{t}_1+\bm{a}_2\otimes\bm{a}_2-a_{22}\bm{I})\right.\notag\\
  &\left.-\bm{s}_2\otimes\bm{t}_1-\bm{t}_1\otimes\bm{s}_2\right)\text{,}\\
  \bm{D}^{22}=&\frac{1}{a}\left(\beta_{\alpha\beta}(3\bm{t}_2\otimes\bm{t}_2+\bm{a}_1\otimes\bm{a}_1-a_{11}\bm{I})\right.\notag\\
  &\left.-\bm{s}_1\otimes\bm{t}_2-\bm{t}_2\otimes\bm{s}_1\right)\text{,}\\
  \bm{D}^{12}=&\frac{1}{a}\left(\beta_{\alpha\beta}(3\bm{t}_1\otimes\bm{t}_2-2\bm{a}_1\otimes\bm{a}_2+\bm{a}_2\otimes\bm{a}_1+a_{12}\bm{I})\right.\notag\\
  &\left.-\bm{s}_2\otimes\bm{t}_2-\bm{t}_1\otimes\bm{s}_1-\sqrt{a}[\bm{a}_{\alpha,\beta}]\right)\text{,}\\
  \bm{D}^{21}=&\frac{1}{a}\left(\beta_{\alpha\beta}(3\bm{t}_2\otimes\bm{t}_1-2\bm{a}_2\otimes\bm{a}_1+\bm{a}_1\otimes\bm{a}_2+a_{12}\bm{I})\right.\notag\\
  &\left.-\bm{s}_1\otimes\bm{t}_1-\bm{t}_2\otimes\bm{s}_2+\sqrt{a}[\bm{a}_{\alpha,\beta}]\right)\text{,}
\end{align}
where $t_1$, $t_2$, $s_1$, and $s_2$ are defined as
\begin{align}
  \bm{t}_1&=\bm{a}_2\times\bm{a}_3\text{,}\\
  \bm{t}_2&=\bm{a}_3\times\bm{a}_1\text{,}\\
  \bm{s}_1&=\bm{a}_{\alpha,\beta}\times\bm{a}_1\text{,}\\
  \bm{s}_2&=\bm{a}_2\times\bm{a}_{\alpha,\beta}\text{.}
\end{align}
Here we use $[\cdot]$ to represent the cross product matrix of a vector, which is defined as
\begin{equation}
  \left[\begin{pmatrix}v_1\\v_2\\v_3\end{pmatrix}\right]=
  \begin{pmatrix}
    0&-v_3&v_2\\
    v_3&0&-v_1\\
    -v_2&v_1&0
  \end{pmatrix}
  \text{.}
\end{equation}

\paragraph*{The pseudo Hessian.}
In the calculation of the Hessian matrix, the term $\partial^2\beta_i/\partial\bm{q}_J\partial\bm{q}_I$ takes up the most time.
Thanks to the low magnitude of this term, we can subtract it to construct an inexact Hessian matrix.
Although this compromise causes the number of Newton steps to increase, it can reduce the time consumption per step.

\end{document}